\documentclass[fleqn,onecolumn,usenatbib]{mnras}
\usepackage[T1]{fontenc}
\usepackage{ae,aecompl}

\usepackage{graphicx}	% Including figure files
\usepackage{amsmath}	% Advanced maths commands
\usepackage{amssymb}	% Extra maths symbols

\usepackage{deluxetable}

\renewcommand {\deg}   {\mbox{$^\circ$}}

\newcommand   {\kms}   {\mbox{km\,s$^{-1}$}}
\renewcommand {\ga}    {\mbox{\rlap{\hbox{\lower5pt\hbox{$\sim$}}}\hbox{$>$}}}
\renewcommand {\la}    {\mbox{\rlap{\hbox{\lower5pt\hbox{$\sim$}}}\hbox{$<$}}}

\title[Stellar Counterparts]{The population of Galactic centre filaments III:\\ 
candidate radio and stellar sources}

\author[Yusef-Zadeh, R. Arendt,   M. Wardle]{
F. Yusef-Zadeh$^1$\thanks{E-mail: zadeh@northwestern.edu}, 
 R. G. Arendt$^{2}$,  M. Wardle$^3$, I. Heywood$^4$, \& W. Cotton$^5$ 
\\
$^{1}$ Department of Physics and Astronomy Northwestern University, Evanston, IL 60208\\
$^2$Code 665, NASA/GSFC, 8800 Greenbelt Road, Greenbelt, MD 20771, USA \& UMBC/CRESST 2\\
$^{3}$School of Mathematical and Physical Sciences,  Research Centre for Astronomy, Astrophysics\\
and Astrophotonics, Macquarie University, Sydney NSW 2109, Australia\\
$^{4}$Royal Astronomical Society, Burlington House, Piccadilly, London W1J 0BQ, UK\\
and Department of Physics and Electronics, Rhodes University, PO Box 94, Makhanda, 6140, South Africa\\
$^5$ National Radio Astronomy Observatory, Charlottesville, VA, USA}

\date{Accepted XXX. Received YYY; in original form ZZZ}

\pubyear{2019}

\begin{document}
\label{firstpage}
\pagerange{\pageref{firstpage}--\pageref{lastpage}}
\maketitle
%%%%%%%%%%%%%%%%%%%%%%%%%%%%%%%%%%%%%%%%%%%%%%%%%%%%%%%%
\def\msol{\hbox{$\hbox{M}_\odot$}}
\def\lsol{\hbox{$\hbox{L}_\odot$}}
\def\kms{km s$^{-1}$}
\def\Blos{B$_{\rm los}$}
\def\etal   {{\it et al.}}                     % et al
\def\psec           {$.\negthinspace^{s}$}
\def\pasec          {$.\negthinspace^{\prime\prime}$}
\def\pdeg           {$.\kern-.25em ^{^\circ}$}
\def\degree{\ifmmode{^\circ} \else{$^\circ$}\fi}
\def\ut #1 #2 { \, \textrm{#1}^{#2}} % \ut unit p  unit^p 
\def\u #1 { \, \textrm{#1}}          % \u unit     unit
\def\nH {n_\mathrm{H}}
\def\ddeg   {\hbox{$.\!\!^\circ$}}              % Degrees over dot
\def\deg    {$^{\circ}$}                        % Degrees symbol
\def\le     {$\leq$}                            % <=
\def\sec    {$^{\rm s}$}                        % Second of time
\def\msol   {\hbox{$M_\odot$}}                  % Solar mass
\def\i      {\hbox{\it I}}                      % italic I
\def\v      {\hbox{\it V}}                      % italic V
\def\dasec  {\hbox{$.\!\!^{\prime\prime}$}}     % Arcseconds over dot
\def\asec   {$^{\prime\prime}$}                 % Arcseconds symbol
\def\dasec  {\hbox{$.\!\!^{\prime\prime}$}}     % Arcseconds over dot
\def\dsec   {\hbox{$.\!\!^{\rm s}$}}            % Second over dot
\def\min    {$^{\rm m}$}                        % Minutes of time
\def\hour   {$^{\rm h}$}                        % Hours of time
\def\amin   {$^{\prime}$}                       % Arcminutes symbol
\def\lsol{\, \hbox{$\hbox{L}_\odot$}}
\def\sec    {$^{\rm s}$}                        % Second of time     
\def\etal   {{\it et al.}}                     % et al.
\def\la{\lower.4ex\hbox{$\;\buildrel <\over{\scriptstyle\sim}\;$}}
\def\ga{\lower.4ex\hbox{$\;\buildrel >\over{\scriptstyle\sim}\;$}}

\begin{abstract} 
Recent MeerKAT radio continuum observations of the Galactic center at 20 cm show  a large population of nonthermal radio filaments (NRFs) in 
the inner few hundred pc of the Galaxy. We have selected a sample of 57  radio 
sources, mainly compact objects,   in the MeerKAT mosaic image that appear to be associated with NRFs. 
The selected sources are  about 4 times  the number of radio point sources associated with filaments than would be expected by random chance.
Furthermore, an apparent  correlation between bright IR stars and NRFs is inferred from their similar latitude distributions, suggesting that they both 
co-exist  within the 
same region. To examine if compact radio sources are related to compact IR sources, we have used archival 2MASS, and {\em Spitzer} data to make 
spectral energy distribution of individual stellar sources coincident or close to  radio sources. 
We provide a catalogue of radio and IR sources for  future   detailed 
observations to investigate a potential 3-way physical association between NRFs, compact radio and IR stellar sources.
 This association 
is suggested by models in which NRFs are cometary tails produced by the interaction of a  large-scale nuclear outflow  with stellar wind bubbles  in the 
Galactic center. 
\end{abstract}

%The cosmic ray ionization rate in the Galactic center is two to three orders of magnitudes higher than in the 
%Galactic disk, implying that the cosmic ray pressure is sufficient to launch a large-scale nuclear wind from this region. 

%{\em Spitzer}, {\em Chandra}, {\em XMM}) 

%Our analysis suggests that the associated stellar candidates are typically dusty red giants, with strong 24 $\mu$m emission.
%that extend above and below the Galactic plane, suggestive of an outflow that is the relic of past activity. The spatial distributions of the bipolar 
%radio/X-ray structure and the filaments suggest a causal connection between the two.  The high cosmic ray pressure, known from multiple measurements, 
%drives a large-scale wind away from the Galactic plane creating the bipolar emission. The interaction of the wind with obstacles, such as stellar wind 
%bubbles, creates magnetized cometary tails that may explain the unusual characteristics of NRFs, a subject that we are focusing in this proposal.

%The origin  of the Galactic center NRFs, which  have only been detected at radio and X-rays.  
 
\keywords{ISM: magnetic fields,  ISM:  cosmic rays, radiation mechanisms: non-thermal, plasmas}

%%%%%%%%%%%%%%%%%%%%%%%%%%%%%%%%%%%%%%%%%%%%%%%%%%

%%%%%%%%%%%%%%%%% BODY OF PAPER %%%%%%%%%%%%%%%%%%
\section{Introduction} 

The inner thousand light years of our Galaxy is unusually active compared to the solar neighborhood. The  dormant supermassive black hole, Sgr A*,  
is  surrounded by hot young stars with powerful winds that irradiate the nearby interstellar gas. There is circumstantial evidence of two past 
explosions 
within the last few million years. First, vast $\gamma$-ray emitting lobes filled with relativistic particles emerge from the Galactic center and extend 
above and below the disk for many thousands of light years (i.e., the Fermi bubble)  \citep{su10}. Second, a bipolar MeerKAT
radio bubble filled by coronal X-ray emitting hot gas emerges from the inner few hundred light years of the Galaxy. 
The radio bubble hosts hundreds of mysterious magnetized radio filaments. The filaments occupy a region  consistent 
with the  radio bubble, suggesting a causal association between the two \citep{heywood19}. 
Theoretical models explain the Fermi 
and MeerKAT bubbles by invoking a correlated burst of supernovae 
or an explosion from the black hole 
as it swallows a cloud \citep{crocker11,yang13}.\\

The high cosmic-ray flux in the Galactic center suggests global injection of relativistic particles produced as a result of a relic of a past activity by Sgr A*,
due to starburst activity producing multiple
supernova explosions \citep{heywood19}.
One scenario for the origin of the bipolar radio/X-ray features considers high cosmic ray pressure driving large-scale winds and expanding the medium 
away from the Galactic plane. It has become apparent 
that relativistic particles permeate the central molecular zone (CMZ) at levels a thousand times that in the solar neighborhood 
\citep{geballe99,oka05,indriolo12,goto14,lepetit16,oka19}. 
This  provides  a significant source of pressure when compared to thermal  pressure in interstellar space at the 
Galactic center. High cosmic-ray pressure drives winds in the nuclei of galaxies \citep{everett08,everett10, ruskowski17,zweibel17} and play a 
role in feedback, limiting star formation and the growth of the central supermassive black holes by transferring their momentum and energy into the 
surrounding medium. 
The high cosmic-ray flux in the Galactic center suggests global injection of relativistic particles produced as a result of a relic of a past activity by Sgr A*,
due to star-burst activity producing multiple supernovae explosions \citep{heywood19}.
The interaction of a cosmic-ray driven wind with stellar wind bubbles create cometary tails that could explain the energetic 
nonthermal radio filaments found throughout the Galactic center 
\citep{shore99,bicknell01,zadeh19,zadeh22a}. 
In this scenario, a compact radio source is expected to be located  along or at one end of NRFs where the  wind 
interacts with mass-losing stars.\\ 

Another model suggests a time-dependent injecting source,  such as a pulsar,  
crossing spatially intermittent magnetic bundles  and generates filamentary structure \citep{thomas20}. 
This model predicts the filaments run  perpendicular 
to the direction of the motion, unlike cometary tail model in which  compact sources inject particles 
in the direction opposite  to  their  motion \citep{shore99,zadeh19}.\\ 

Motivated by these scenarios,  we provide a list of candidate radio and stellar sources that appear to be associated with NRFs. Candidate radio 
sources are selected only from 
43 groupings of 174 magnetized  filaments that run parallel to each other and are separated from each other by mean spacing of 
$16''$ (0.64pc) \citep{zadeh22b}.
We selected  the  most spectacular  groupings of  filaments ranging from pairs to dozens in \citep{zadeh22a}.  We defined  a group of filaments or a grouping
having  similar  orientations, similar curvature or bending,  spatially close to each other,  and in some cases,
converge to a point,  shift sideways together,
and change direction coherently. These characteristics
implied  that they are parts of the same system of filaments with  a common  origin. We avoided groups of filaments close to the Galactic plane because of confusing thermal 
sources in this region.
The groups of filaments we selected are distinct from single  filaments along the line of sight
at large physical distances from each other but which appear close in projection. The analysis of single filaments,  which may comprise of multiple filaments in future 
higher resolution observations,  are postponed elsewhere.
We present radio 
and infrared images of each of the 57 candidate sources as well as the  spectral energy distributions (SEDs) of their IR counterparts.  The purpose of 
this list is to motivate 
further high sensitivity, spatial resolution radio and IR observations to examine a 3-way spatial correlation and to test the cometary tail model of the 
origin of the filaments.

\section{MeerKAT and VLA Observations} 

Details of MeerKAT and the Karl G. Jansky Very Large Array (VLA) observations are found in \cite{heywood19,heywood22}. 
Figure 1 shows a mosaic MeerKAT image of the inner few degrees of the Galactic center  (1$'$ corresponds to 2.4 pc at the Galactic center distance) 
with a FWHM $\sim 4''$ resolution. VLA obtained higher resolution images of the region toward two fields centered on Sgr C and Sgr A. 
Thus, only few sources  have been covered with these limited high-resolution measurements. 
VLA observation was carried out at L-band (1–2 GHz) with the array in the most extended A-configuration and centered at Sgr C (J2000 17$^{\rm 
h}$44$^{\rm m}$35$\fs$0, $-29^\circ 29'\,00\farcs0$). The initial flagging and reference calibration was performed using the VLA casa pipeline8 and 
processed with the wsclean multi-scale clean algorithm with a resolution of $\sim1''$. Further details on the use of the VLA data can be found in 
\cite{heywood22}.

\section{Results and Discussion} 

\subsection{Stellar and NRF distributions} 

To examine the possible association of NRFs and stellar sources on a much larger scale, Figure  2a,b display 
the histograms  of magnetized  filaments longer than 66$''$ 
and  of 
bright IR stars \citep{ramirez08},  
as traced in 4  {\em Spitzer}  bands,
 as a function of Galactic latitudes. Bright stars are selected mainly to avoid   confusion limits and minimize extinction. 
A length of 66$''$ is selected because it is a good constraint for exclusion of thermal filamentary features vs. NRFs, \citep{zadeh22a}.  
The  similar scale heights of the filaments and stellar sources indicate their Galactic population
is suggestive  of a  spatial correlation 
between the two. In addition, both  distributions tend to peak at  slightly negative latitudes at 24$\mu$m, suggesting that  both populations are possibly associated 
with each other. 
However, robust statistical tests, e.g. the two-sample KS test, simply confirm that there are real differences between the distributions in Figures 2a and 2b.
That's no surprise given that some fraction of the stars will be "local" while all of the NRFs are at the Galactic center. 
To investigate further this weak correlation further, we will next  examine the spatial relationship  between compact radio  and IR stellar sources 
as well as the SEDs of IR sources closest to the peaks of compact radio sources.

%Past studies have not shown a clear evidence of a   physical association between individual radio filaments and other Galactic center 
%objects. This is mainly because of the  narrow widths of the filaments with small deviations from straight line. 
%If there is a true correlation on such a large scale, the 
%it suggests that stellar sources interact with NRFs. 
%the nuclear wind is interacting with stellar sources and 
%is responsible for  formation of  a  cometary tail behind a moving stellar source.  

\subsection{Compact radio and stellar sources} 

We have identified a sample of small resolved radio sources and unresolved compact  radio sources in the MeerKAT image of the Galactic center that appear to coincide 
with one end of or along 
NRFs. Examination of modified images in which  all radio sources are artificially shifted with respect to the filaments\footnote{Tests were done by shifting point source 
images by $\pm 64$ pixels ($69''$) in $l$ and $b$ (4 independent tests), and then recounting the number of apparent associations between the shifted sources and the NRFs.},  
indicates that only $\sim$1/4 of 
the apparent correlations between NRFs and compact sources can be attributed to random coincidence.
A single filament with a length of 50pc and width of 0.1pc has a volume of $\sim$0.5  pc$^3$
 whereas the large-scale volume of the central molecular zone (CMZ) 
is roughly  $\sim10^7$  pc$^3$, giving a  small filling factor  $\sim10^{-3}$ for NRFs.  Given the large stellar density of $\sim10^6$ IR  sources 
in the CMZ \citep{ramirez08}, the average number of stellar sources  within the volume of a filament is expected to be about one. 
In addition, there are many striking examples 
of a single filament splitting into two prongs at a junction where a resolved compact radio source is located, suggestive of a flow of plasma 
along the 
filaments \citep{zadeh22b}.  
So, it is possible that a source,  such as a mass-losing star, is responsible for injecting cosmic-ray particles into 
the filaments  or acting as an obstacle  redirecting plasma flow \citep{zadeh19,zadeh22a}.
In this picture mass-losing stars such as  the red giants are impacted  by a cosmic-ray driven wind, creating 
shocks at the wind-wind interface, and pushing the shocked stellar wind into the tail, thus, providing  additional acceleration of cosmic ray particles \citep{zadeh19}.

%The contribution of foreground red giants may  not be  significant for chance alignment, thus suggesting  that 
%a physical  interaction between the two populations is possible.   

Not all  filaments show a compact radio  and infrared candidate source at their  ends, so the cometary model may not be viable to explain their origin. 
It is possible that this subclass of  filaments could be explained by an alternative model in which a compact radio source,  like a pulsar crossing the filaments, injects   
cosmic ray  particles intermittently. In this picture, no IR  counterpart  is expected. A more detailed study of radio compact sources near this subclass of 
filaments need to 
be  done in the future to test this model. \\ 

%Some of the filaments can be as 
%long as 50 pc in length with typical width 0.1-0.2 pc. The long filamentary structures 
%occupy significant 
%volume of the ISM and will encounter mass-losing stars in a densely populated region of the inner Galaxy. 

To examine whether   compact radio  sources have stellar counterparts, 
we have used archival 2MASS and {\em Spitzer} (IRAC and MIPS), data to examine the spectral 
energy distribution 
(SED) of individual stellar sources that lie closest to the  compact radio sources. 
We have also checked the Gaia catalog and found  14 of the GALCEN sources having  Gaia counterparts within $2''$,  12 of which are within $1''$.
These sources are likely foreground sources. In some cases, there may be coincidences of unrelated sources. 
Table 1  tabulates the source number, the positions of 
radio and infrared sources in  
Galactic coordinates $(l,b)$, the Gaia source offset, 
the separation between the peak radio and infrared sources
and the peak intensity of the compact radio sources in columns 1 to 7, respectively.  The mosaic image was blanked below 10 $\mu$Jy beam$^{-1}$ \citep{heywood22}. 
The last column shows the associated 
groups of filaments named  in \cite{zadeh22b}.\\

We present the SEDs of 
the nearest  
GALCEN catalogue \citep{ramirez08} IR source to each of the
 radio sources, as listed in Table 1.
Figure 3 shows a close up view of all 57 compact sources at radio and IR and 1.25 - 24 
$\mu$m (2MASS+IRAC+MIPS) SEDs of stellar sources. 
In spite of the lack of full wavelength coverage, Figure~3 shows a sample illustrating two  different classes of SEDs that may be IR counterparts to
compact radio sources. 
In one class, 'Red Giant SEDs' show distinct dip at 4.5 $\mu$m (see Fig. 3j, t, v,  oo and  ww). The dip is caused by the CO fundamental, 
which 
strengthens with
decreasing temperature \citep{engelke06}. These sources are not detected at 24 $\mu$m. 
Other sources do exhibit 24 $\mu$m emission. A subset of these are visibly extended
in the images and have very red SEDs (Fig. 3u, y, cc). Thus these are
likely young stellar objects and H II regions. The remainder (Fig. 3b, c, bb,vv)
are point-like and
typically brighter than  those sources with red giant SEDs 
These stars may be dusty AGB stars or supergiants \citep{reiter15}. 
It is possible that all
red giants have 24 $\mu$m emission, but that it is only sufficiently bright to
be detected by MIPS in these brighter classes of giants.
At the 
shorter wavelengths, the 
IRAC data are confusion limited. So spatial coincidence of IR and radio sources at these wavelengths alone is not a strong indication of actual 
association of the sources.\\

%As with the association between the radio sources and the NRFs, we examined the likelihood of random coincidence between 24 $\mu$m  sources and the 
%compact radio sources. Again, in our estimation,  we find that there are $\sim4$ times as many associations as would be expected from random chance. 

Some images appear to show 
cometary tails traced by the filaments (see Fig. 3i).
An interpretation of the  head-tail morphology is 
that the streaming of cosmic  rays from the Galactic plane interacts with 24~$\mu$m sources and create a tail oriented in the  
directions away from the source motion and the nuclear wind.

\section{Summary}
This short 
paper provides a catalogue of the positions of radio and infrared sources that  lie along groups of nonthermal radio filaments that have recently been 
identified. These sources could potentially be 
associated with radio filaments. 
We have displayed radio and IR images of 57 filaments as well as the SEDs of stellar sources coincident with radio sources.  
A true physical 
association will support a picture in which the compact radio sources are a byproduct of a large-scale  nuclear wind  interacting with stellar 
systems, thus  producing cometary tails that trace  NRFs.
Near- to mid- IR SEDs  illustrate  the various classes of sources that appear to be
associated with the radio filaments. A trend we notice 
is the SEDs of mass-losing  dusty red giants with strong 24 $\mu$m emission, or in some cases young, massive stars. 
Future high resolution radio and infrared 
observations of the candidate sources are needed to establish their true association with NRFs, thus testing  cometary tail model.

\section*{Acknowledgments}
Work by R.G.A. was supported by NASA under award number 80GSFC21M0002. FYZ is partially
supported by the grant AST-0807400 from the the National Science Foundation. The MeerKAT
telescope is operated by the South African Radio Astronomy Observatory, which is a
facility of the National Research Foundation, an agency of the Department of Science and
Innovation. The National Radio Astronomy Observatory is a facility of the National Science Foundation operated under cooperative agreement by Associated 
Universities, Inc.

%\vfill\eject

%Table 2
\begin{deluxetable}{cccccccccll}
\tabletypesize{\scriptsize}
\tablewidth{0pt}
\tablecaption{The positions of  infrared sources closest to compact radio sources, both of which  could be 
associated with groups of nonthermal radio filaments. 
}
\tablehead{
\colhead{} &
\multicolumn{2}{c}{Radio Selection} & & 
\multicolumn{2}{c}{Closest IRAC/MIPS Counterpart} & 
\colhead{Gaia Source Offset} &
\colhead{20cm Peak Intensity} &
\colhead{Associated Filament}
\\
\cline{2-3}\cline{5-6}
\colhead{Number} &
\colhead{$l$ (deg)} &
\colhead{$b$ (deg)} & &
\colhead{$l$ (deg)} &
\colhead{$b$ (deg)} & 
\colhead{(arcsec)} & 
\colhead{(arcsec)} & 
\colhead{(mJy beam$^{-1}$)} &
\colhead{(Group Names)}
}
\startdata
  1 &    358.73269 &     -0.22866 & &    358.73234 &     -0.23110 &    -  &   8.9 & 0.20 & Horseshoe\\
  2 &    358.77559 &      0.45383 & &    358.77572 &      0.45470 &     0.2 &  3.2 & 0.03 & Pelican \\
  3 &    358.79424 &      0.47940 & &    358.79452 &      0.47898 &     0.4 &   1.8 &0.07 &  ''  \\
  4 &    358.79832 &      0.46780 & &    358.79845 &      0.46830 &     0.3 &  1.9 &0.14 &   '' \\

  5 &    359.07250 &      0.73548 & &    359.07256 &      0.73526 &     - & 0.8 & 3.30 & Arrow \\
  6 &    359.11972 &     -0.26479 & &    359.11941 &     -0.26523 &     - &  1.9 & 0.57 & Snake \\
  7 &    359.13242 &     -0.20015 & &    359.13314 &     -0.20011 &     - &  2.6 & 1.10 & Snake \\
  8 &    359.21100 &     -0.08500 & &    359.21045 &     -0.08541 &     - &  2.5 & 0.73 & Candle \\

  9 &    359.21388 &     -0.09965 & &    359.21327 &     -0.09990 &    - &   2.4 & 0.35 &  '' \\
 10 &    359.22389 &     -0.13610 & &    359.22442 &     -0.13603 &    - &   1.9 &0.20  &  '' \\
 11 &    359.32086 &     -0.15780 & &    359.31964 &     -0.15809 &    - &   4.5 &1.40  &   Hummingbird  \\
 12 &    359.32139 &     -0.43090 & &    359.32211 &     -0.43014 &    - &   3.8 &0.26  &   Sausage \\

 13 &    359.32353 &     -0.42952 & &    359.32478 &     -0.43018 &     0.9 &  5.1 &0.24 &    '' \\
 14 &    359.34395 &     -0.41610 & &    359.34418 &     -0.41606 &     - &  0.8 &0.60  &   '' \\
 15 &    359.40080 &     -0.23797 & &    359.40081 &     -0.23823 &     - &  1.0 & 0.01 &   '' \\
 16 &    359.40494 &     -0.25914 & &    359.40420 &     -0.25935 &     0.2  &  2.8 &1.41 &    ''\\

 17 &    359.41575 &     -0.70605 & &    359.41495 &     -0.70514 &   - &    4.4 &0.20 &   Feather \\
 18 &    359.41879 &     -0.58340 & &    359.41981 &     -0.58245 &    - &   5.0 &0.04 &     '' \\
 19 &    359.42024 &     -0.66049 & &    359.42136 &     -0.66162 &    - &   5.7 &0.07 &     '' \\
 20 &    359.42288 &     -0.66401 & &    359.42366 &     -0.66269 &    - &   5.5 &0.41 &     '' \\

 21 &    359.43856 &      0.00595 & &    359.43854 &      0.00541 &     1.1 &  1.9 &6.2 &      Sgr C\\
 22 &    359.43580 &      0.11272 & &    359.43867 &      0.10242 &     - &  2.9 &0.01 &     French Knife\\
 23 &    359.45286 &     -0.03331 & &    359.45338 &     -0.03338 &     - &  1.9 &7.20 &     Sgr C \\  
 24 &    359.45476 &     -0.05466 & &    359.45423 &     -0.05583 &    - &  4.6 &1.01 &      '' \\

 25 &    359.46664 &     -0.17061 & &    359.46723 &     -0.17126 &       - & 3.2 &3.3  &      '' \\
 26 &    359.47531 &      0.12672 & &    359.47524 &      0.12599 &      - & 2.6 &0.20 &       Bent Harp\\
 27 &    359.48496 &      0.12052 & &    359.48514 &      0.11940 &      - & 4.1 &1.80 &       '' \\  
 28 &    359.49136 &      0.12141 & &    359.49023 &      0.12055 &      - &  5.1 &0.01 &        ''\\
 29 &    359.52174 &      0.09697 & &    359.52146 &      0.09694 &      - & 1.0 &0.90 &        ''\\
 30 &    359.55750 &      0.14030 & &    359.55749 &      0.13849 &     - &  6.5 &0.50 &         Ripple \\

 31 &    359.72920 &     -0.81139 & &    359.72907 &     -0.81095 &      - & 1.6 &0.32 &         Knot \\
 32 &    359.73372 &     -0.80122 & &    359.73448 &     -0.80124 &      - &  2.7 &0.12 & ''\\
 33 &    359.74574 &     -0.78398 & &    359.74498 &     -0.78400 &      - & 2.8 &1.00 & ''\\
 34 &    359.74829 &     -0.78059 & &    359.74785 &     -0.78135 &      0.3 &  3.2 &0.54 & ''\\

 35 &    359.75764 &      0.19774 & &    359.75782 &      0.19785 &     0.4 &  0.8 &0.78 &  Flamingo\\
 36 &    359.80790 &      0.11685 & &    359.80773 &      0.11774 &      - & 3.3 &0.31 & ''\\
 37 &    359.83844 &      0.37514 & &    359.83913 &      0.37489 &      - & 2.7 &0.18 & Harp \\
 38 &    359.84121 &      0.36571 & &    359.84095 &      0.36649 &      - & 3.0 &0.17 & ''\\

 39 &    359.84411 &      0.35841 & &    359.84324 &      0.35919 &      - & 4.2 &0.23 & ''\\
 40 &    359.84814 &      0.29777 & &    359.84743 &      0.29717 &      - & 3.3 &0.04 & ''\\
 41 &    359.85073 &      0.29309 & &    359.84970 &      0.29374 &      - & 4.4 &0.20 & ''\\
 42 &    359.87391 &     -0.26430 & &    359.87404 &     -0.26446 &      - & 0.7 &1.30 & Cleaver Knife\\

 43 &    359.88580 &      0.10155 & &    359.88586 &      0.10101 &    - &   2.0 &5.60 &  Harp \\
 44 &    359.92107 &     -0.28271 & &    359.92023 &     -0.28300 &    - &   3.2 &0.27 &  Cleaver Knife\\
 45 &    359.94984 &     -0.25940 & &    359.95039 &     -0.25939 &    - &   2.0 &0.73 &   ''\\
 46 &    359.99238 &     -0.55465 & &    359.99260 &     -0.55314 &    0.3 &   5.5 &0.08 &  Comet \\

 47 &      0.02333 &     -0.62574 & &      0.02343 &     -0.62668 &     0.05 &  3.4 &0.07 & ''\\
 48 &      0.02952 &     -0.65987 & &      0.02986 &     -0.66006 &     1.9 &  1.4 &0.29 &  ''\\
 49 &      0.03434 &     -0.66120 & &      0.03486 &     -0.66023 &     - &  4.0 &0.28 & ''\\
 50 &      0.04531 &     -0.66785 & &      0.04583 &     -0.66676 &     - &  4.4 &0.25 & ''\\

 51 &      0.22457 &      0.84297 & &      0.22419 &      0.84330 &      0.8 (1.3)  &  1.8 &0.04 & Space Shuttle\\
 52 &      0.22605 &      0.79178 & &      0.22651 &      0.79185 &       - & 1.7 &0.25 & ''\\
 53 &      0.26408 &     -0.19683 & &      0.26361 &     -0.19778 &     - &  3.8 &0.64 &  Porcupine\\
 54 &      0.42008 &     -0.29663 & &      0.41970 &     -0.29718 &      - & 2.4 &0.41&  Contrail\\

 55 &      0.42075 &     -0.29048 & &      0.42099 &     -0.28975 &      - & 2.8 &0.39 & ''\\
 56 &      0.44269 &     -0.32407 & &      0.44244 &     -0.32376 &      0.4 & 1.4 &0.17& ''\\
 57 &      0.69688 &      0.14827 & &      0.69703 &      0.14805 &      0.8 &  1.0 &0.75 & Bent Fork\\
\enddata
\end{deluxetable}

\label{tab:table2}

\section{Data Availability}

All the data including  VLA and MeerKAT  that we used here are available online and are not proprietary.
We have reduced and calibrated these data and are available if  requested.

%%%%%%%%%%%%%%%%%%%%%%%%%%%%%%%%%%%%%%%%%%%%%%%%%%



\bibliographystyle{mnras}
%\bibliography{myrefs} % if your bibtex file is called example.bib

%%%%%%%%%%%%%%%%%%%%%%%%%%%%%%%%%%%%%%%%%%%%%%%%%%
\vfill\eject

\begin{figure}
\center
\includegraphics[width=7.0in]{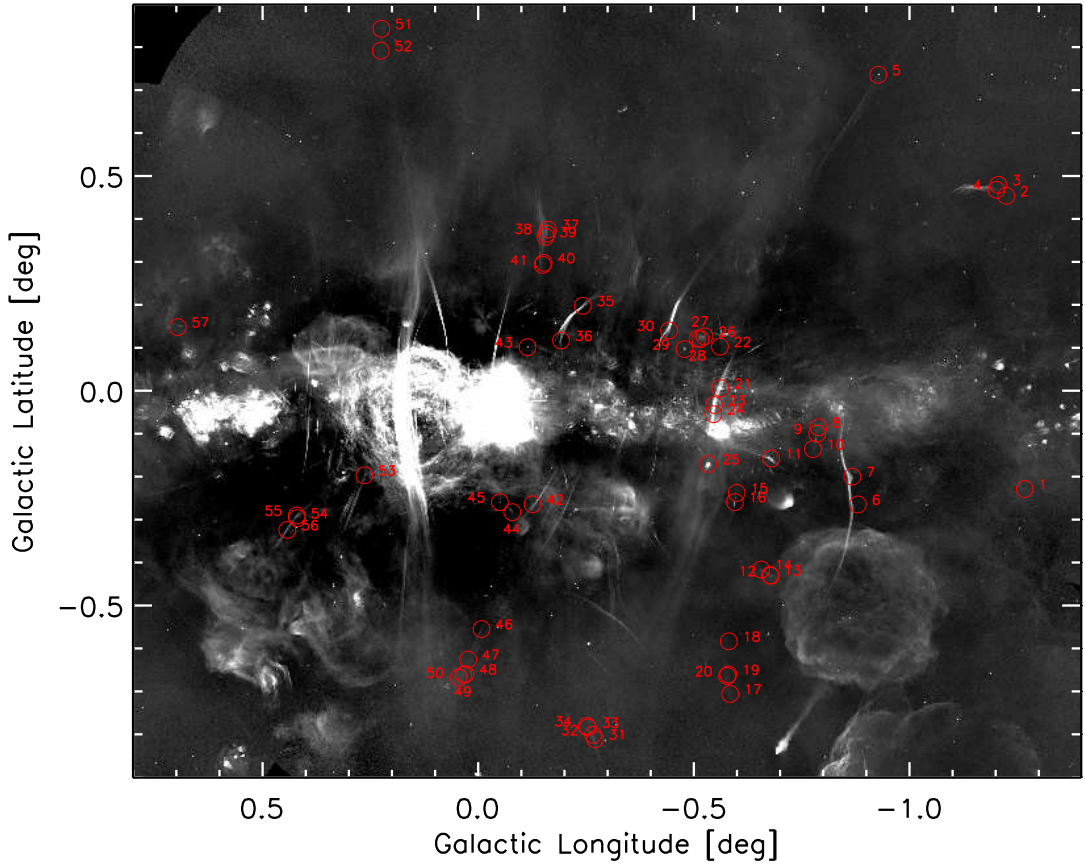}
\caption{
A 1.28 GHz mosaic image of the  Galactic center region \citep{heywood22}. Red circles show 
the positions of compact radio sources with possible  infrared counterparts. A blow-up of these sources are 
shown in Figures 3 (a) to (eee).  There are 57 circles  corresponding to the number of sources listed in Table 1. 
}
\label{f:fig1}
\end{figure}

%\begin{figure}[ht!]
\begin{figure}
\center
 \includegraphics[width=5.50in]{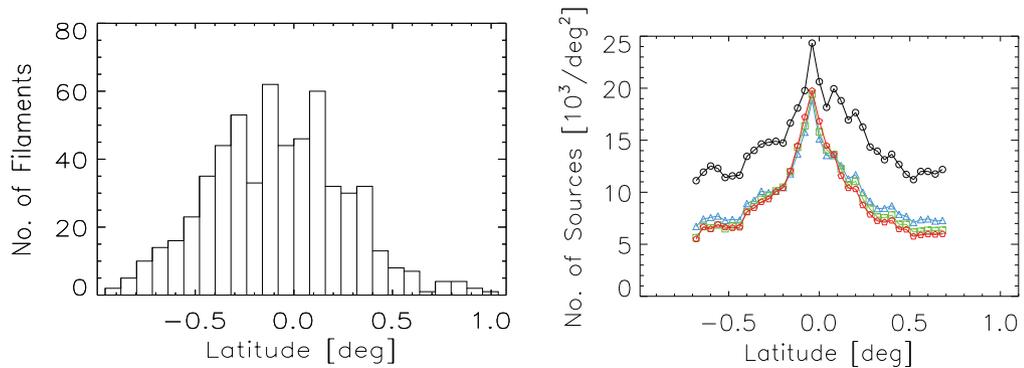}
 \caption{
{\it (a) Left}
A histogram of the number of NRFs is shown as a function of Galactic latitude ($b$)
for  filament lengths  longer than
$66''$  AND confined to within Galactic latitude $\ |l| < 0.45^\circ$ (which excludes most of thermal features), based on MeerKAT data.
{\it (b) Right}
A histogram of the number of bright stars is shown as a function of Galactic latitude using  IRAC data  in all four bands 
IRAC 3.6, 4.5, 8 $\mu$m (red,green,blue), and  MIPS 24$\mu$m (black). These plots are constructed from data presented in \citep{ramirez08}.
}
\end{figure}

%%%%%%%%%%%%%%%%%%%%%%%%%%%%%%%%%%%%%%%%%%%%%%%%%%

%\vfill\eject

%\addtocounter

%{figure}{-1}
\begin{figure}
\center
\includegraphics[]{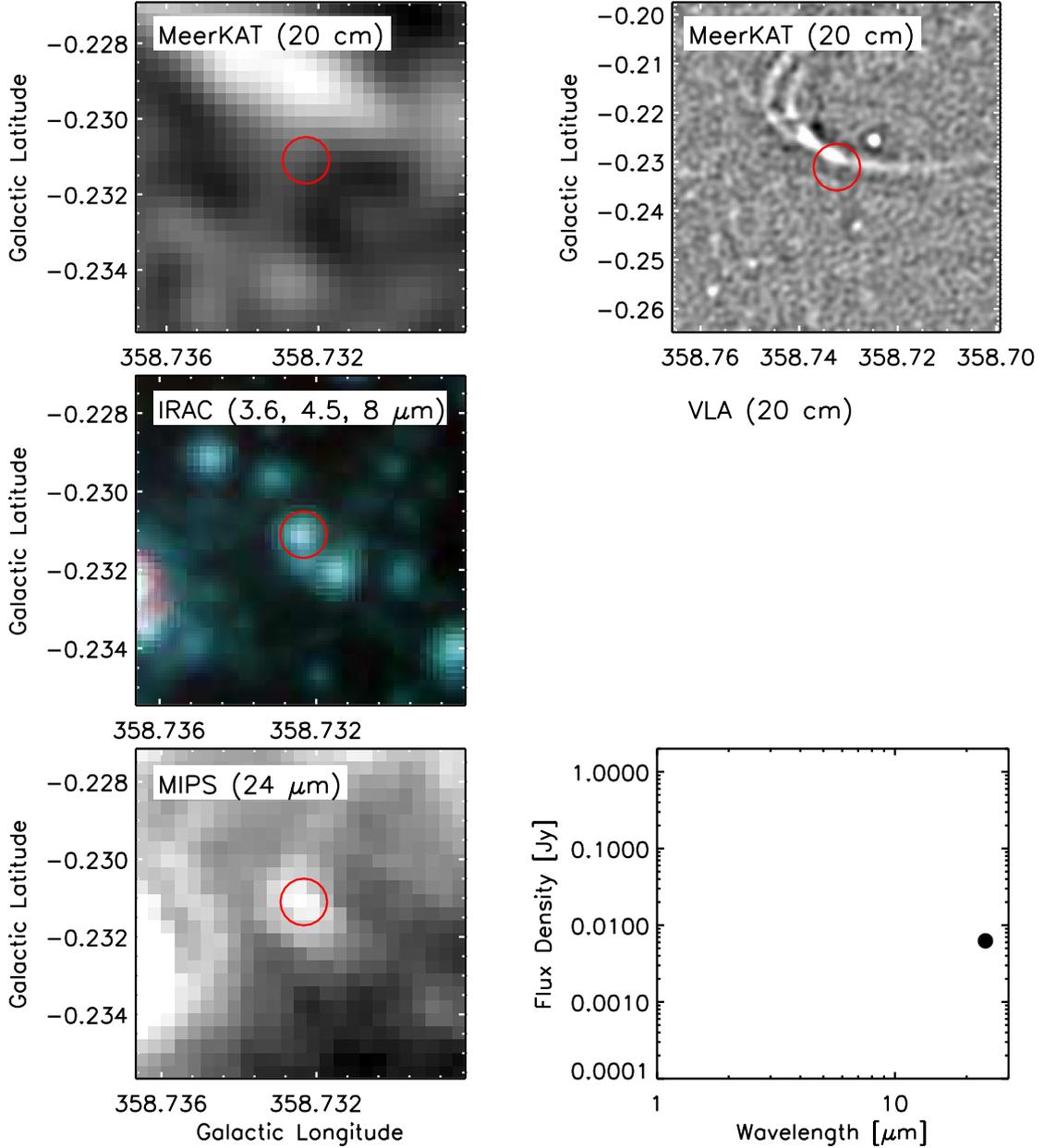}
\caption{
{\it (a)}   
Each candidate radio source numbered 1 to 57 in Table 1 are shown in 6 separate panels. 
The left column shows a narrow view of, a composite 
narrow view of 
IRAC 3.6, 4.5, 8 $\mu$m images, and a narrow view of MIPS 24$\mu$m 
images from top to bottom, respectively. 
The right column shows a wider view of MeerKAT image images (filtered to highlight filaments, \citep{zadeh22a}), 
unfiltered VLA image (if 
available), and the SED
using 2MASS+IRAC+MIPS data, 
(black symbols: \citep{ramirez08,hinz09}). 
Red stars indicate measurements from the GLIMPSE II Spring '08 Archive 
\citep{benjamin03}. Error bars are plotted for all measurements, but
are generally smaller than the symbol size.  2MASS 95\% confidence upper limits are now plotted as downward 
triangles. Upper  limits 
are not stated in the other catalogs on a source by source basis. Red circles in the image panels show the positions where the IR SED, nearest source to the peak radio 
position,  is  measured. 
}
\end{figure}

\addtocounter{figure}{-1}
\begin{figure}
\center
\includegraphics[]{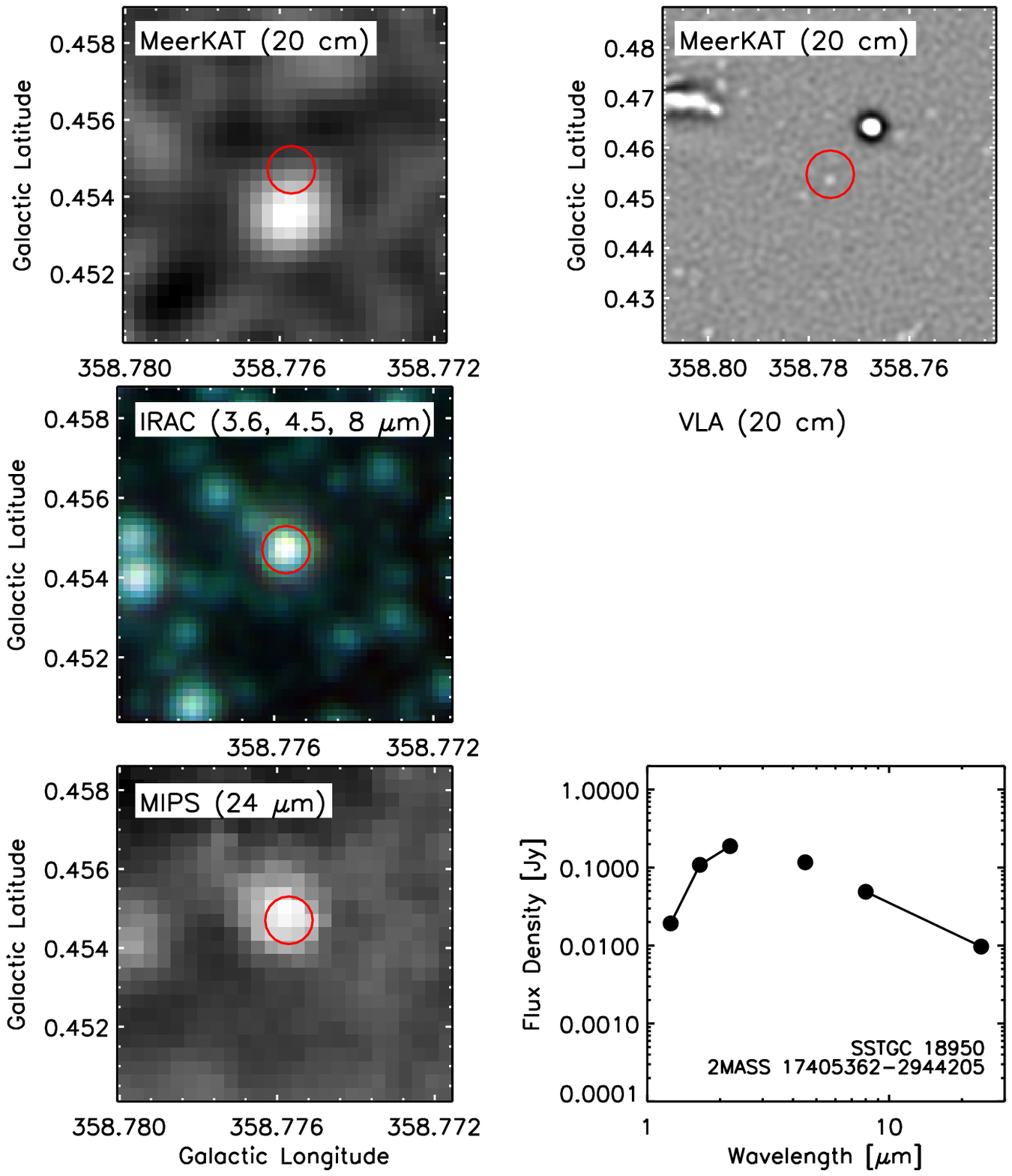}
\caption{
{\it (b)}  Same as Fig. 3a except for source 2 in Table 1. Open circles indicate potentially saturated IRAC measurements.
}
\end{figure}

\addtocounter{figure}{-1}
\begin{figure}
\center
\includegraphics[]{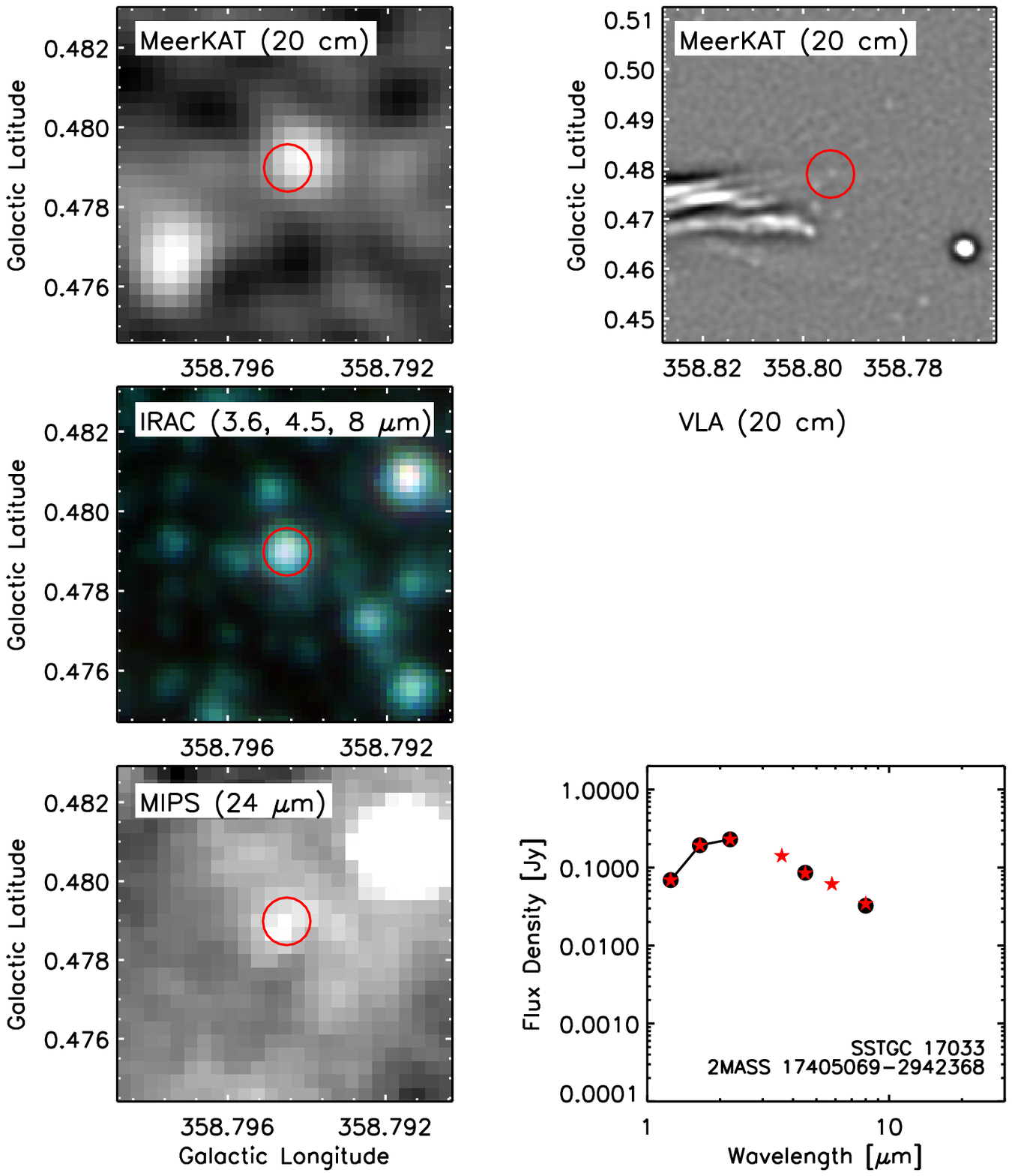}
\caption{
{\it (c)} Same as Fig. 3a except for source 3 Table 1.
}
\end{figure}

\addtocounter{figure}{-1}
\begin{figure}
\center
\includegraphics[]{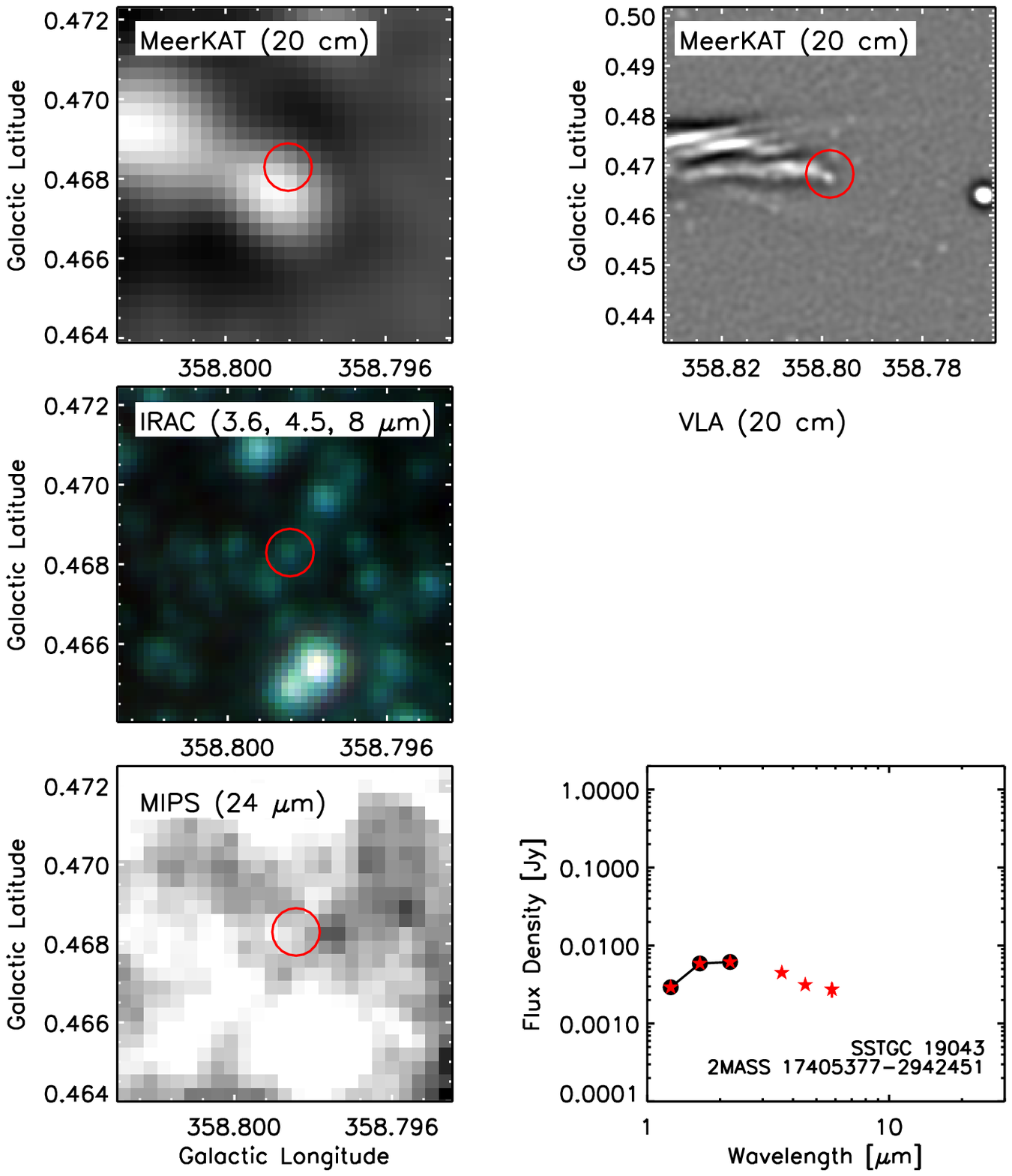}
\caption{
 {\it (d)}  Same as Fig. 3a except for source 4 in Table 1.
}
\end{figure}

\vfill\eject

\addtocounter{figure}{-1}
\begin{figure}
\center
\includegraphics[]{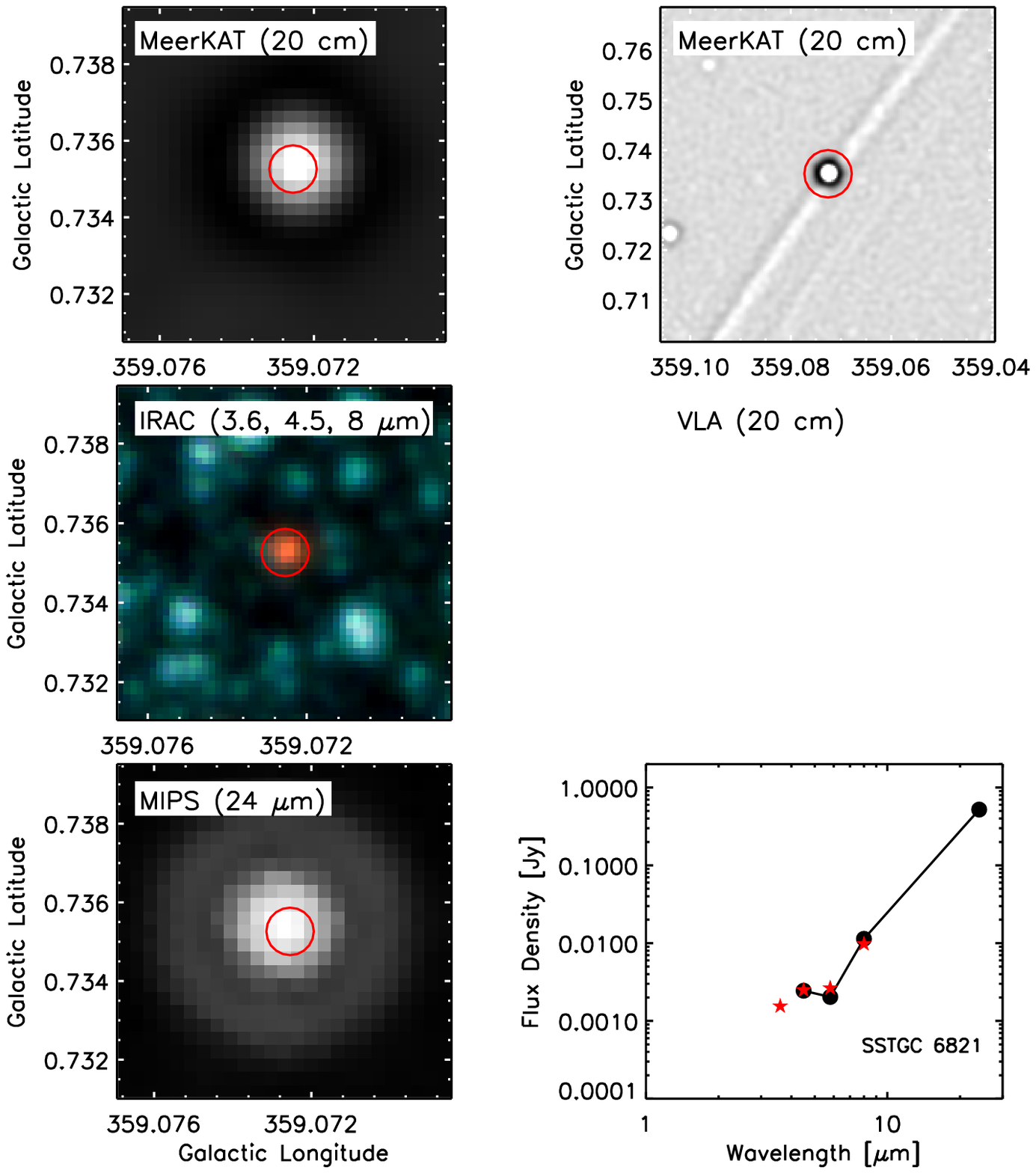}
\caption{
{\it (e)}  Same as Fig. 3a except for source 5 in Table 1.
}
\end{figure}

\addtocounter{figure}{-1}
\begin{figure}
\center
\includegraphics[]{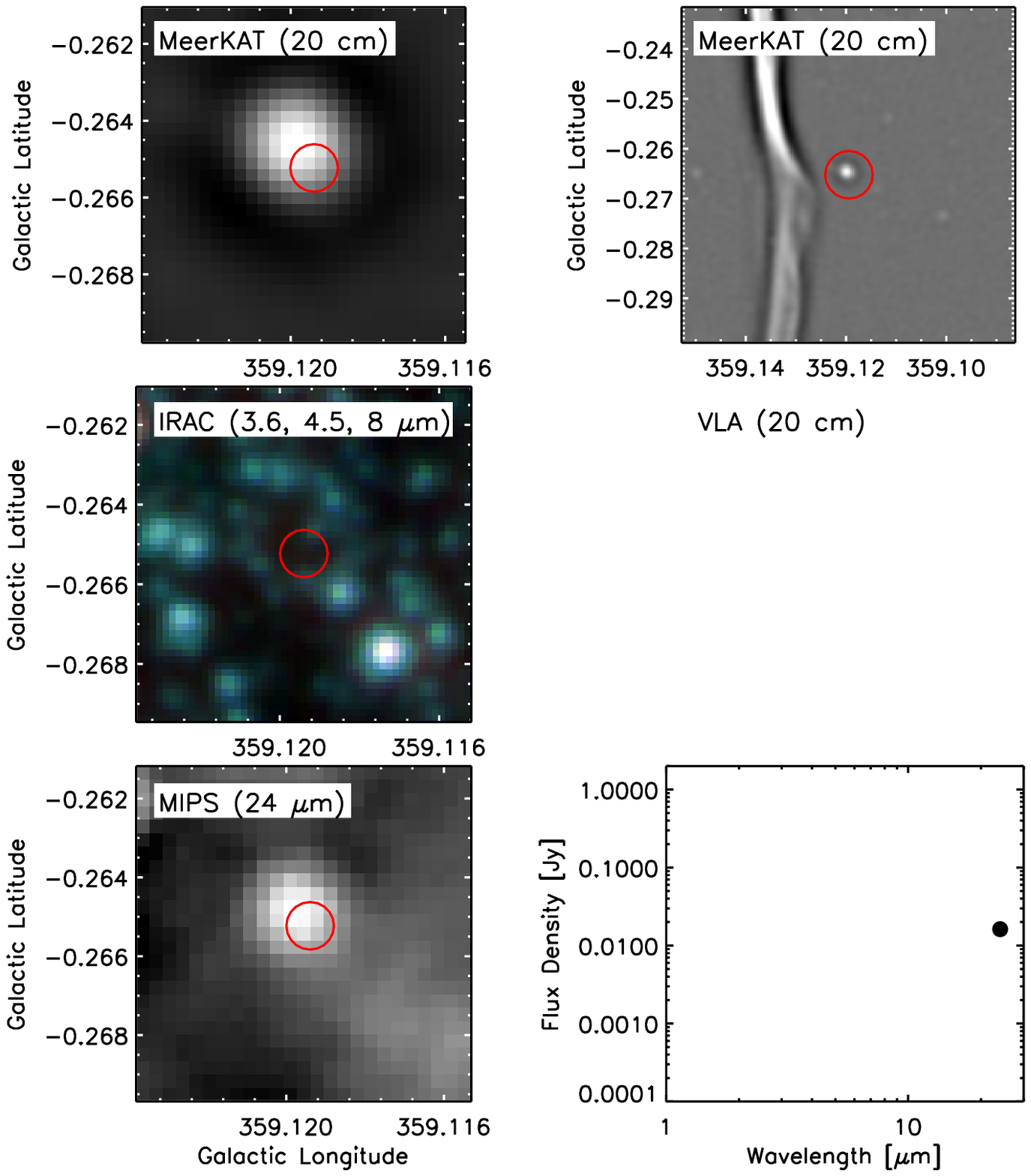}
\caption{
 {\it (f)}  Same as Fig. 3a except for  source 6 in Table 1.
}
\end{figure}

\addtocounter{figure}{-1}
\begin{figure}
\center
\includegraphics[]{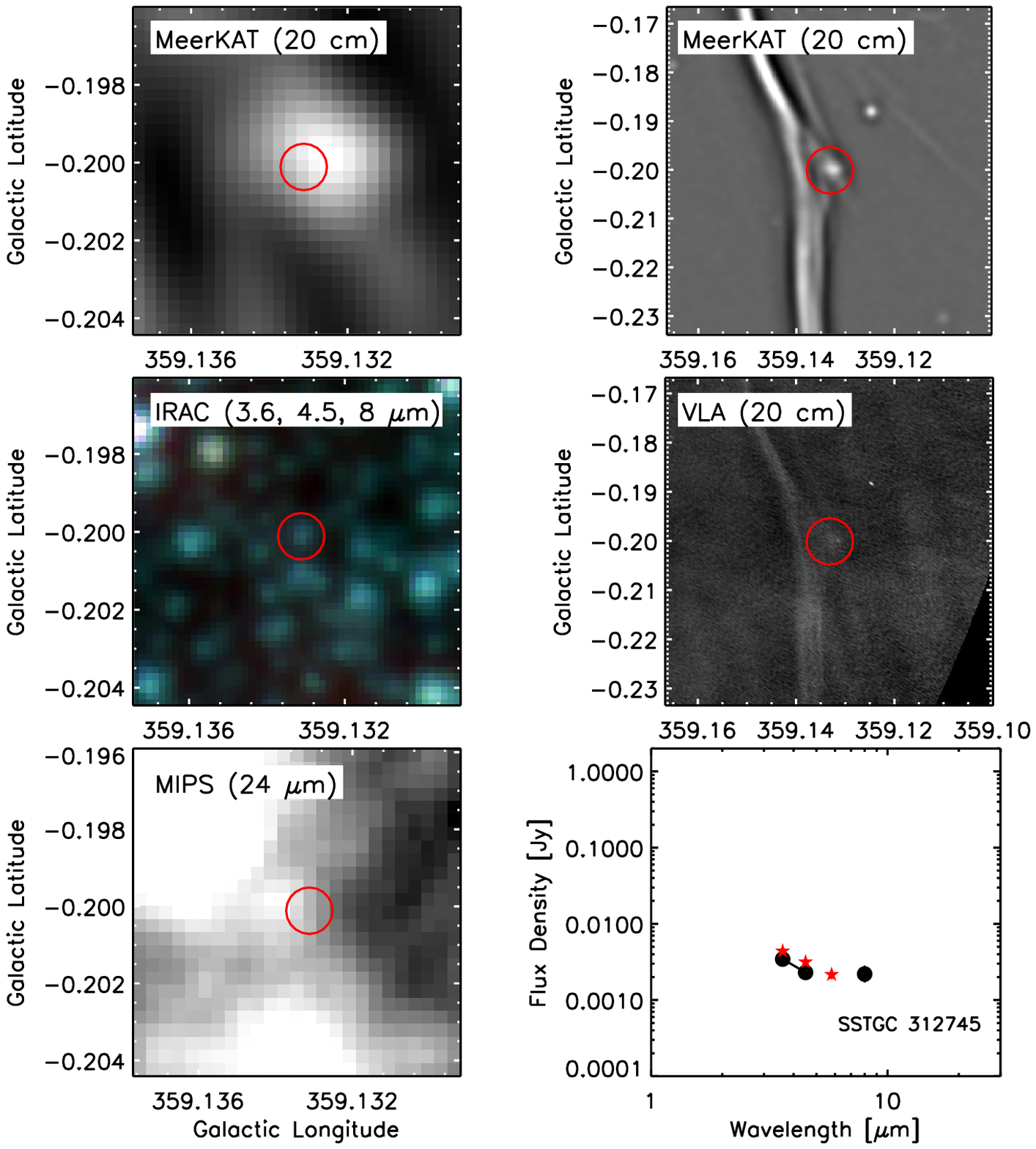}
\caption{
 {\it (g)}  Same as Fig. 3a except for source 7  in Table 1.
}
\end{figure}

\addtocounter{figure}{-1}
\begin{figure}
\center
\includegraphics[]{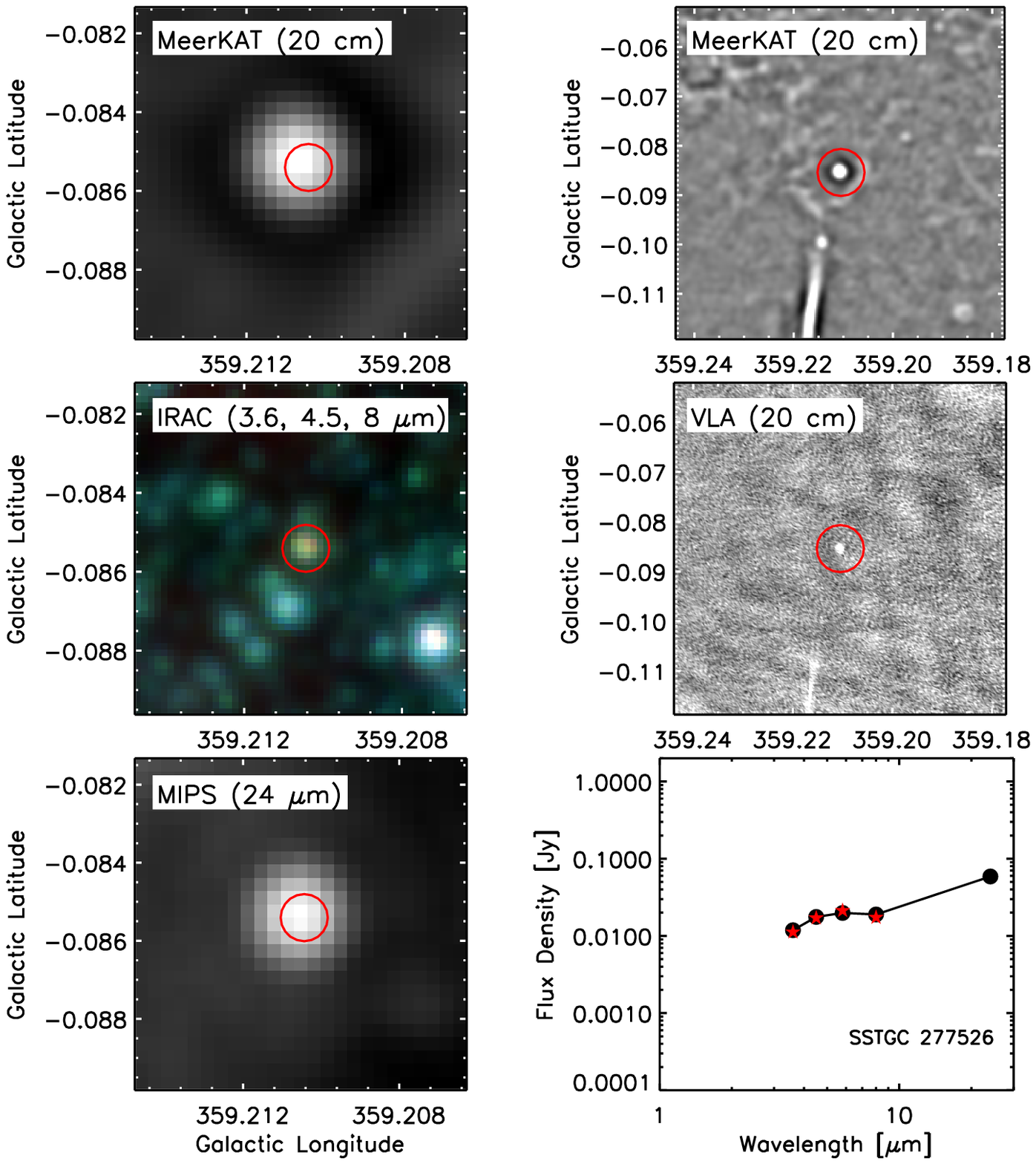}
\caption{
 {\it (h)}  Same as Fig. 3a except for source 8 in Table 1.
}
\end{figure}

\addtocounter{figure}{-1}
\begin{figure}
\center
\includegraphics[]{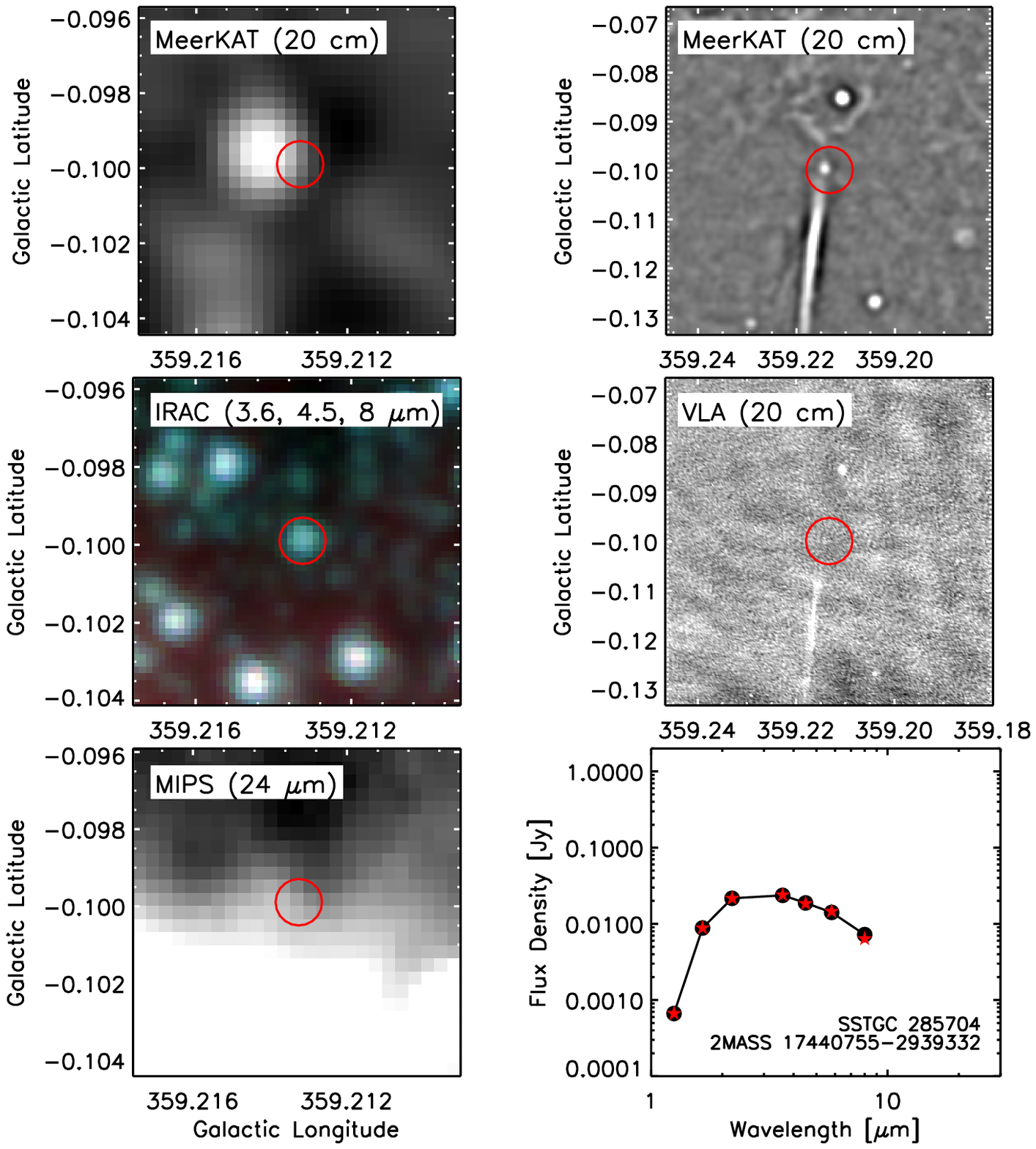}
\caption{
 {\it (i)}  Same as Fig. 3a except for source 9  in Table 1.
}
\end{figure}

\addtocounter{figure}{-1}
\begin{figure}
\center
\includegraphics[]{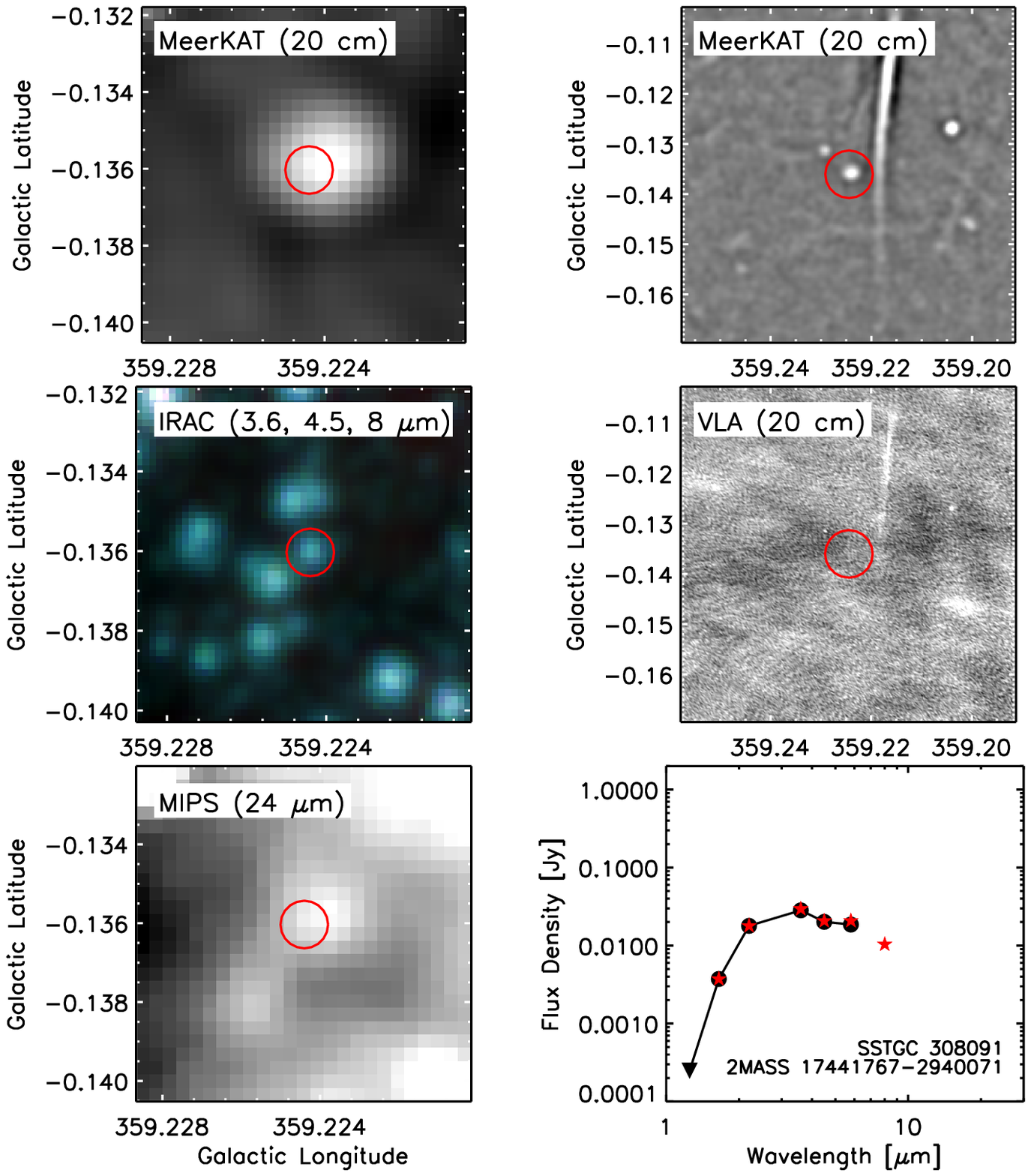}
\caption{
 {\it (j)}  Same as Fig. 3a except for source 10  in Table 1.
}
\end{figure}

\addtocounter{figure}{-1}
\begin{figure}
\center
\includegraphics[]{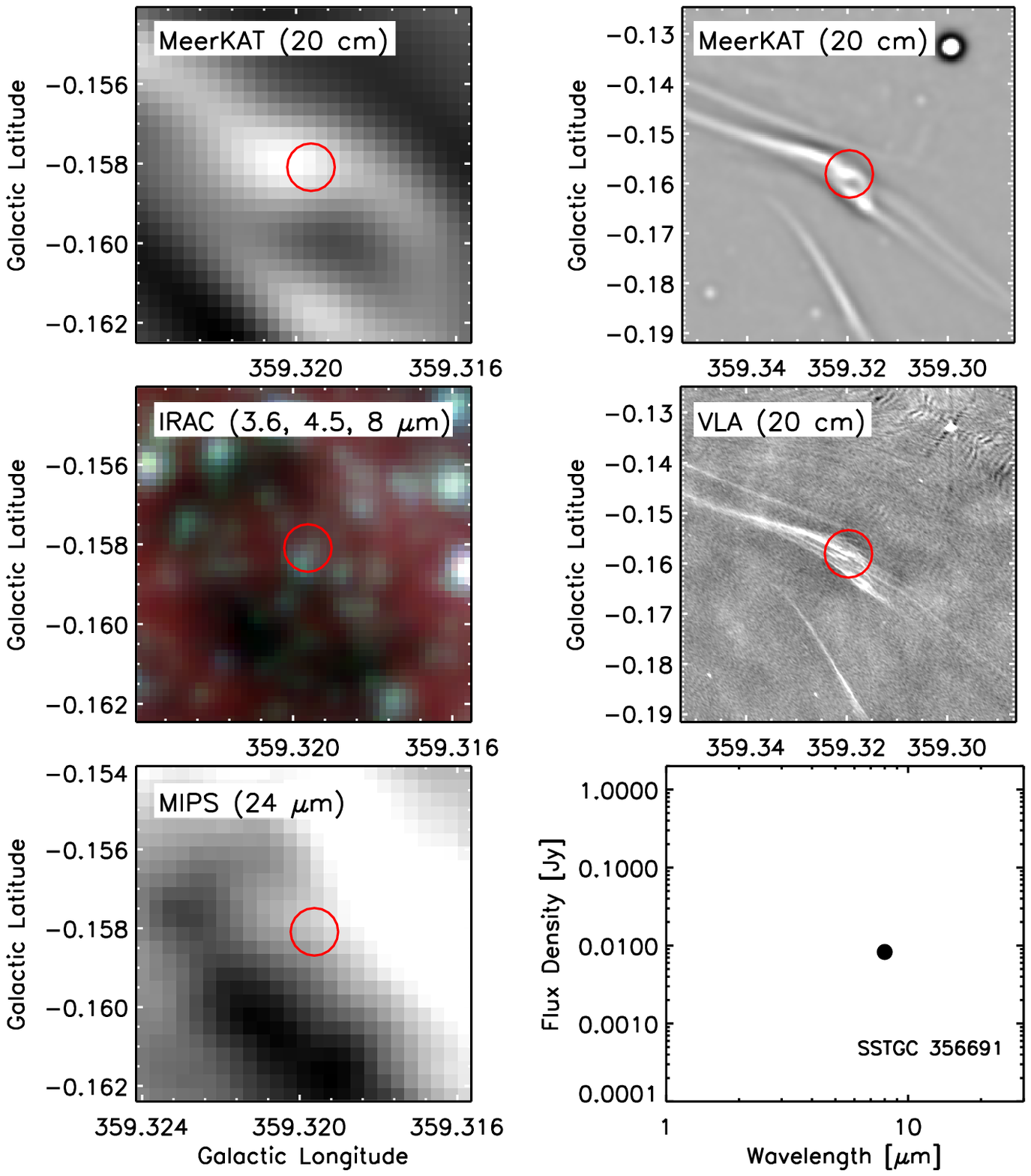}
\caption{
{\it (k)}  Same as Fig. 3a except source 11  in Table 1.
}
\end{figure}

\addtocounter{figure}{-1}
\begin{figure}
\center
\includegraphics[]{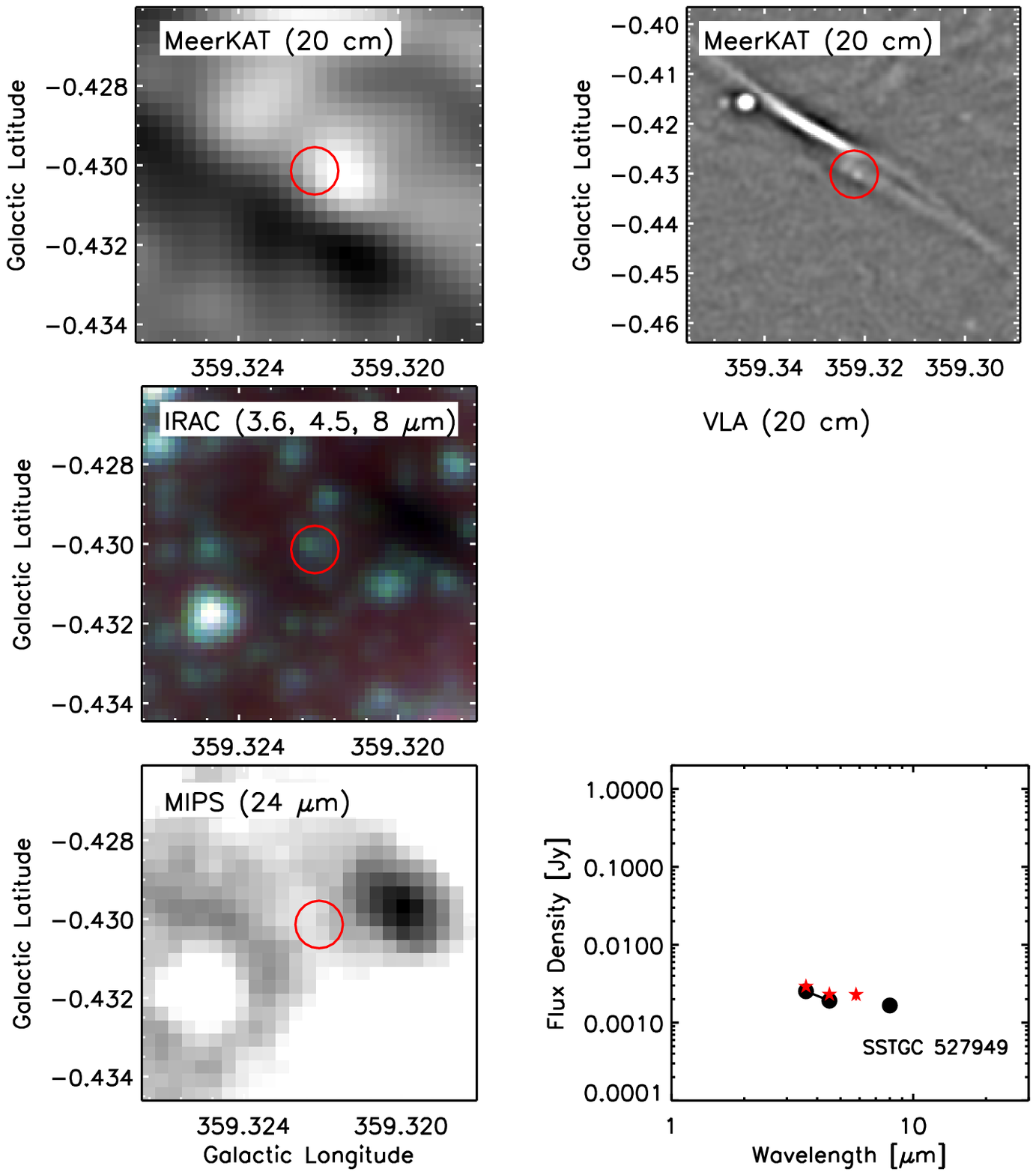}
\caption{
 {\it (l)}  Same as Fig. 3a except source 12  in Table 1.
}
\end{figure}

\addtocounter{figure}{-1}
\begin{figure}
\center
\includegraphics[]{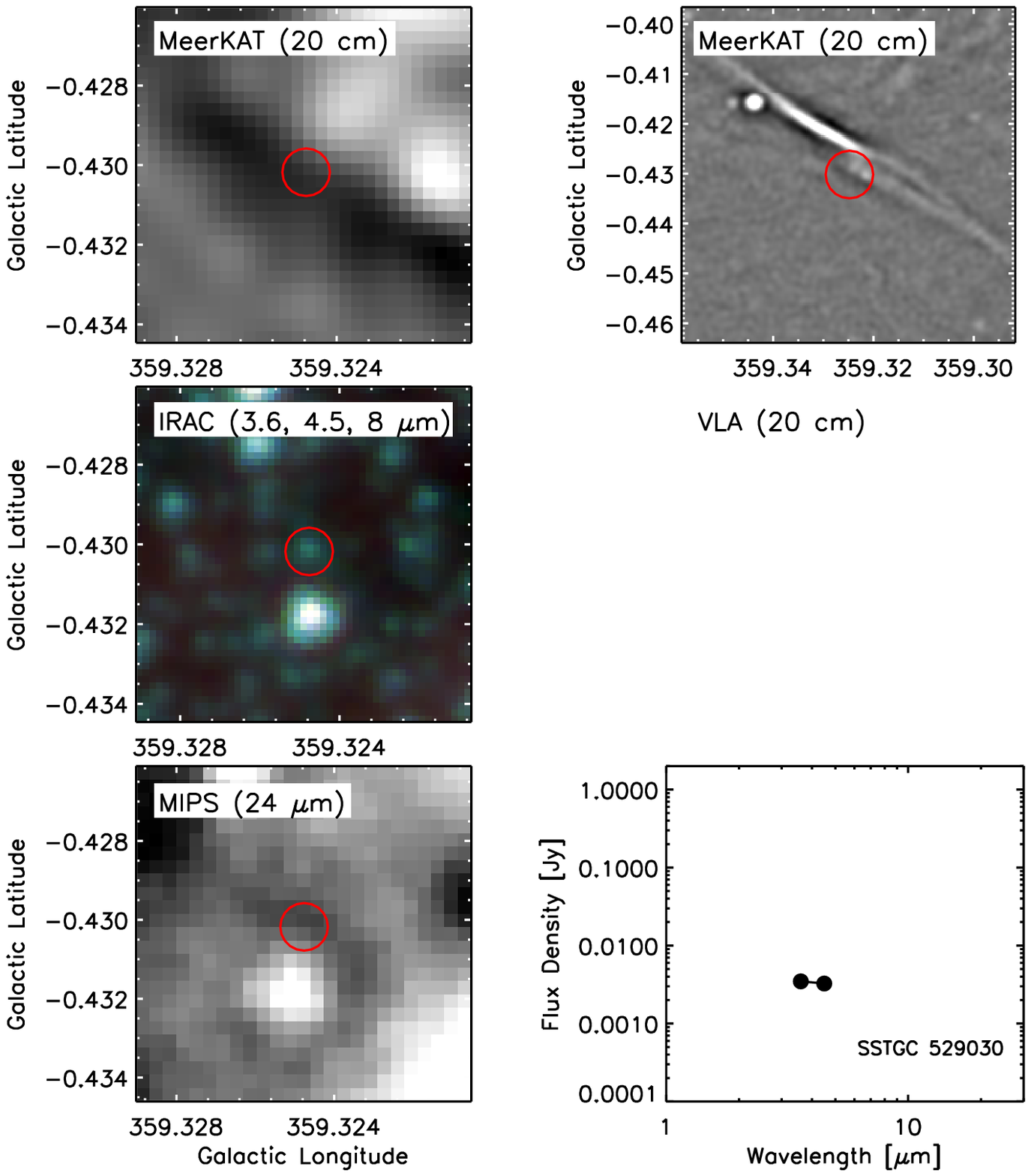}
\caption{
 {\it (m)}  Same as Fig. 3a except source 13  in Table 1.
}
\end{figure}

\addtocounter{figure}{-1}
\begin{figure}
\center
\includegraphics[]{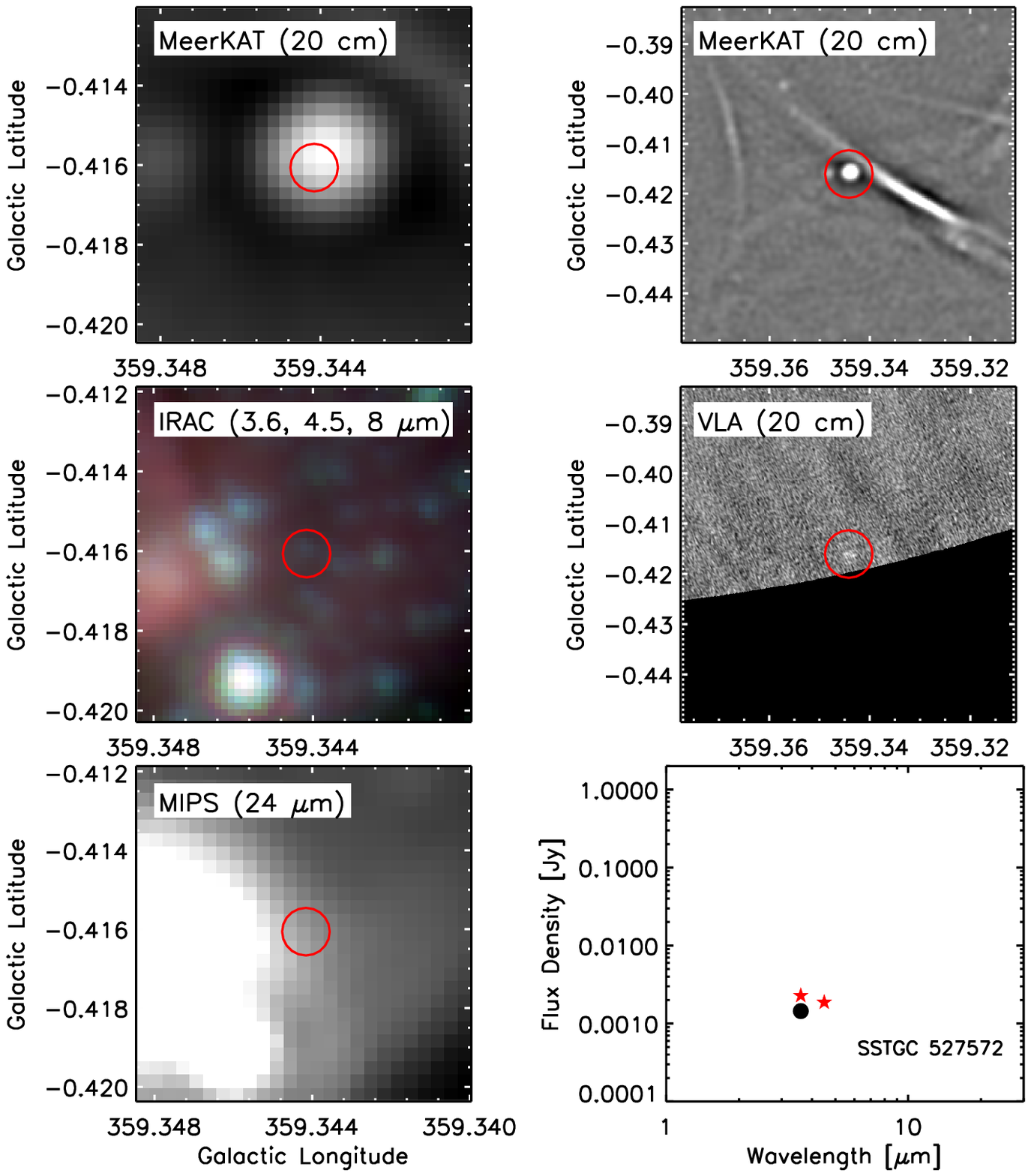}
\caption{
 {\it (n)}  Same as Fig. 3a except source 14  in Table 1.
}
\end{figure}

\addtocounter{figure}{-1}
\begin{figure}
\center
\includegraphics[]{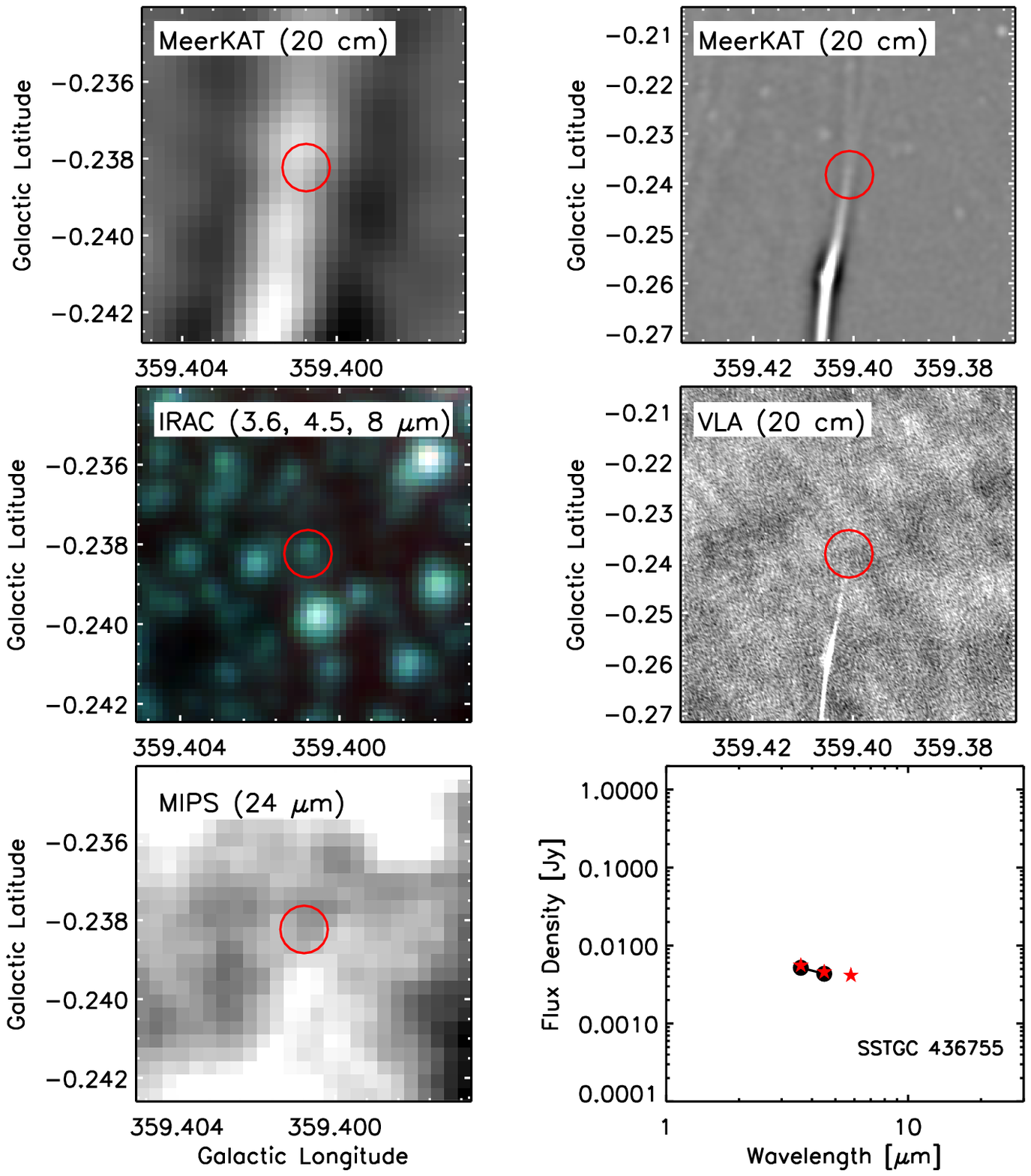}
\caption{
 {\it (o)}  Same as Fig. 3a except source 15  in Table 1.
}
\end{figure}

\addtocounter{figure}{-1}
\begin{figure}
\center
\includegraphics[]{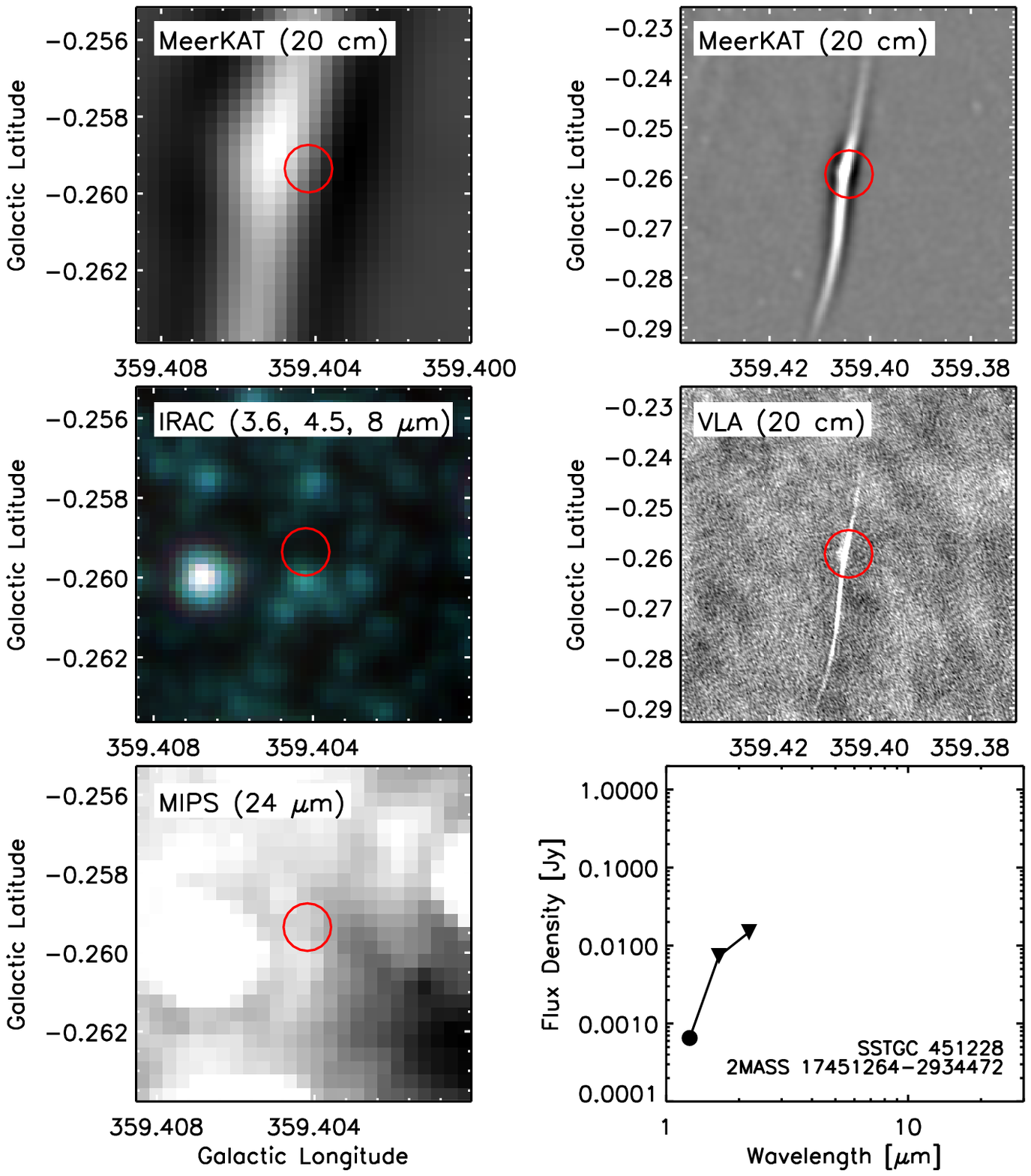}
\caption{
 {\it (p)}  Same as Fig. 3a except source 16  in Table 1.
}
\end{figure}

\addtocounter{figure}{-1}
\begin{figure}
\center
\includegraphics[]{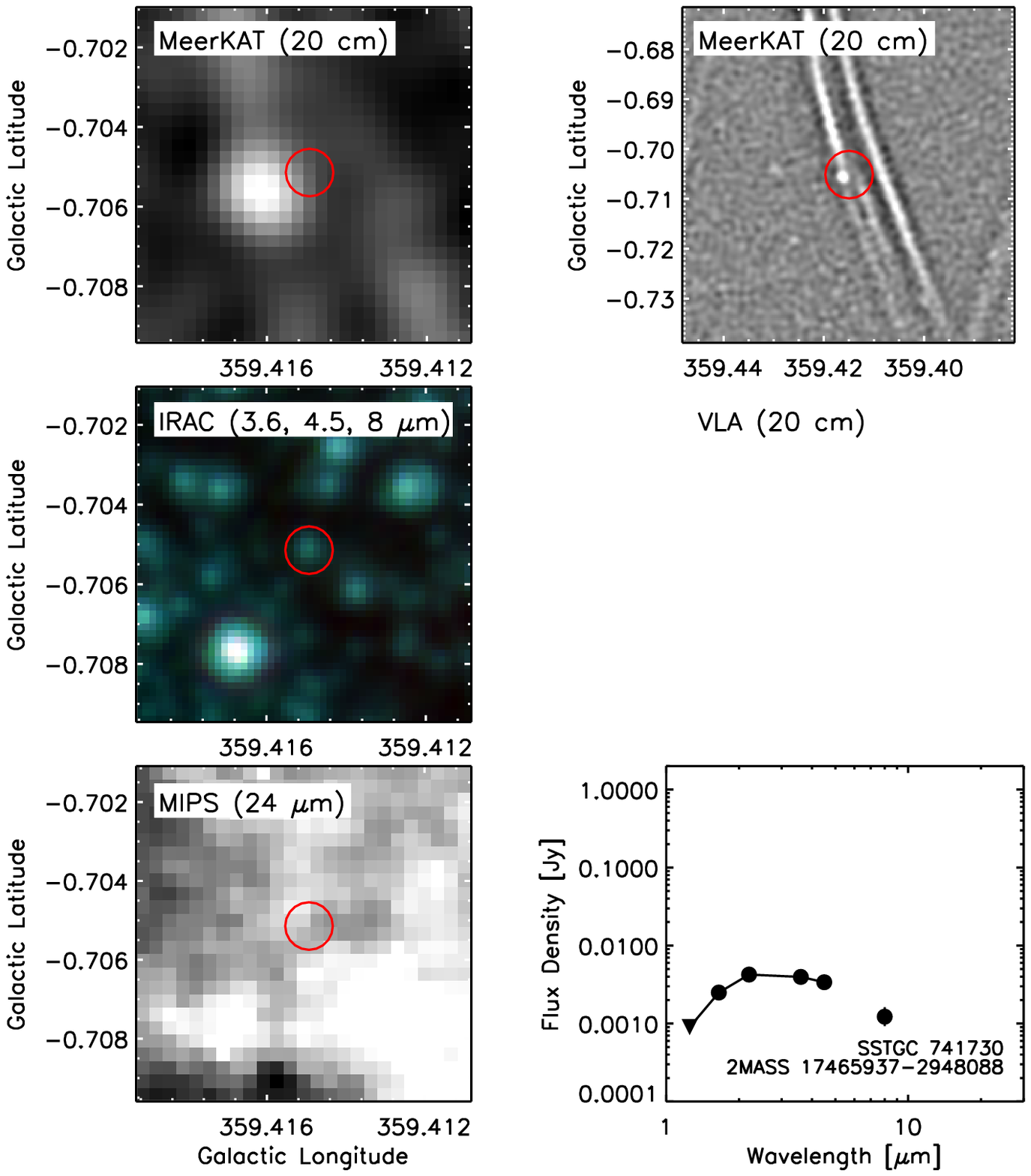}
\caption{
 {\it (q)}  Same as Fig. 3a except source 17  in Table 1.
}
\end{figure}

\addtocounter{figure}{-1}
\begin{figure}
\center
\includegraphics[]{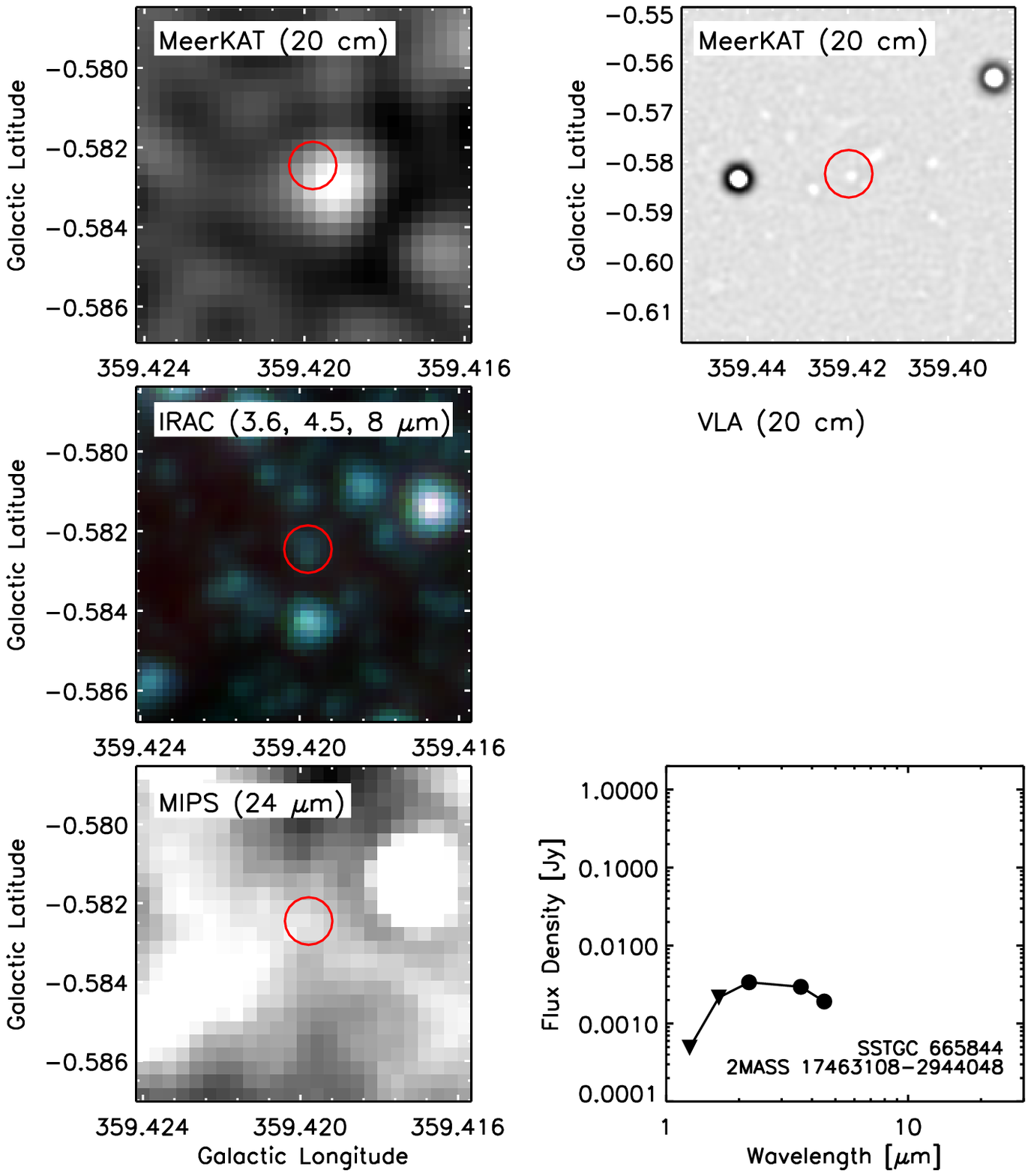}
\caption{
 {\it (r)}  Same as Fig. 3a except source 18  in Table 1.
}
\end{figure}

\addtocounter{figure}{-1}
\begin{figure}
\center
\includegraphics[]{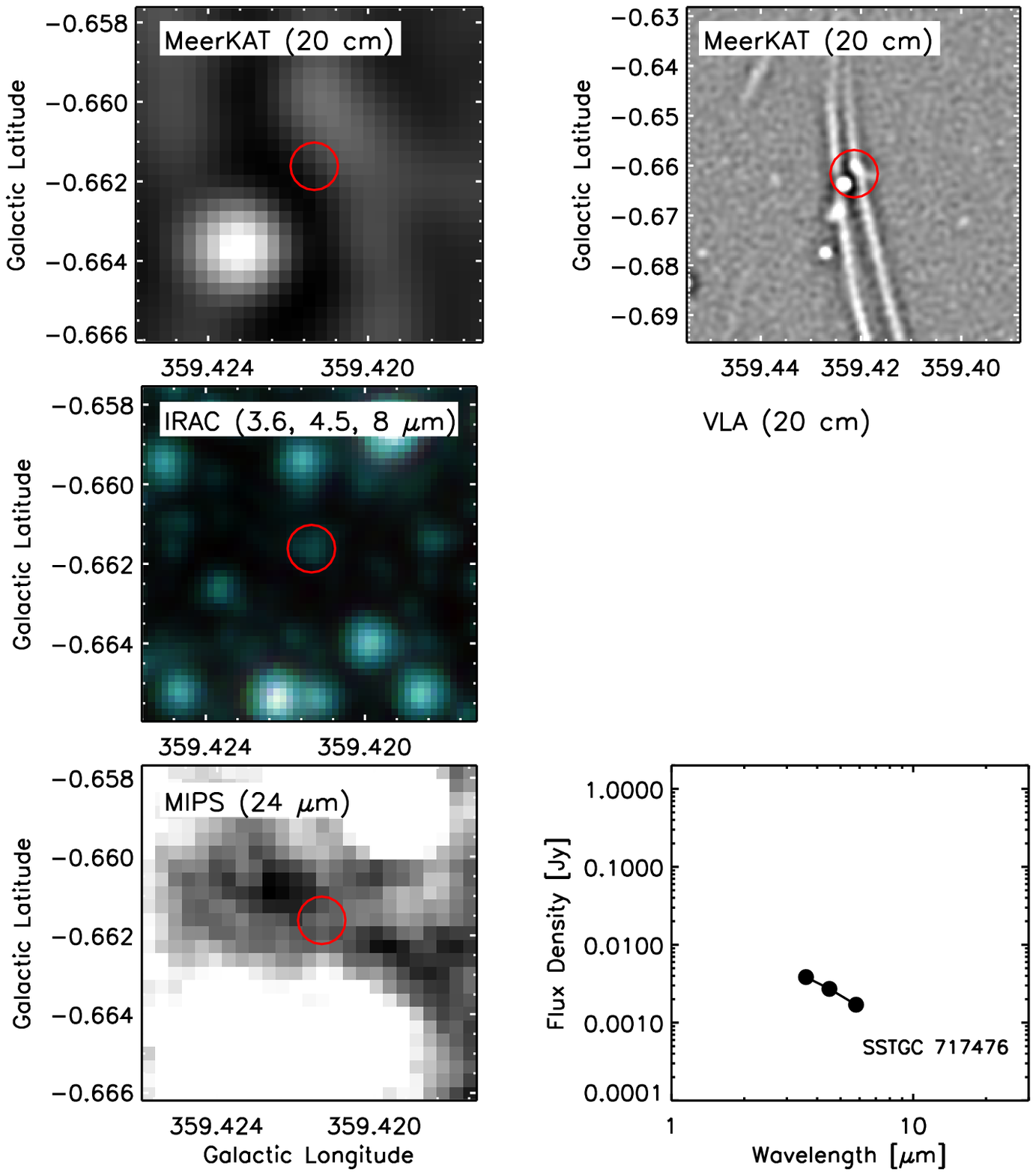}
\caption{
 {\it (s)}  Same as Fig. 3a except source 19 in Table 1.
}
\end{figure}

\addtocounter{figure}{-1}
\begin{figure}
\center
\includegraphics[]{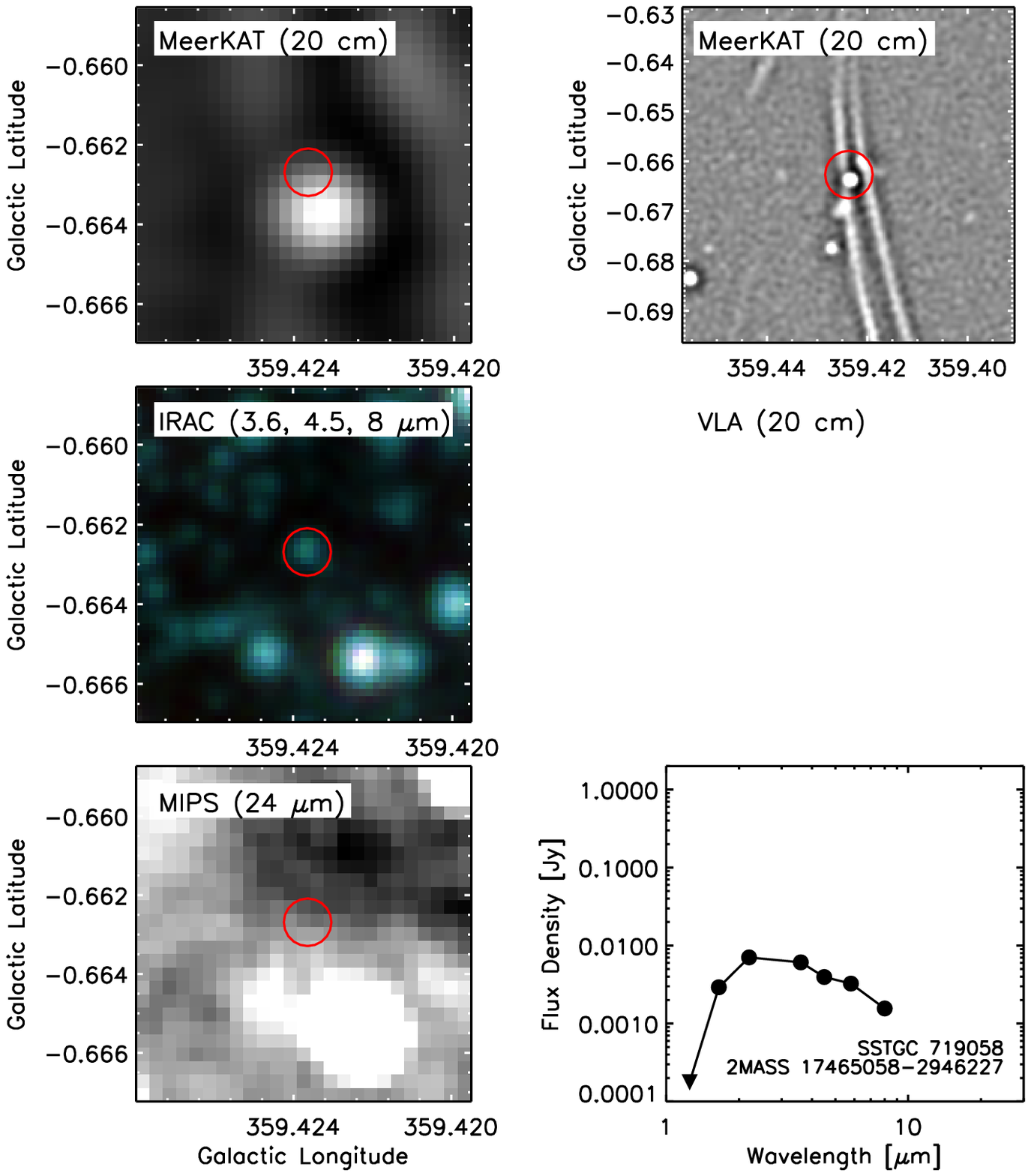}
\caption{
 {\it (t)}  Same as Fig. 3a except source 20   in Table 1.
}
\end{figure}
\addtocounter{figure}{-1}
\begin{figure}
\center
\includegraphics[]{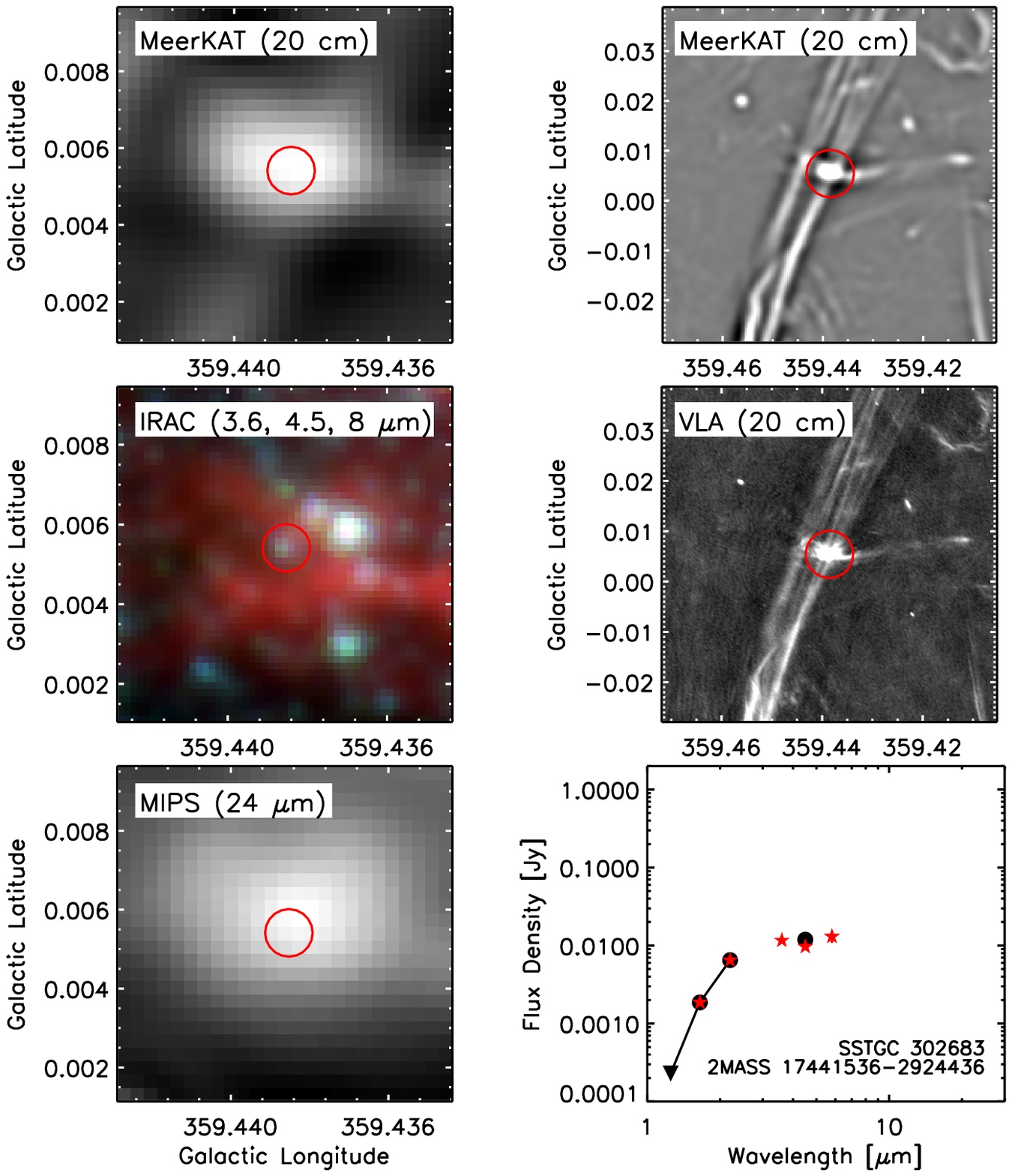}
\caption{
 {\it (u)}  Same as Fig. 3a except source 21   in Table 1.
}
\end{figure}
\addtocounter{figure}{-1}
\begin{figure}
\center
\includegraphics[]{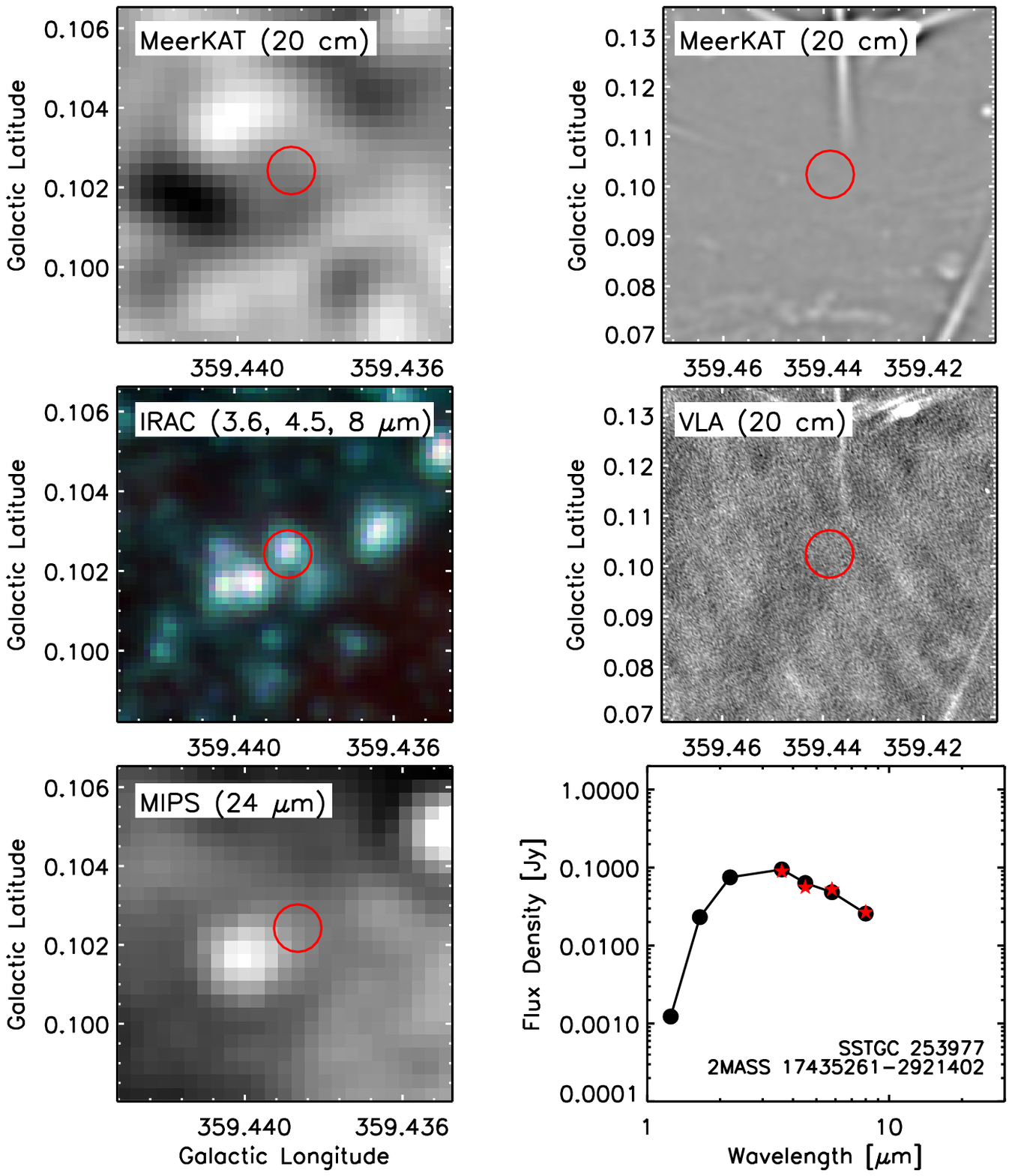}
\caption{
 {\it (v)}  Same as Fig. 3a except source 22  listed in Table 1.
}
\end{figure}
\addtocounter{figure}{-1}
\begin{figure}
\center
\includegraphics[]{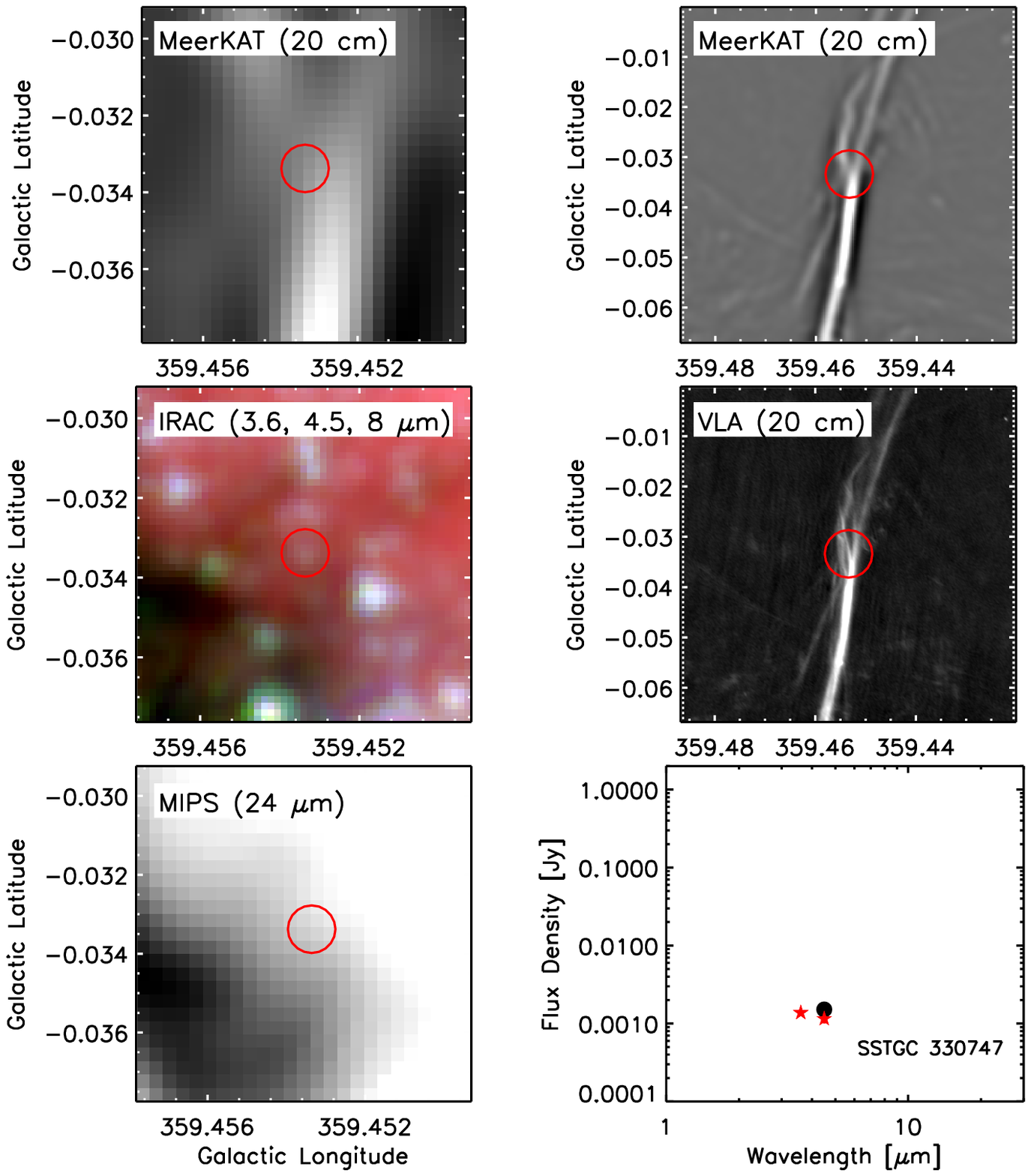}
\caption{
 {\it (w)}  Same as Fig. 3a except source 23  in Table 1.
}
\end{figure}
\addtocounter{figure}{-1}
\begin{figure}
\center
\includegraphics[]{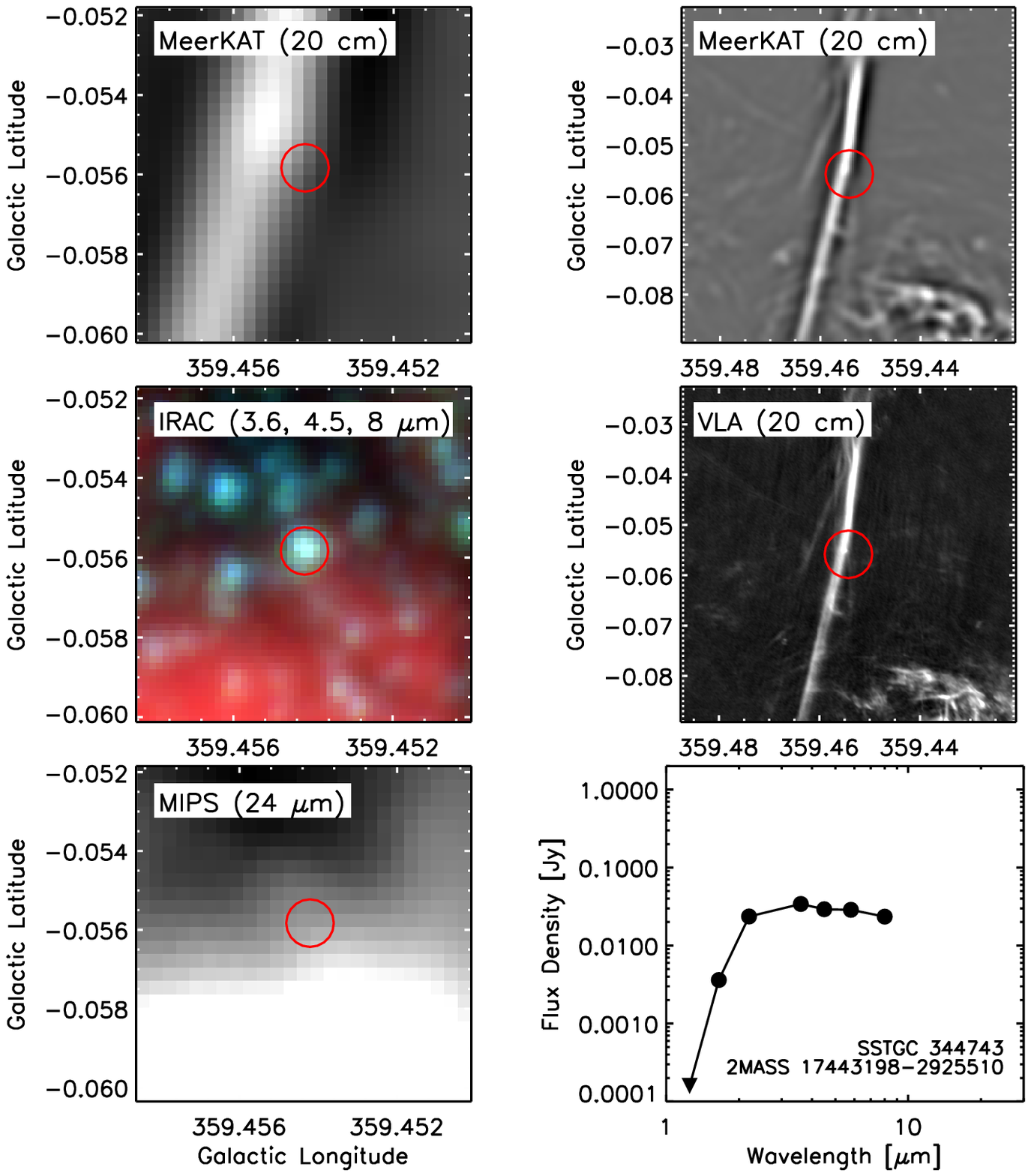}
\caption{
{\it (x)}  Same as Fig. 3a except source 24 in Table 1.
}
\end{figure}

\addtocounter{figure}{-1}
\begin{figure}
\center
\includegraphics[]{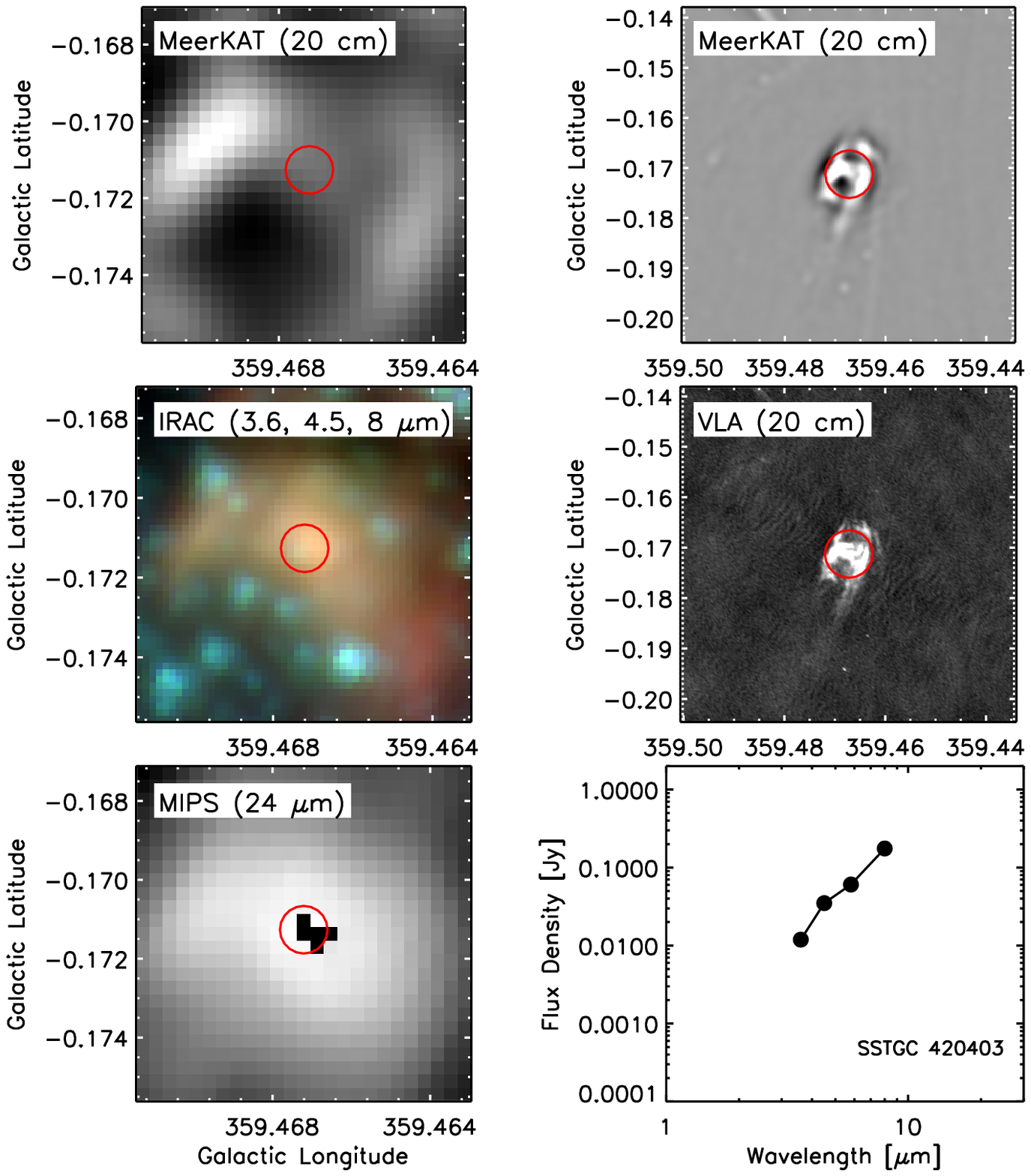}
\caption{
 {\it (y)}  Same as Fig. 3a except source 25 in Table 1.
}
\end{figure}

\addtocounter{figure}{-1}
\begin{figure}
\center
\includegraphics[]{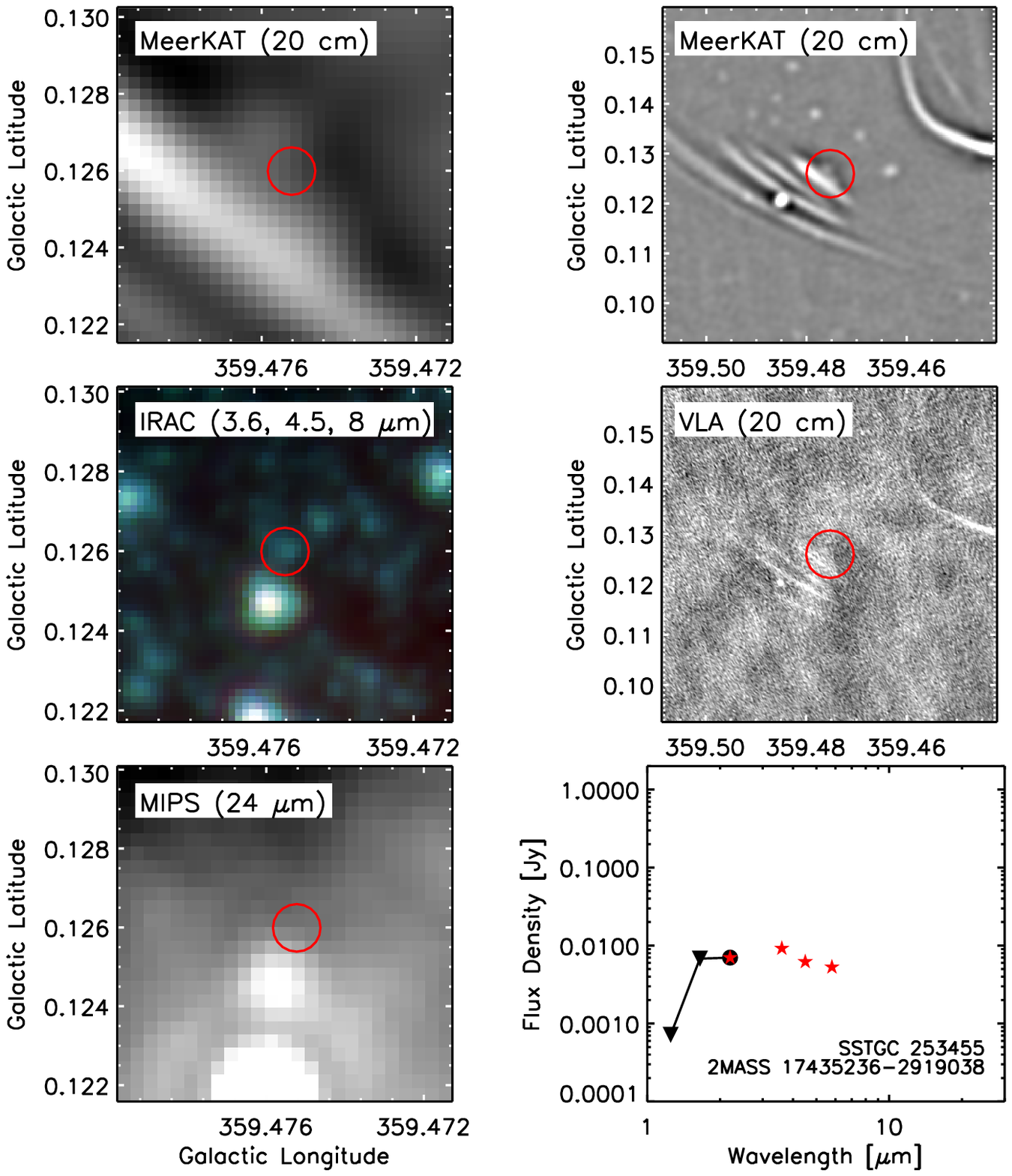}
\caption{
 {\it (z)}  Same as Fig. 3a except source 26  in Table 1.
}
\end{figure}

\addtocounter{figure}{-1}
\begin{figure}
\center
\includegraphics[]{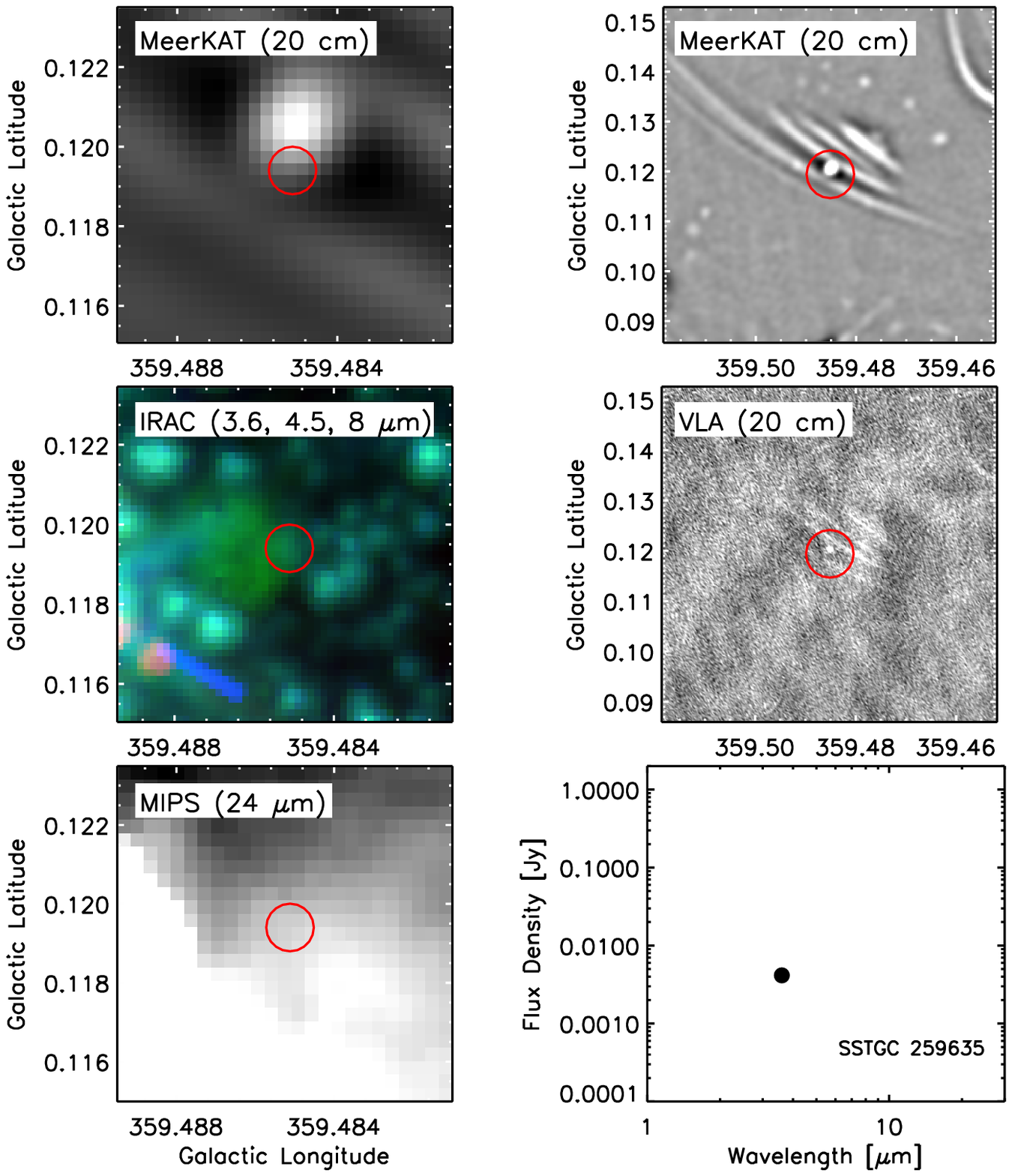}
\caption{
 {\it (aa)}  Same as Fig. 3a except source 27 in Table 1.
}
\end{figure}

\addtocounter{figure}{-1}
\begin{figure}
\center
\includegraphics[]{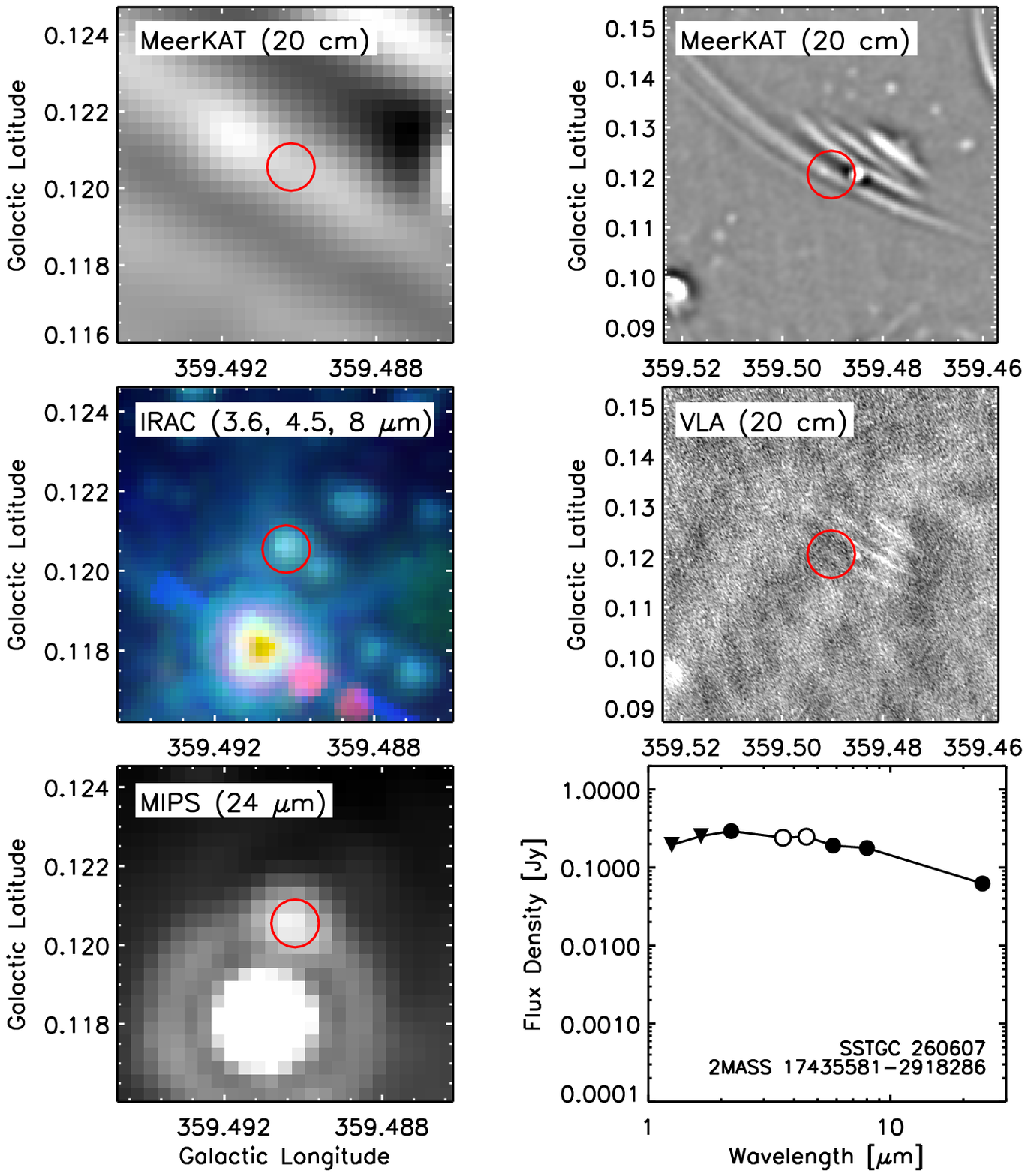}
\caption{
 {\it (bb)}  Same as Fig. 3a except source 28 in Table 1.
Open circles indicate potentially saturated IRAC measurements.
}
\end{figure}
\addtocounter{figure}{-1}
\begin{figure}
\center
\includegraphics[]{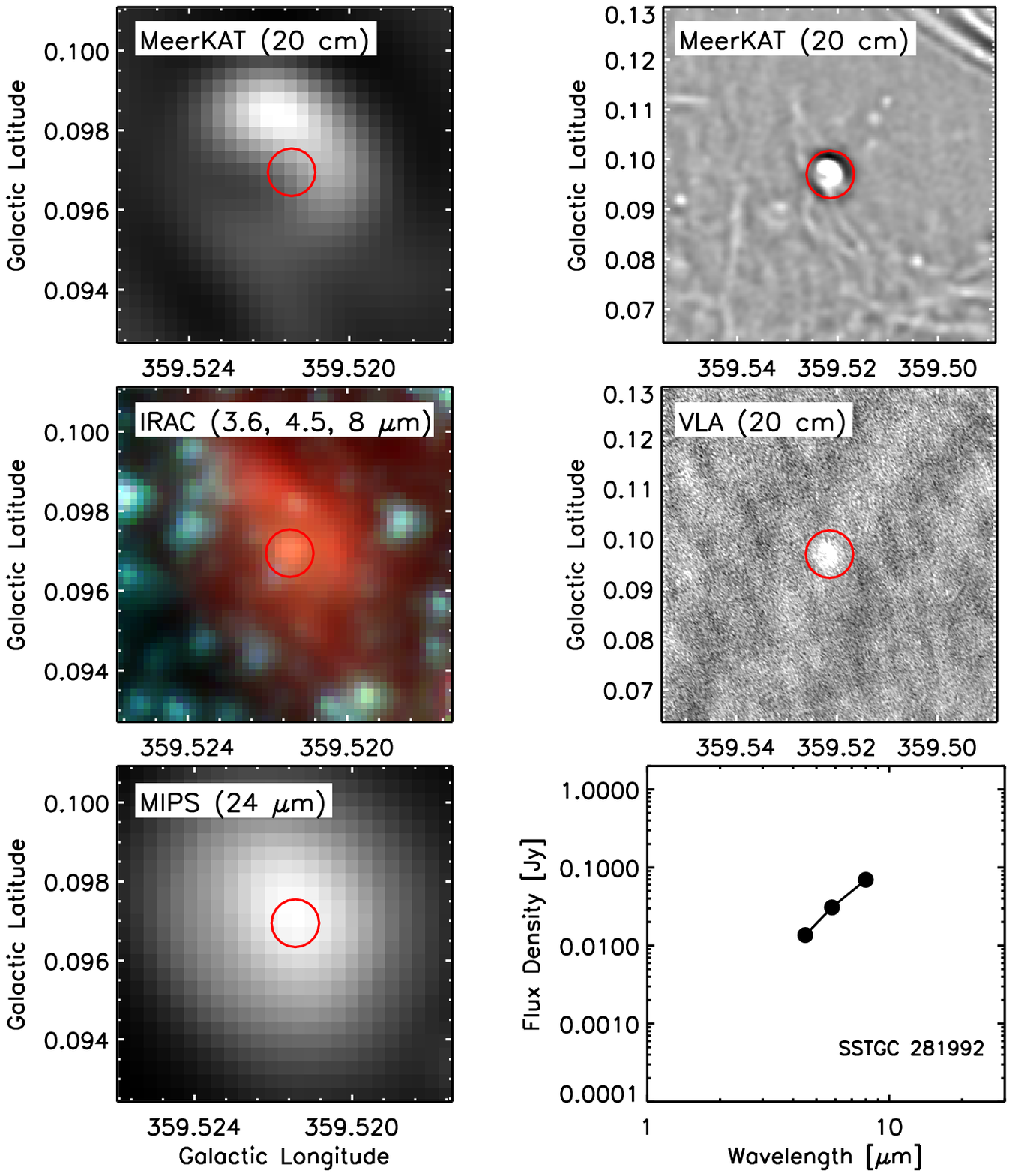}
\caption{
 {\it (cc)}  Same as Fig. 3a except source 29 in Table 1.
}
\end{figure}

\addtocounter{figure}{-1}
\begin{figure}
\center
\includegraphics[]{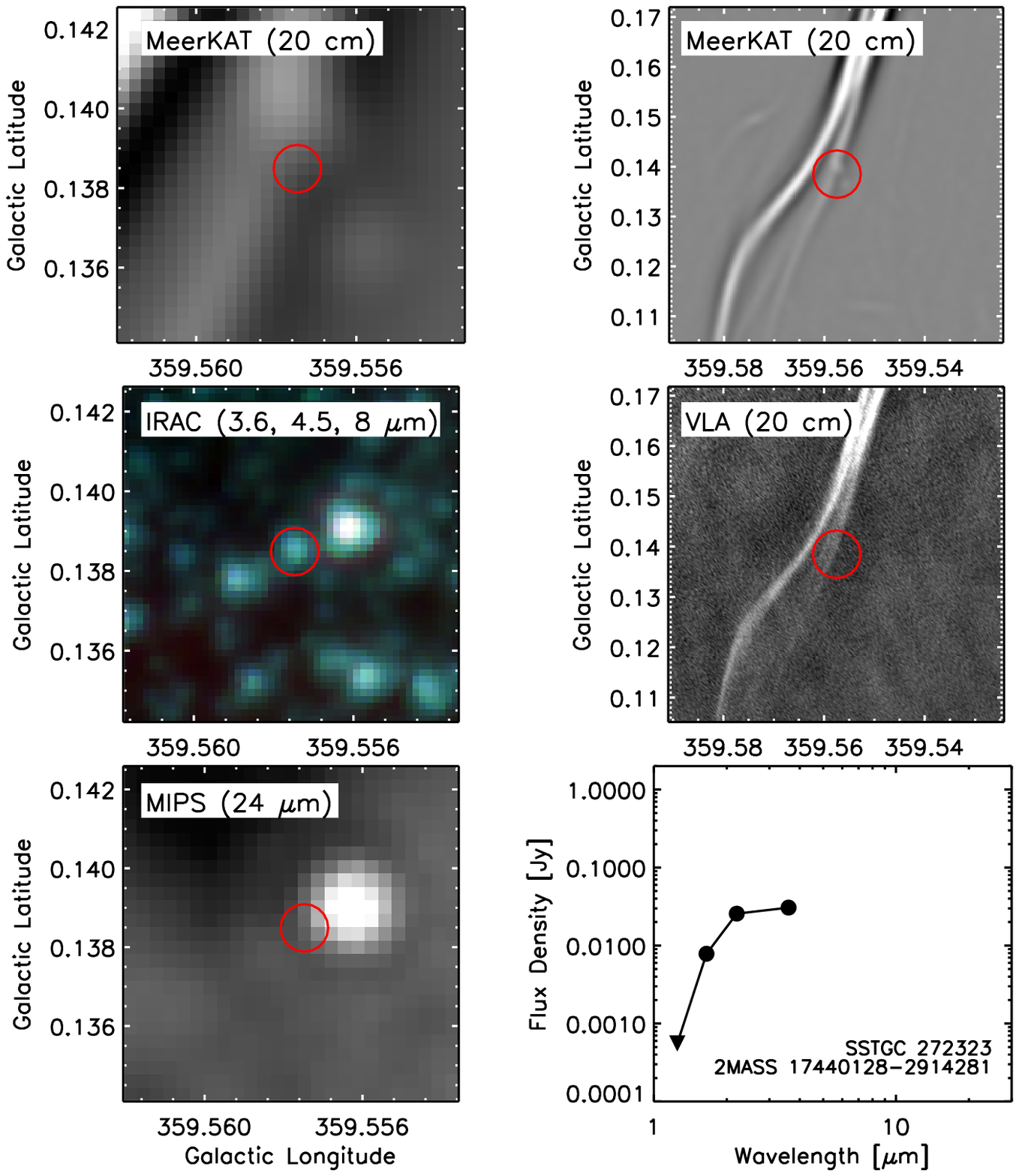}
\caption{
 {\it (dd)}  Same as Fig. 3a except source 30  in Table 1.
}
\end{figure}

\addtocounter{figure}{-1}
\begin{figure}
\center
\includegraphics[]{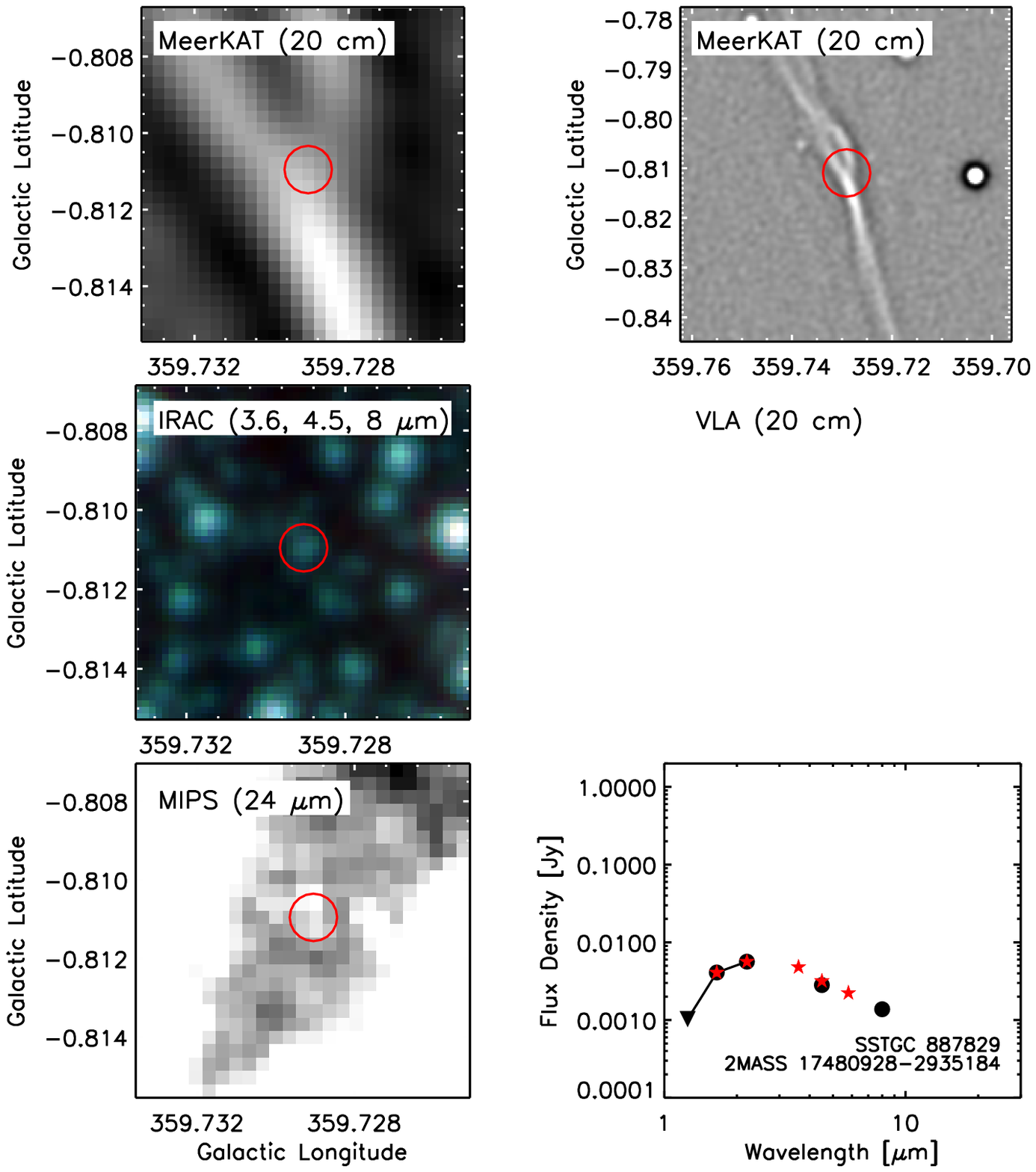}
\caption{
 {\it (ee)}  Same as Fig. 3a except source 31  in Table 1.
}
\end{figure}

\addtocounter{figure}{-1}
\begin{figure}
\center
\includegraphics[]{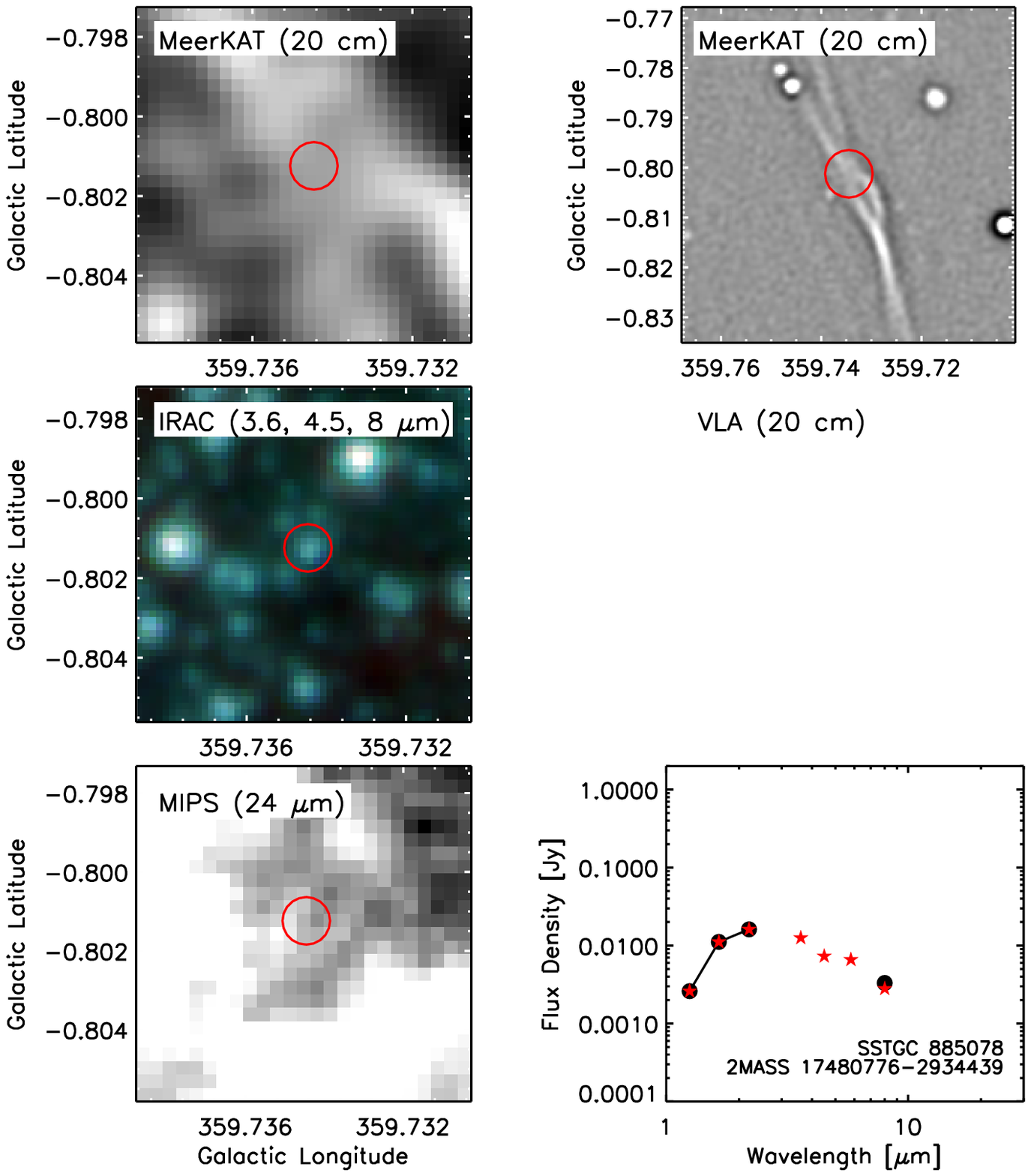}
\caption{
 {\it (ff)}  Same as Fig. 3a except source 32  in Table 1.
}
\end{figure}

\addtocounter{figure}{-1}
\begin{figure}
\center
\includegraphics[]{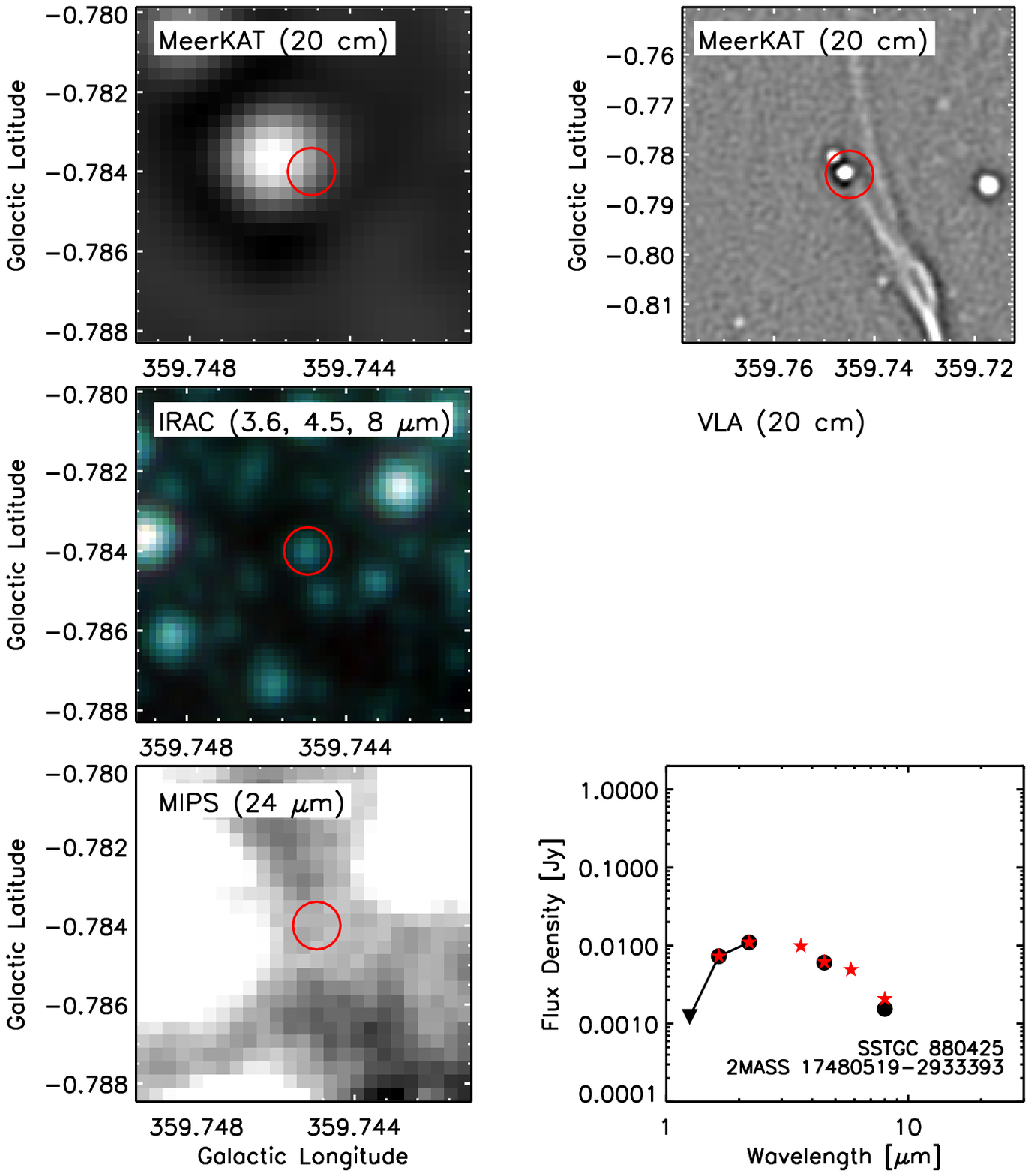}
\caption{
 {\it (gg)}  Same as Fig. 3a except source 33 in Table 1.}
\end{figure}

\addtocounter{figure}{-1}
\begin{figure}
\center
\includegraphics[]{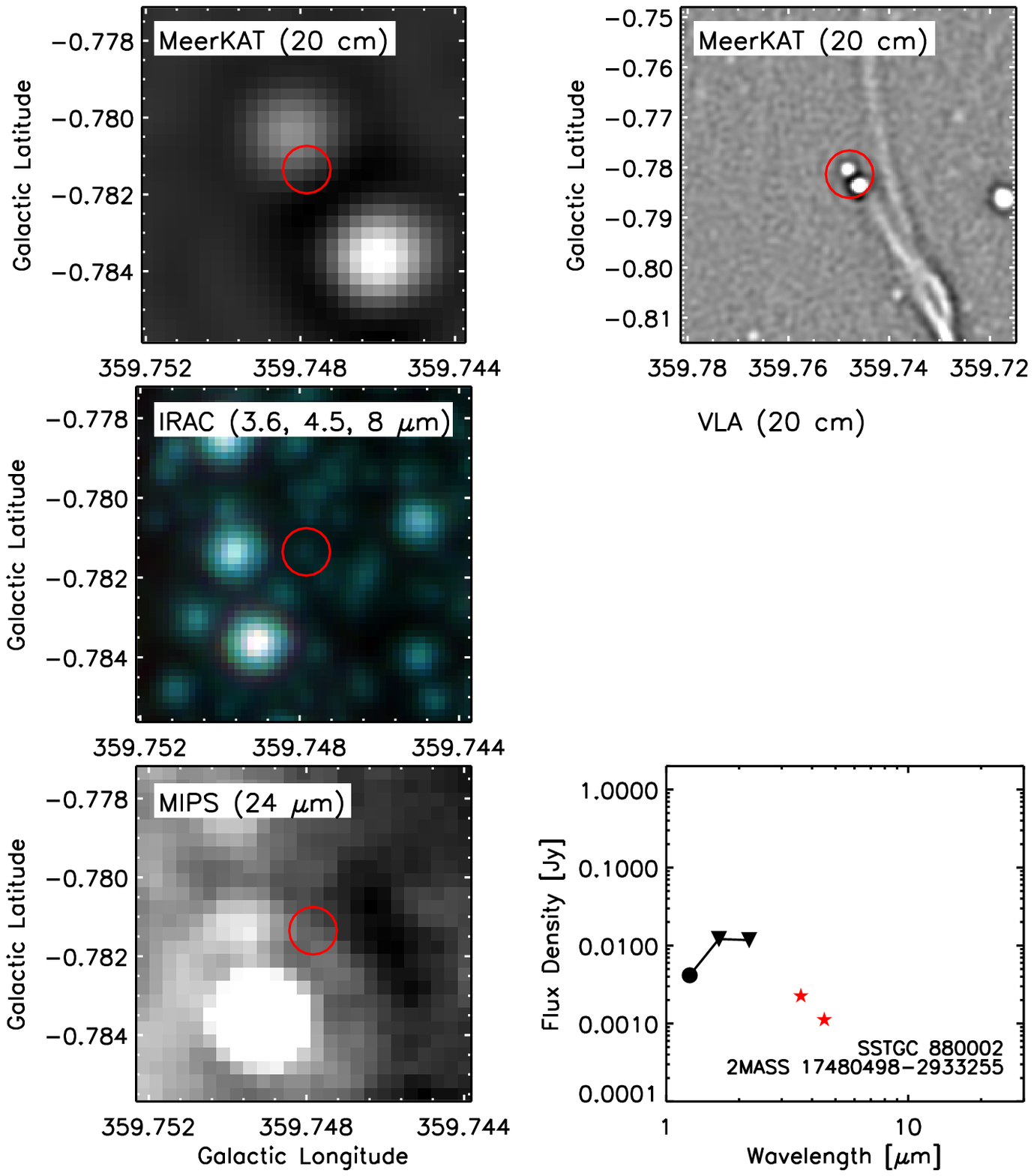}
\caption{
 {\it (hh)}  Same as Fig. 3a except source 34  in Table 1.
}
\end{figure}

\addtocounter{figure}{-1}
\begin{figure}
\center
\includegraphics[]{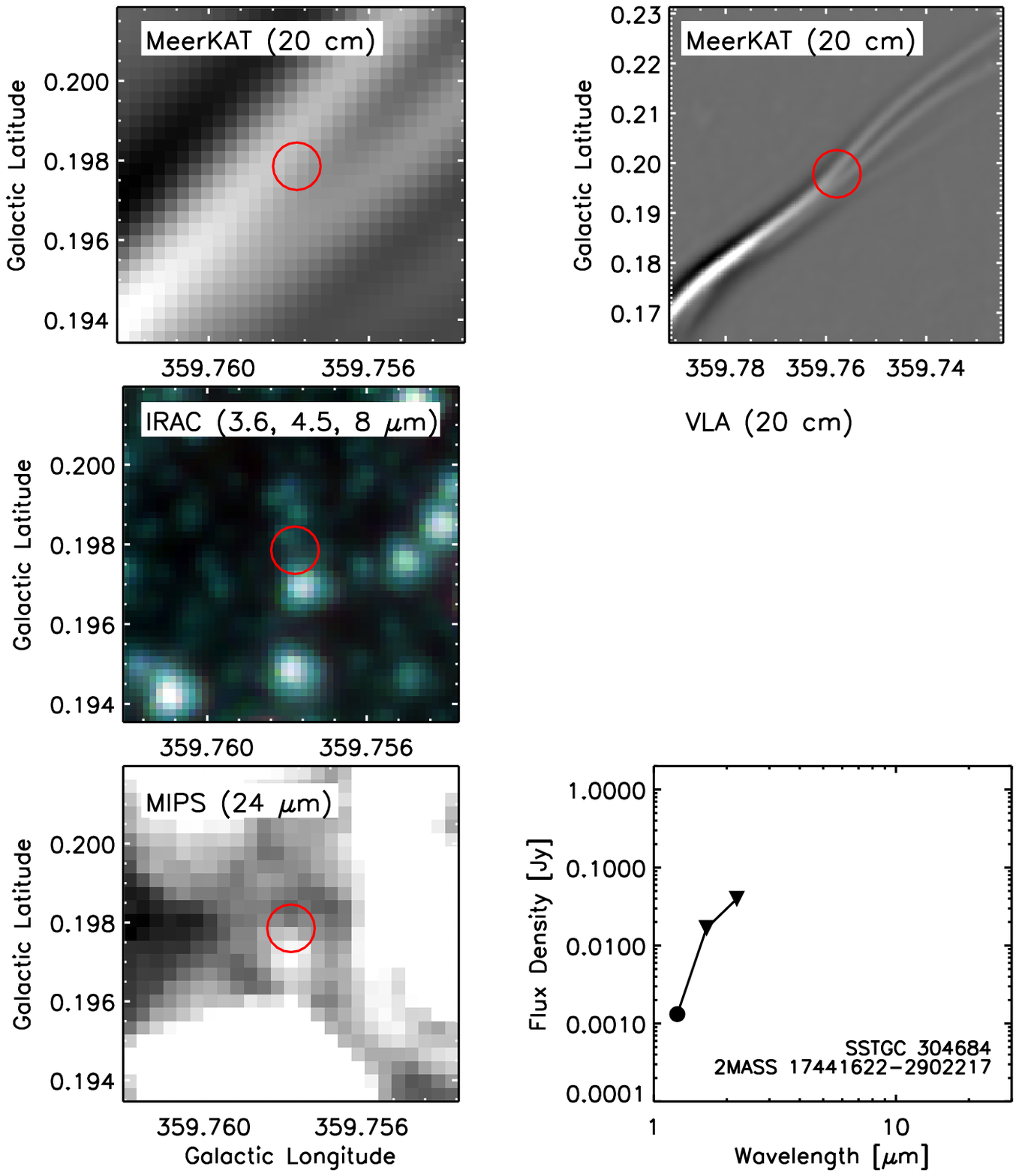}
\caption{
{\it (ii)}  Same as Fig. 3a except source 35  in Table 1.}
\end{figure}

\addtocounter{figure}{-1}
\begin{figure}
\center
\includegraphics[]{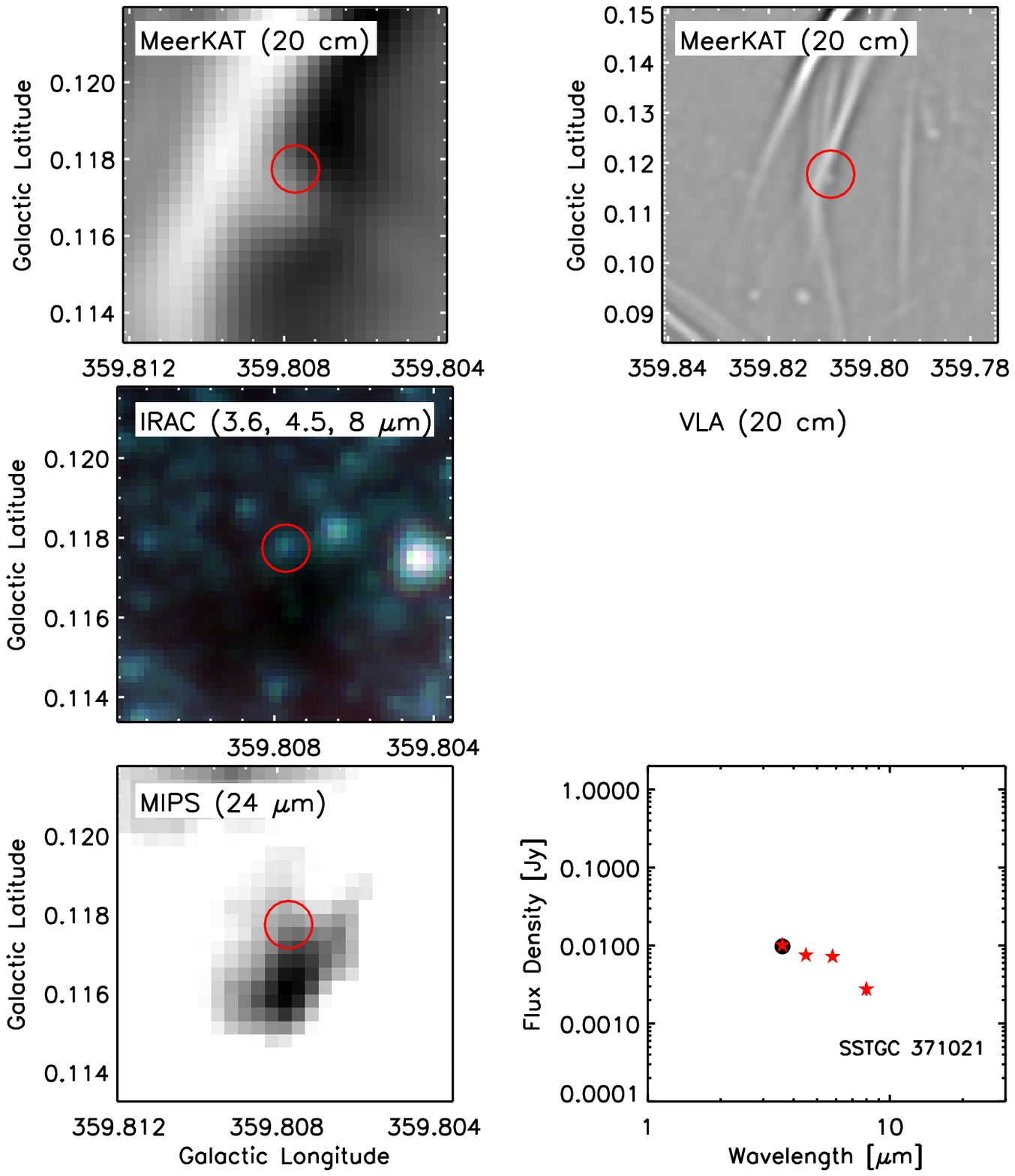}
\caption{
{\it (jj)}  Same as Fig. 3a except source 36 in Table 1.
}
\end{figure}

\addtocounter{figure}{-1}
\begin{figure}
\center
\includegraphics[]{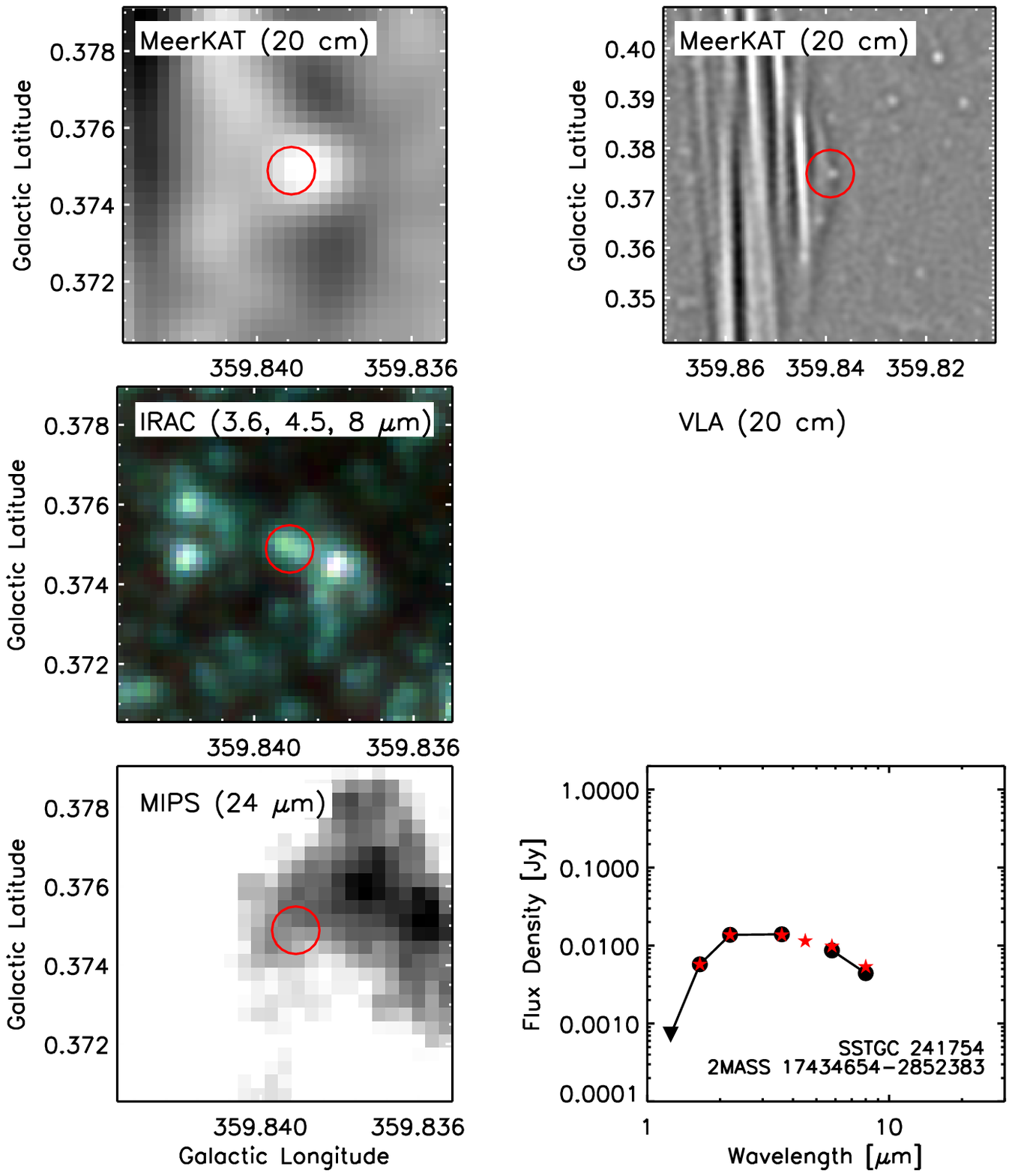}
\caption{
 {\it (kk)}  Same as Fig. 3a except source 37  in Table 1.
}
\end{figure}

\addtocounter{figure}{-1}
\begin{figure}
\center
\includegraphics[]{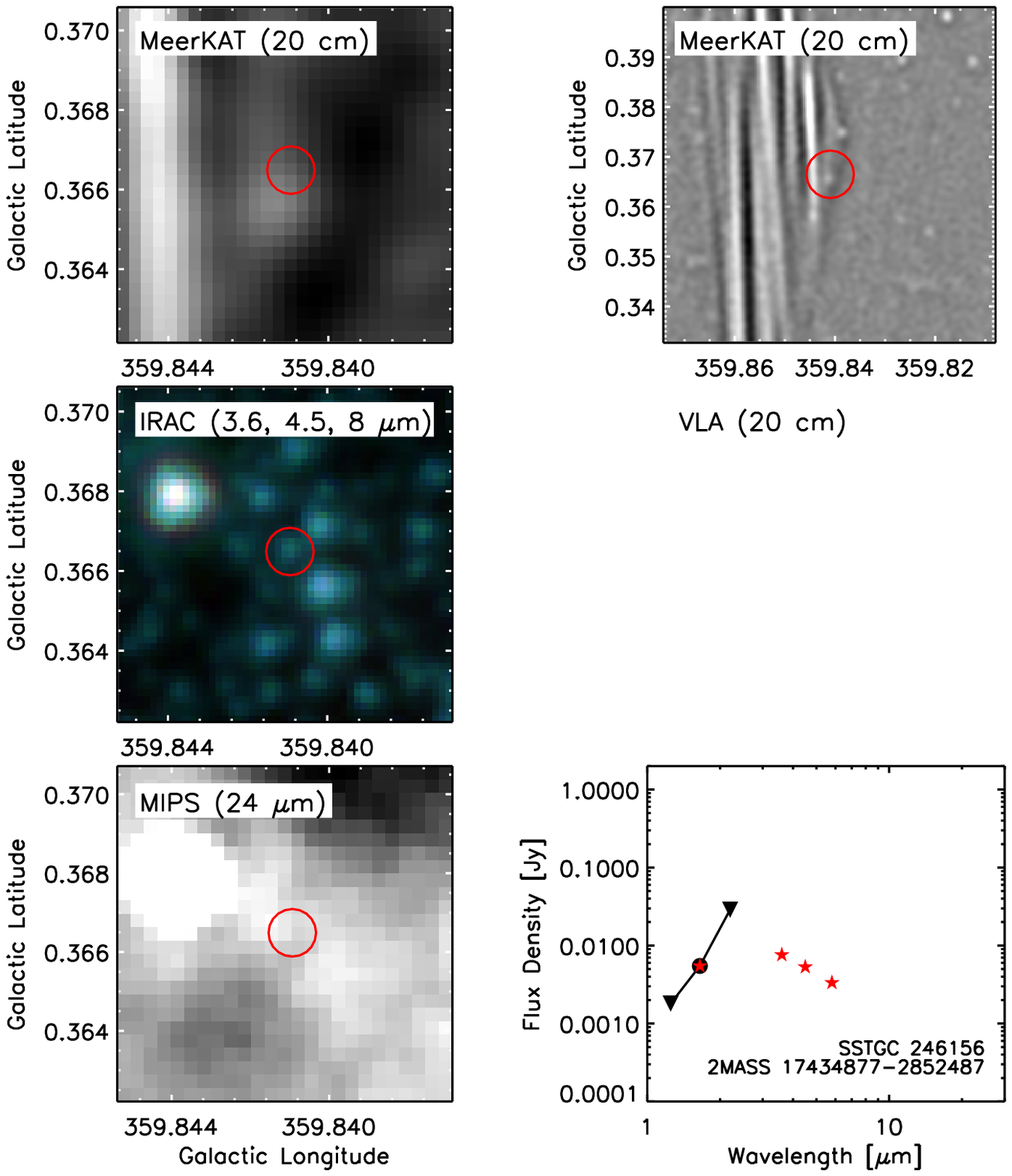}
\caption{
 {\it (ll)}  Same as Fig. 3a except source 38  in Table 1.
}
\end{figure}

\addtocounter{figure}{-1}
\begin{figure}
\center
\includegraphics[]{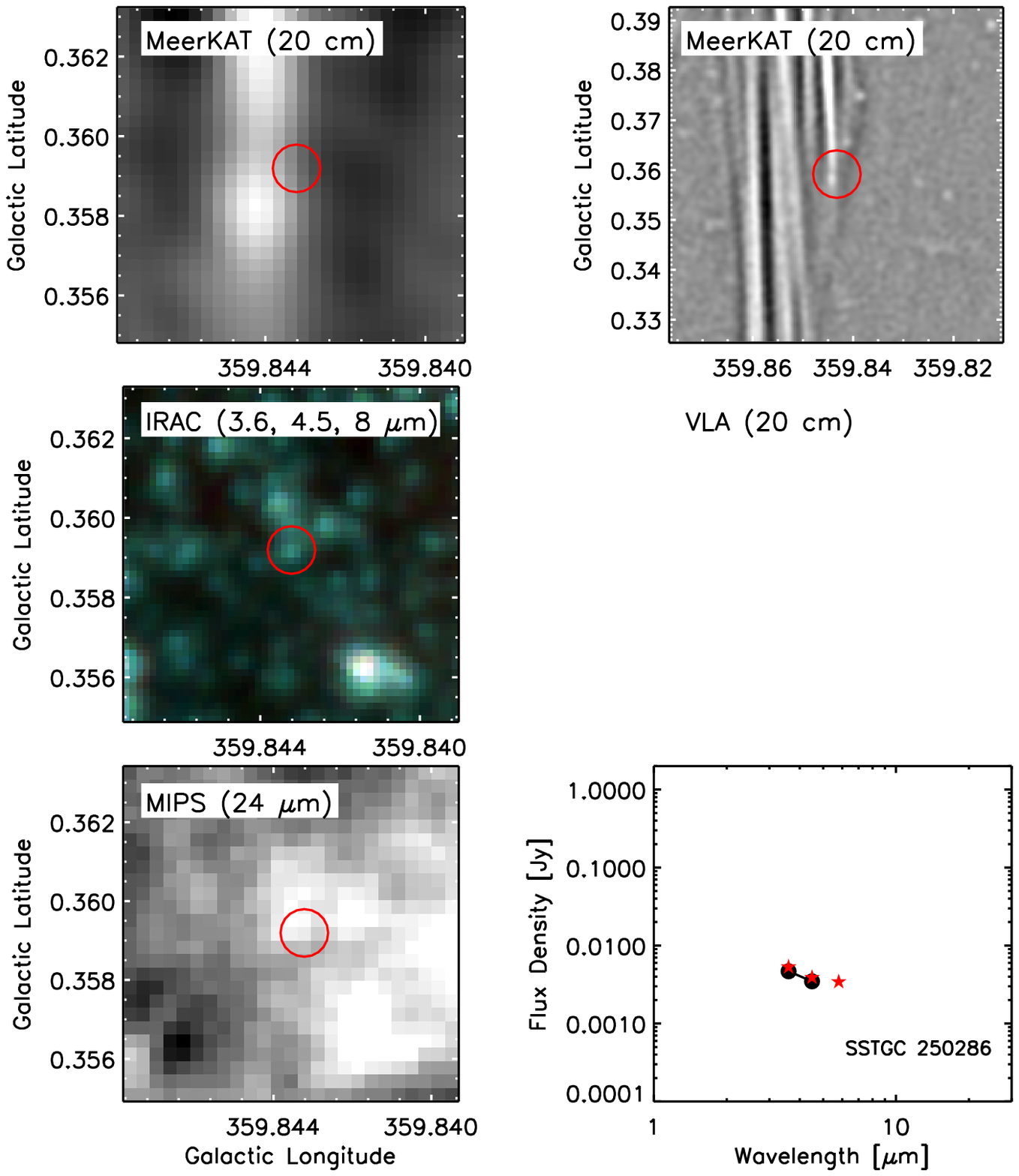}
\caption{
 {\it (mm)}  Same as Fig. 3a except source 39 in Table 1.
}
\end{figure}

\addtocounter{figure}{-1}
\begin{figure}
\center
\includegraphics[]{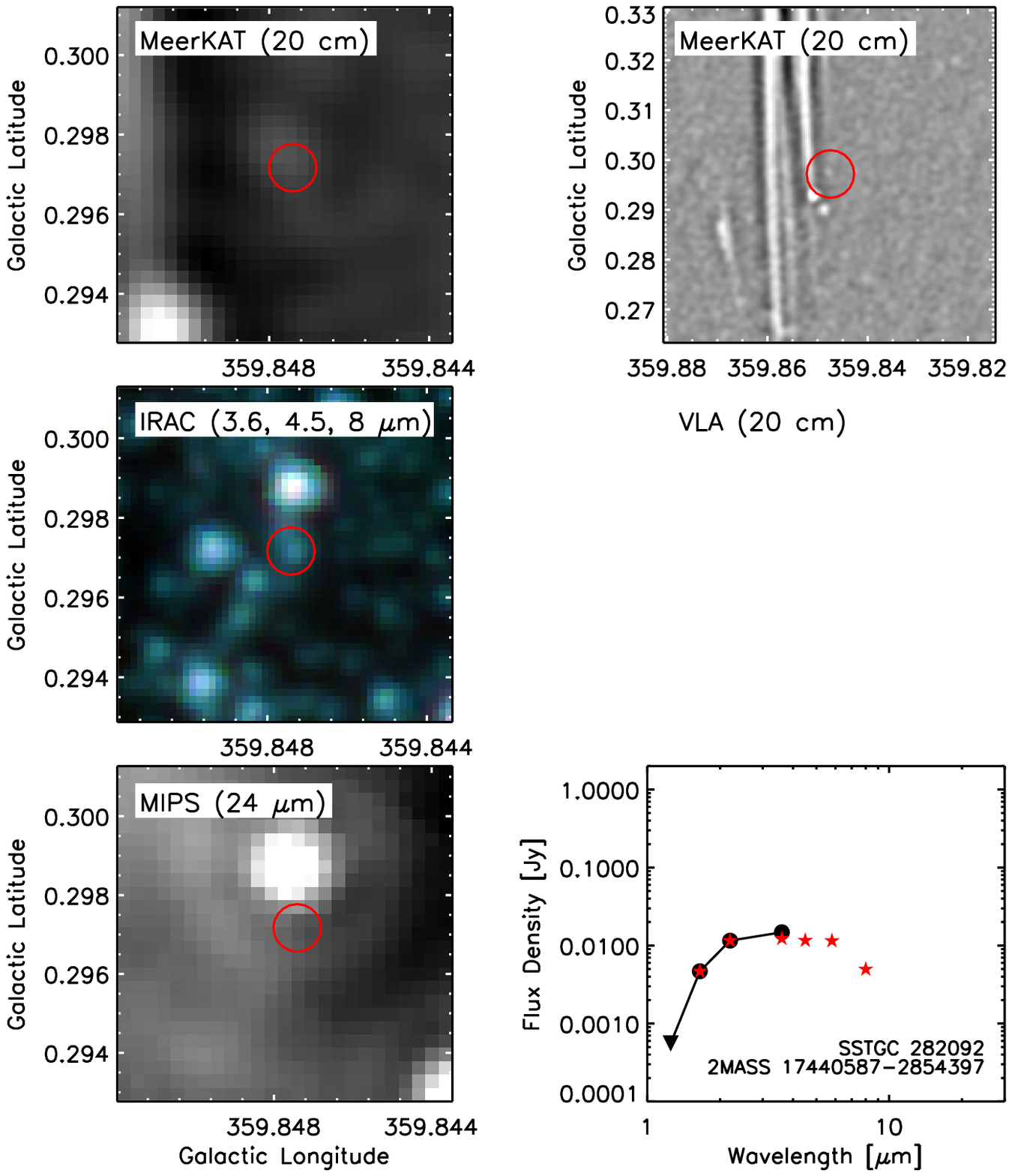}
\caption{
 {\it (nn)}  Same as Fig. 3a except source 40   in Table 1.
}
\end{figure}

\addtocounter{figure}{-1}
\begin{figure}
\center
\includegraphics[]{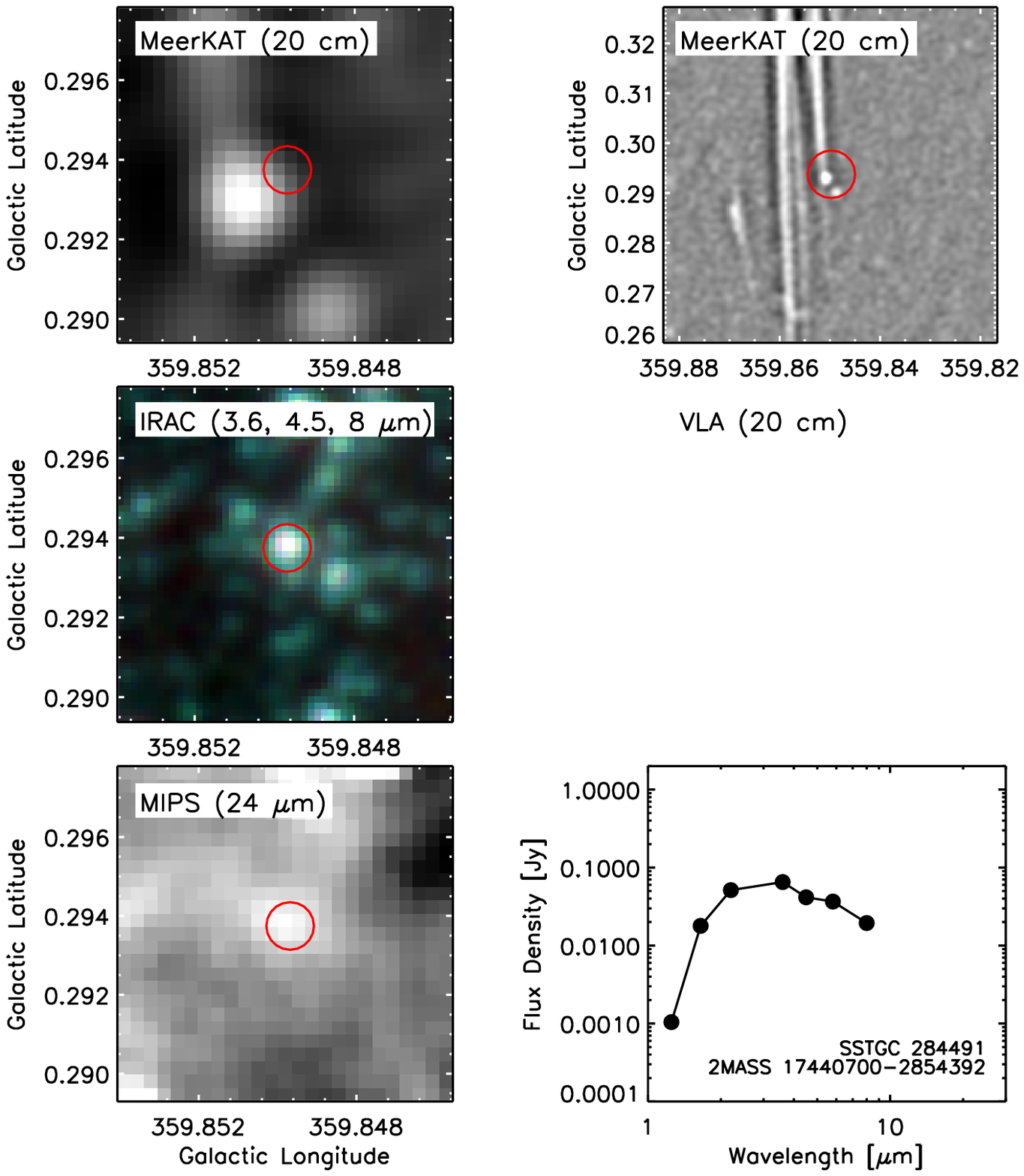}
\caption{
 {\it (oo)}  Same as Fig. 3a except source 41  in Table 1.
}
\end{figure}

\addtocounter{figure}{-1}
\begin{figure}
\center
\includegraphics[]{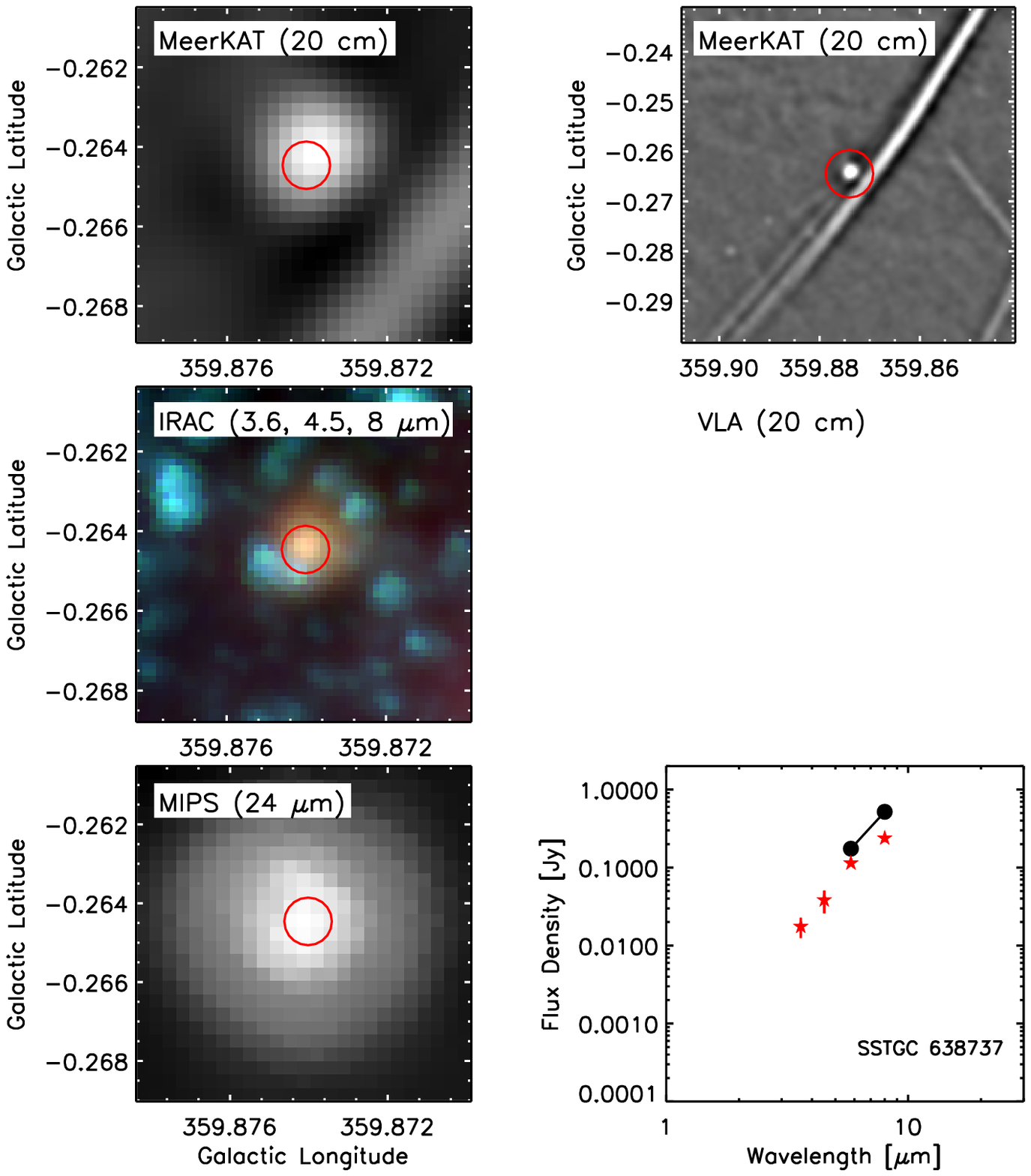}
\caption{
 {\it (pp)}  Same as Fig. 3a except source 42  in Table 1.
}
\end{figure}

\addtocounter{figure}{-1}
\begin{figure}
\center
\includegraphics[]{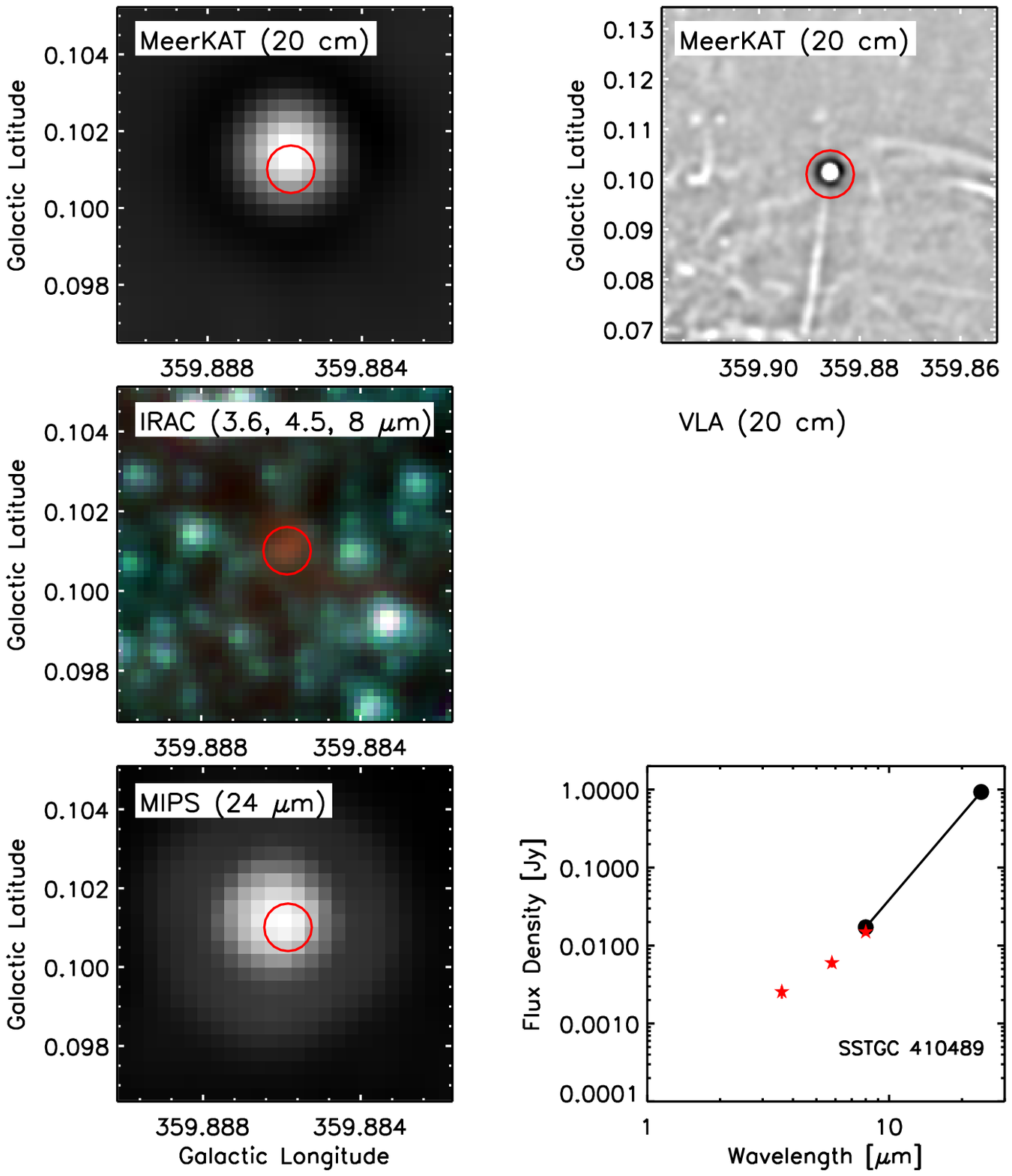}
\caption{
 {\it (qq)}  Same as Fig. 3a except source 43  in Table 1.
}
\end{figure}

\addtocounter{figure}{-1} 
\begin{figure} 
\center 
\includegraphics[]{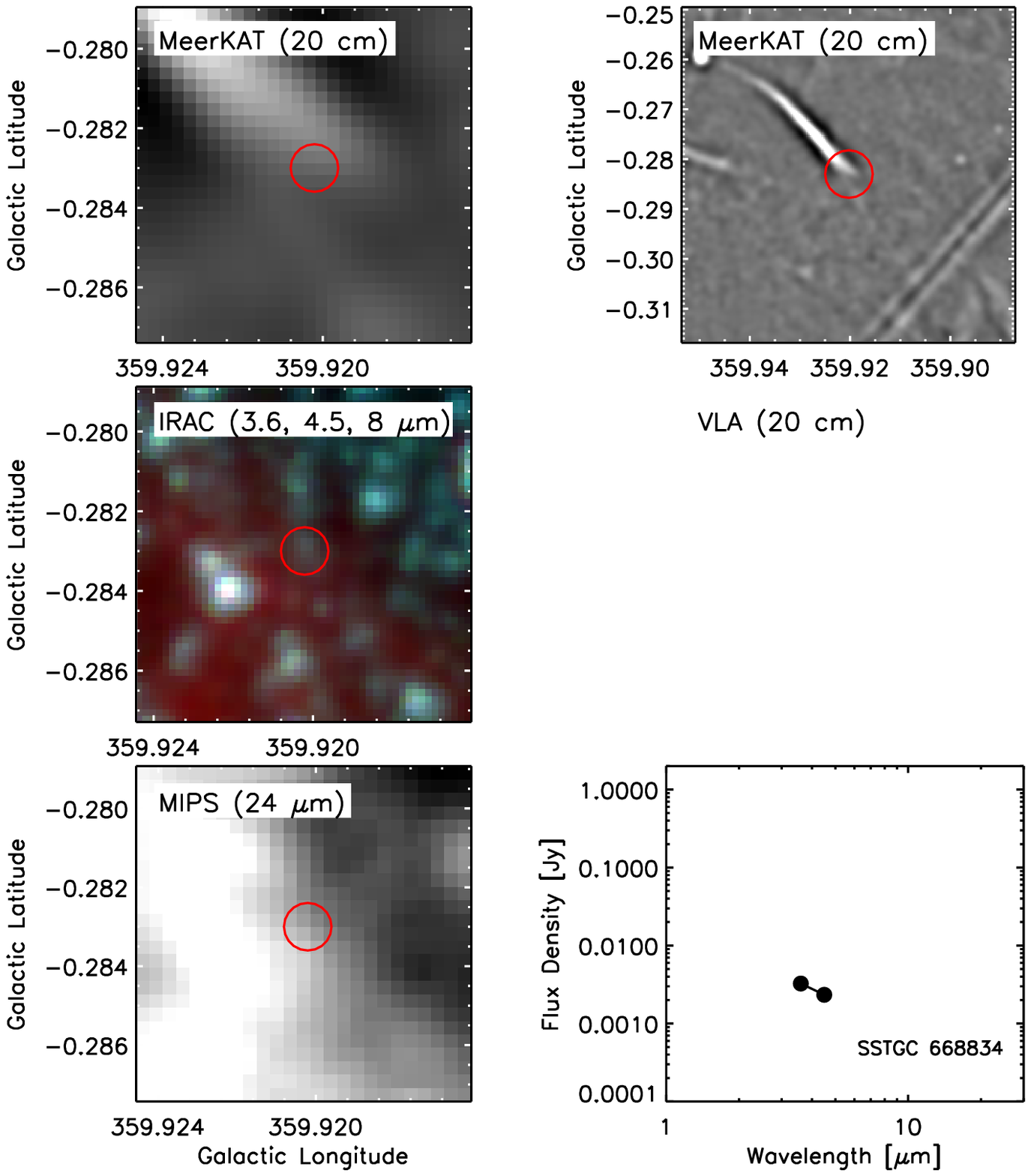}
\caption{
 {\it (rr)}  Same as Fig. 3a except source 44  in Table 1.
}
\end{figure}

\addtocounter{figure}{-1}
\begin{figure}
\center
\includegraphics[]{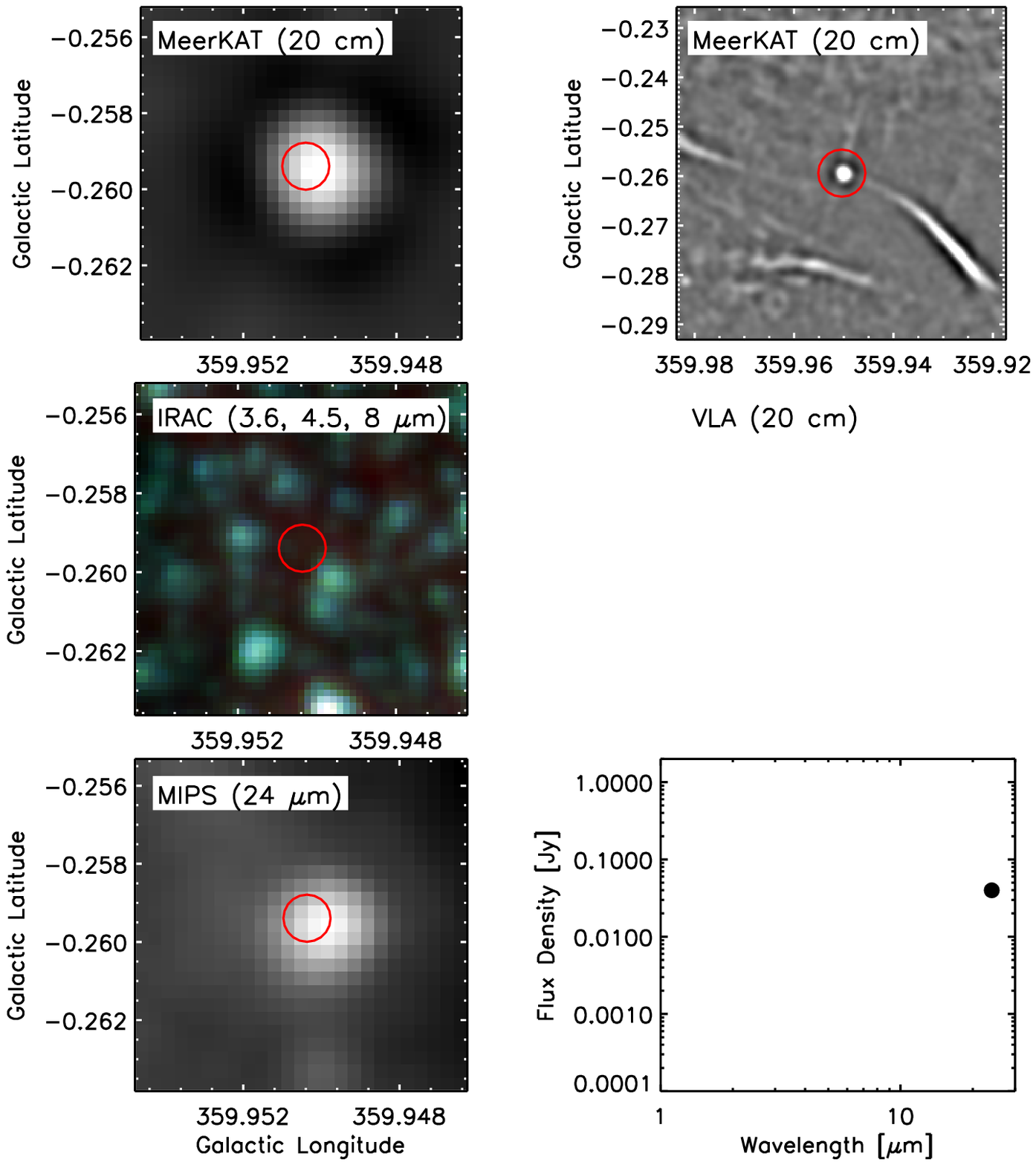}
\caption{
 {\it (ss)}  Same as Fig. 3a except source 45  in Table 1.
}
\end{figure}
\addtocounter{figure}{-1}
\begin{figure}
\center
\includegraphics[]{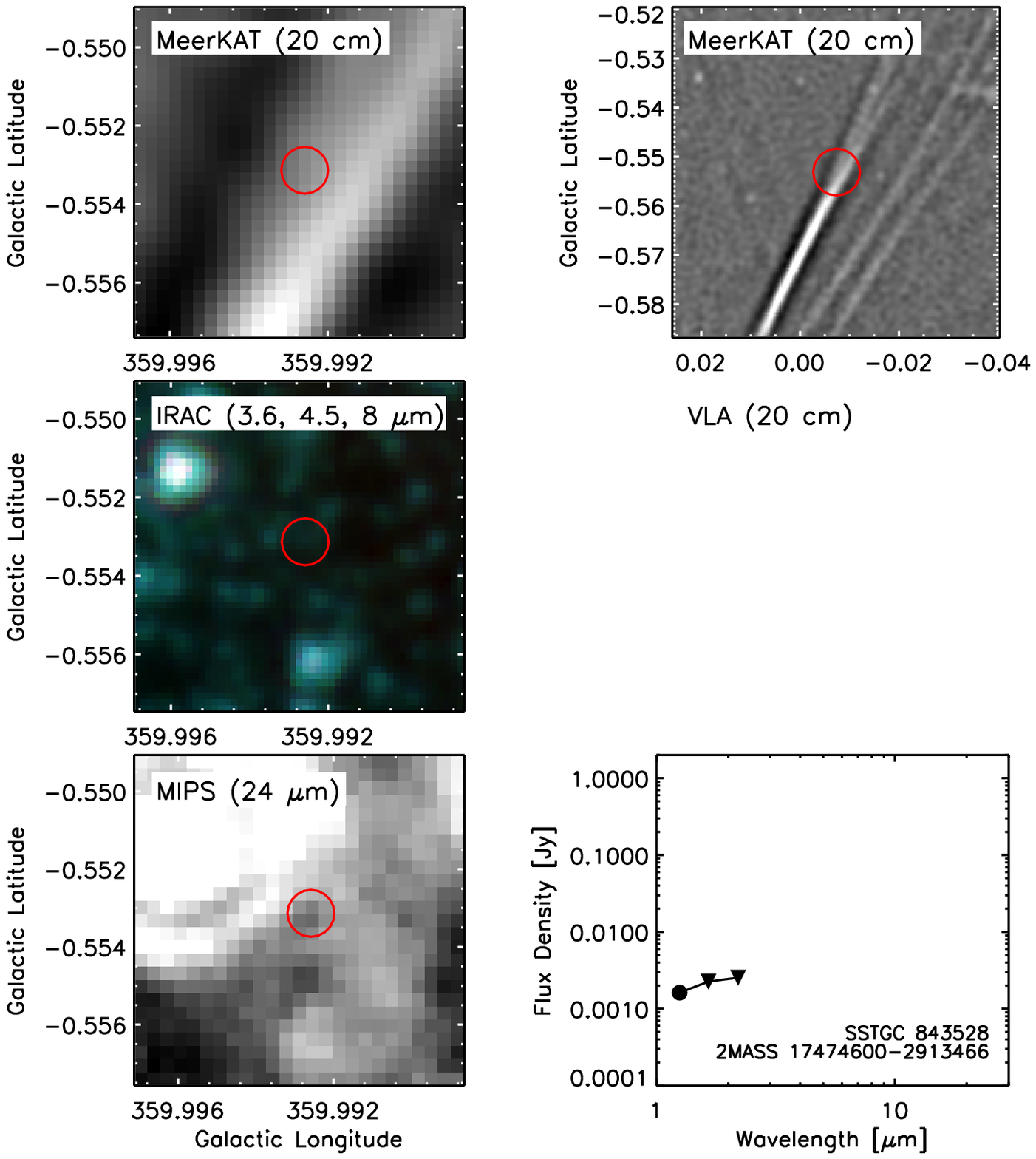}
\caption{
 {\it (tt)}  Same as Fig. 3a except source 46 in Table 1.
}
\end{figure}
\addtocounter{figure}{-1}
\begin{figure}
\center
\includegraphics[]{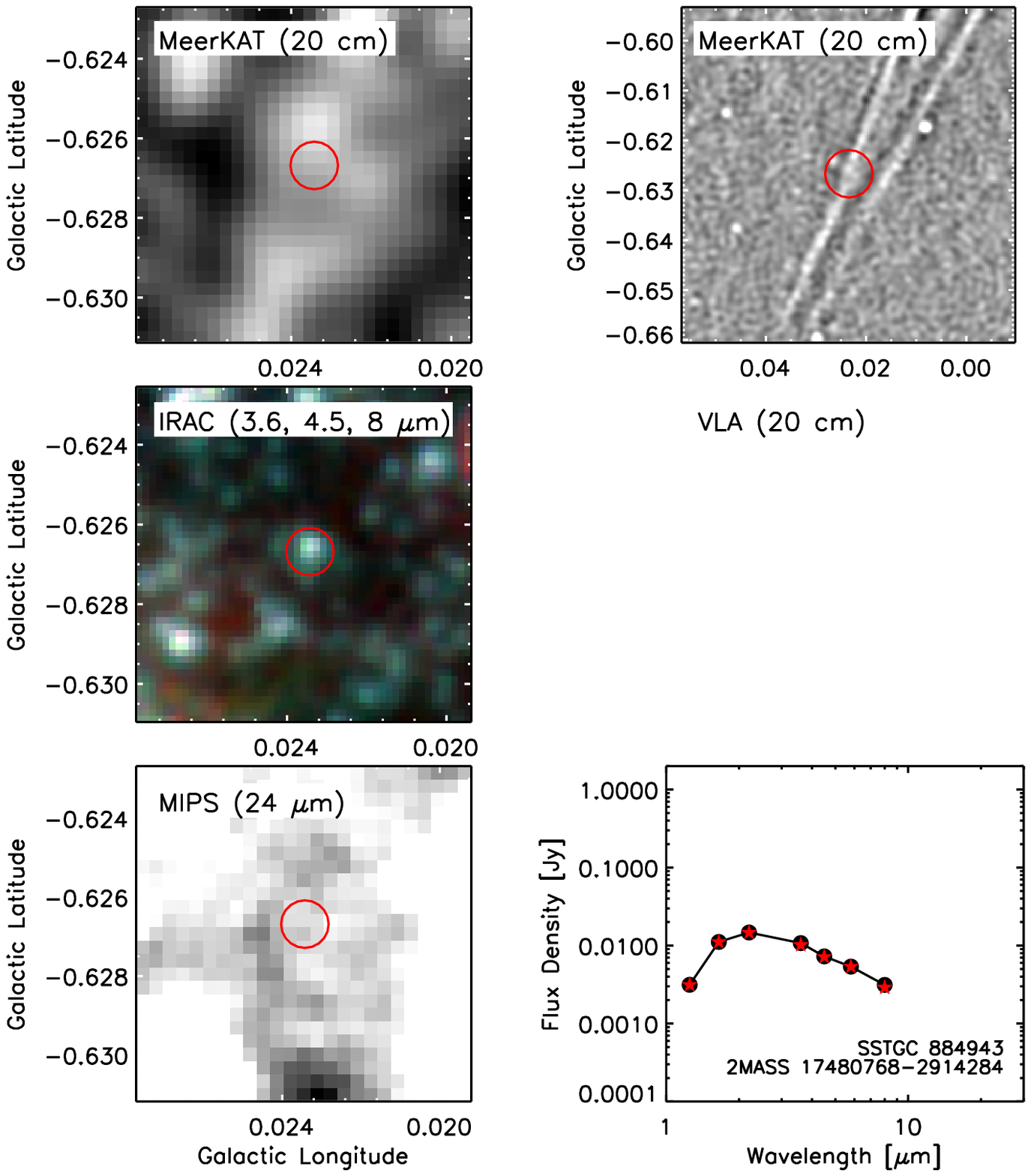}
\caption{
 {\it (uu)}  Same as Fig. 3a except source 47  in Table 1.
}
\end{figure}
\addtocounter{figure}{-1}
\begin{figure}
\center
\includegraphics[]{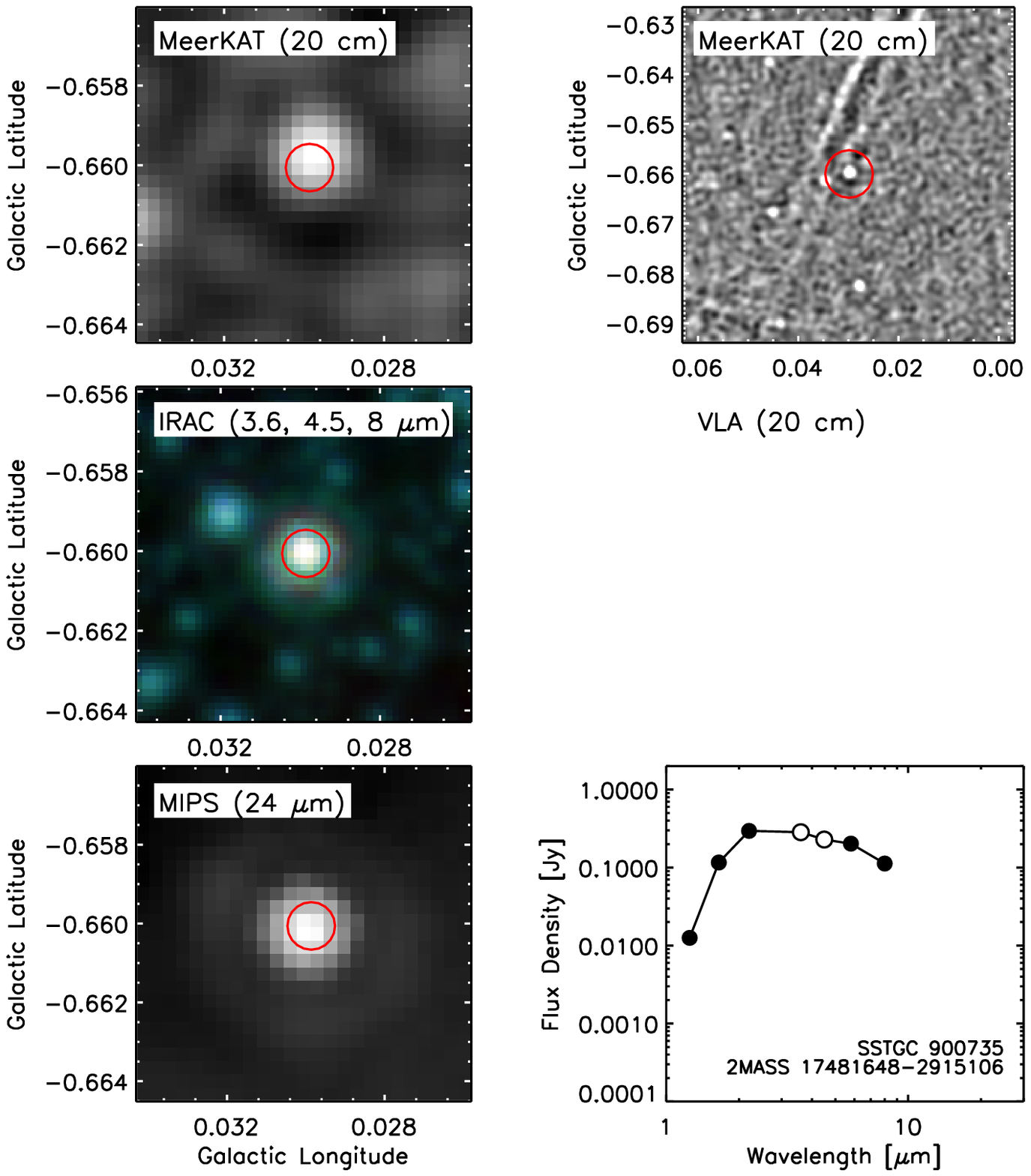}
\caption{
 {\it (vv)}  Same as Fig. 3a except source 48 in Table 1.
Open circles indicate potentially saturated IRAC measurements.
}
\end{figure}

\addtocounter{figure}{-1}
\begin{figure}
\center
\includegraphics[]{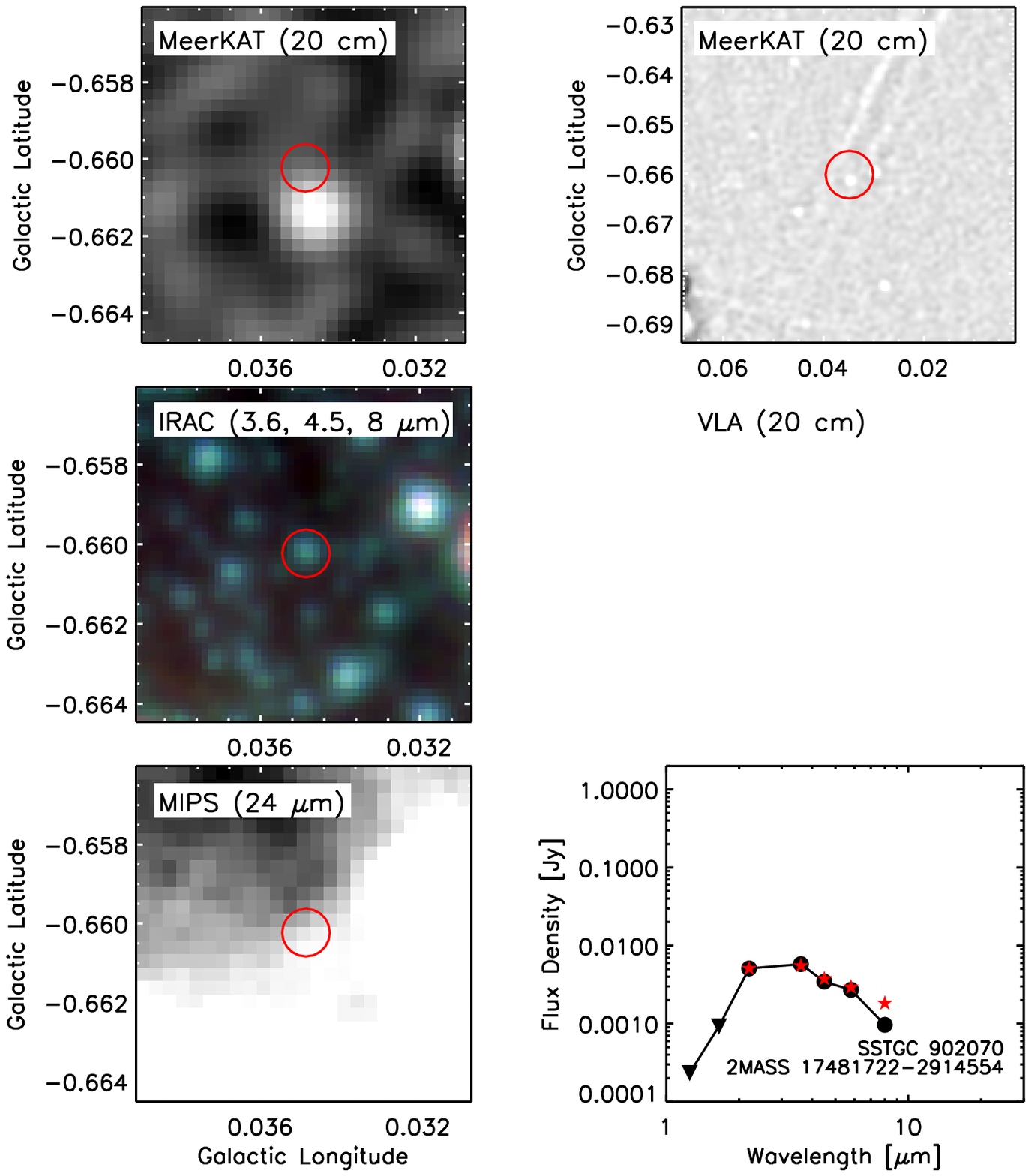}
\caption{
{\it (ww)}  Same as Fig. 3a except source  49  in Table 1.
}
\end{figure}

\clearpage

\addtocounter{figure}{-1}
\begin{figure}
\center
\includegraphics[]{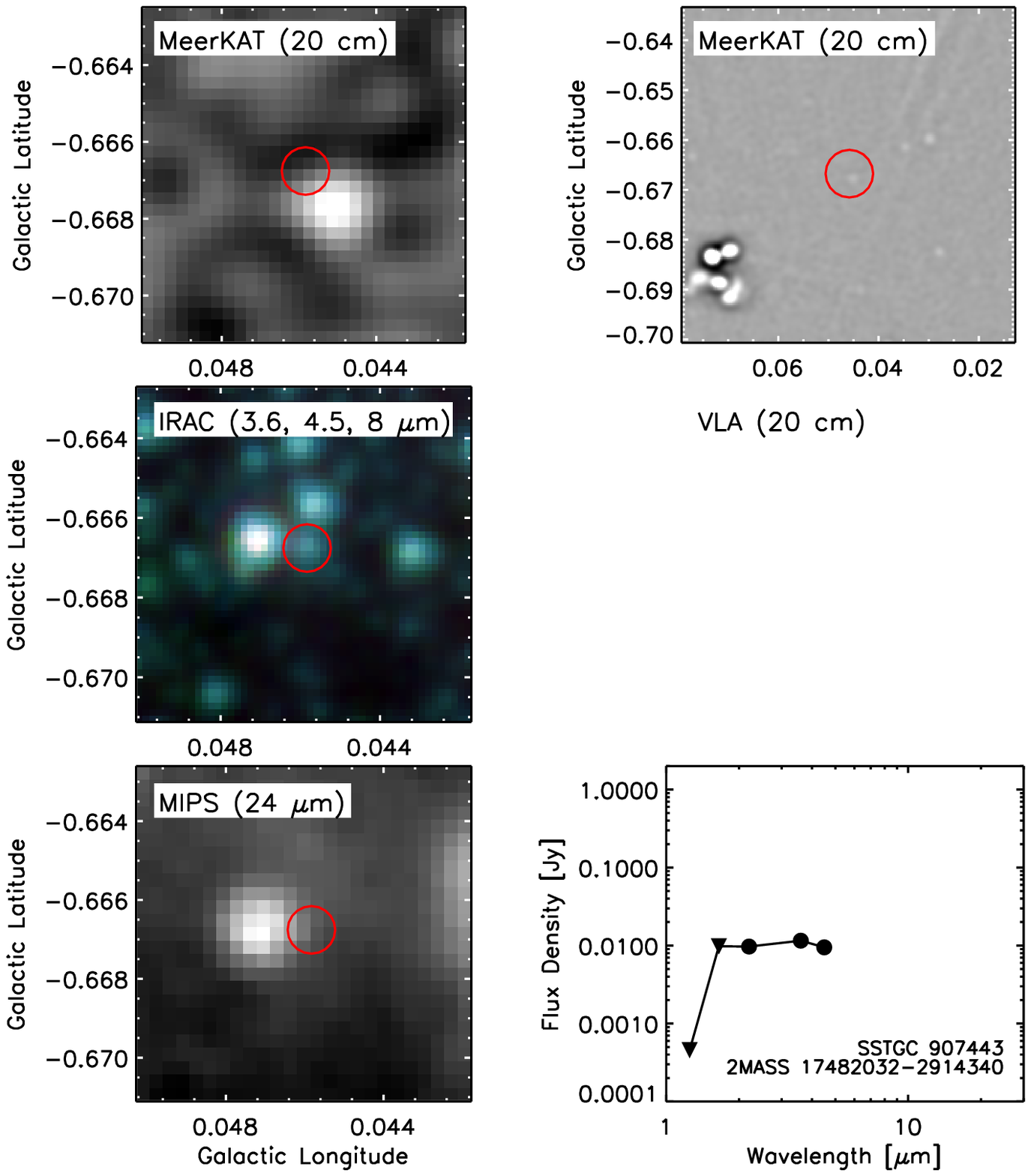}
\caption{
{\it (xx)}  Same as Fig. 3a except source 50 in Table 1.
}
\end{figure}

\addtocounter{figure}{-1}
\begin{figure}
\center
\includegraphics[]{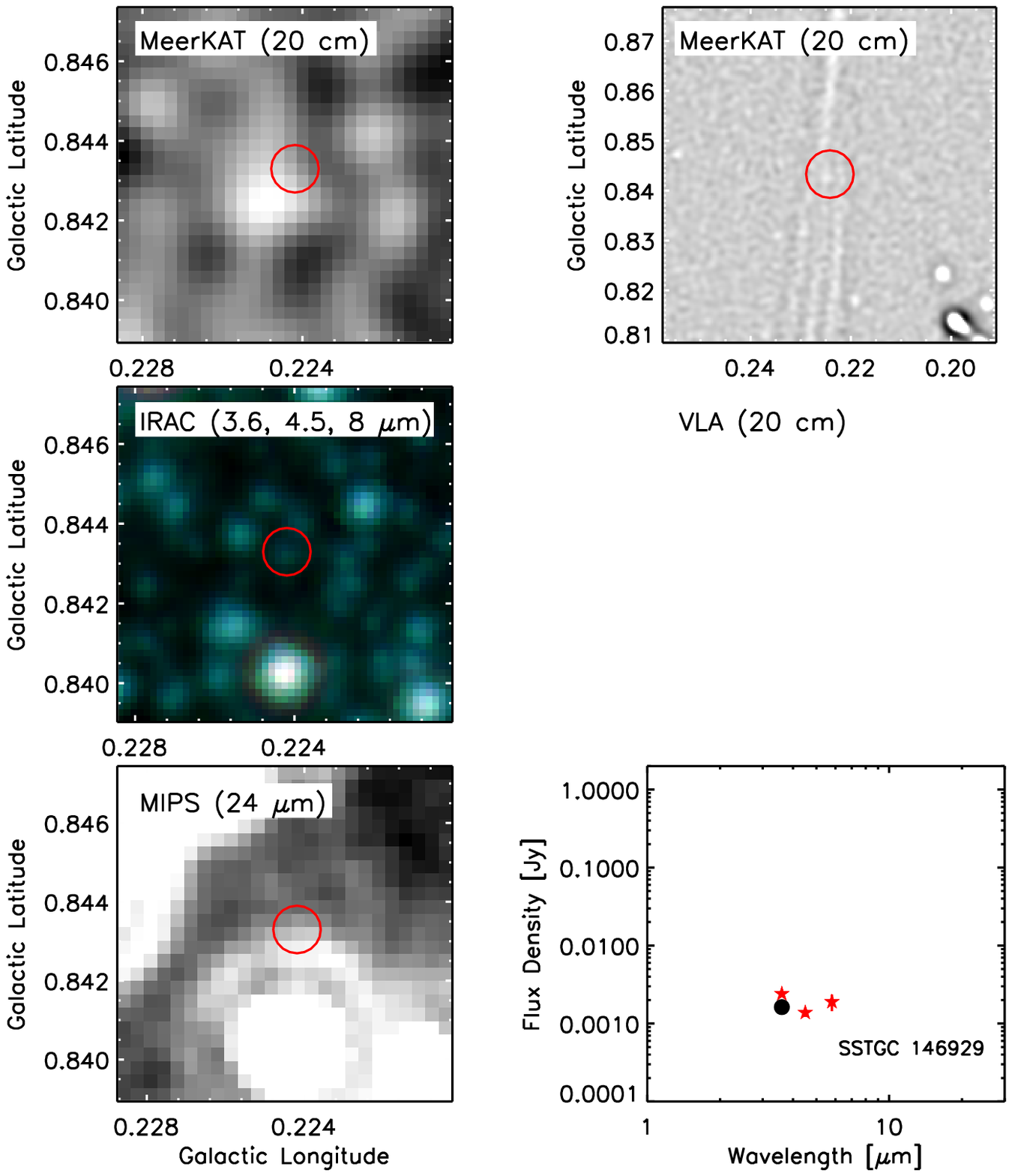}
\caption{
 {\it (yy)}  Same as Fig. 3a except source 51  in Table 1.
}
\end{figure}

\clearpage

\addtocounter{figure}{-1}
\begin{figure}
\center
\includegraphics[]{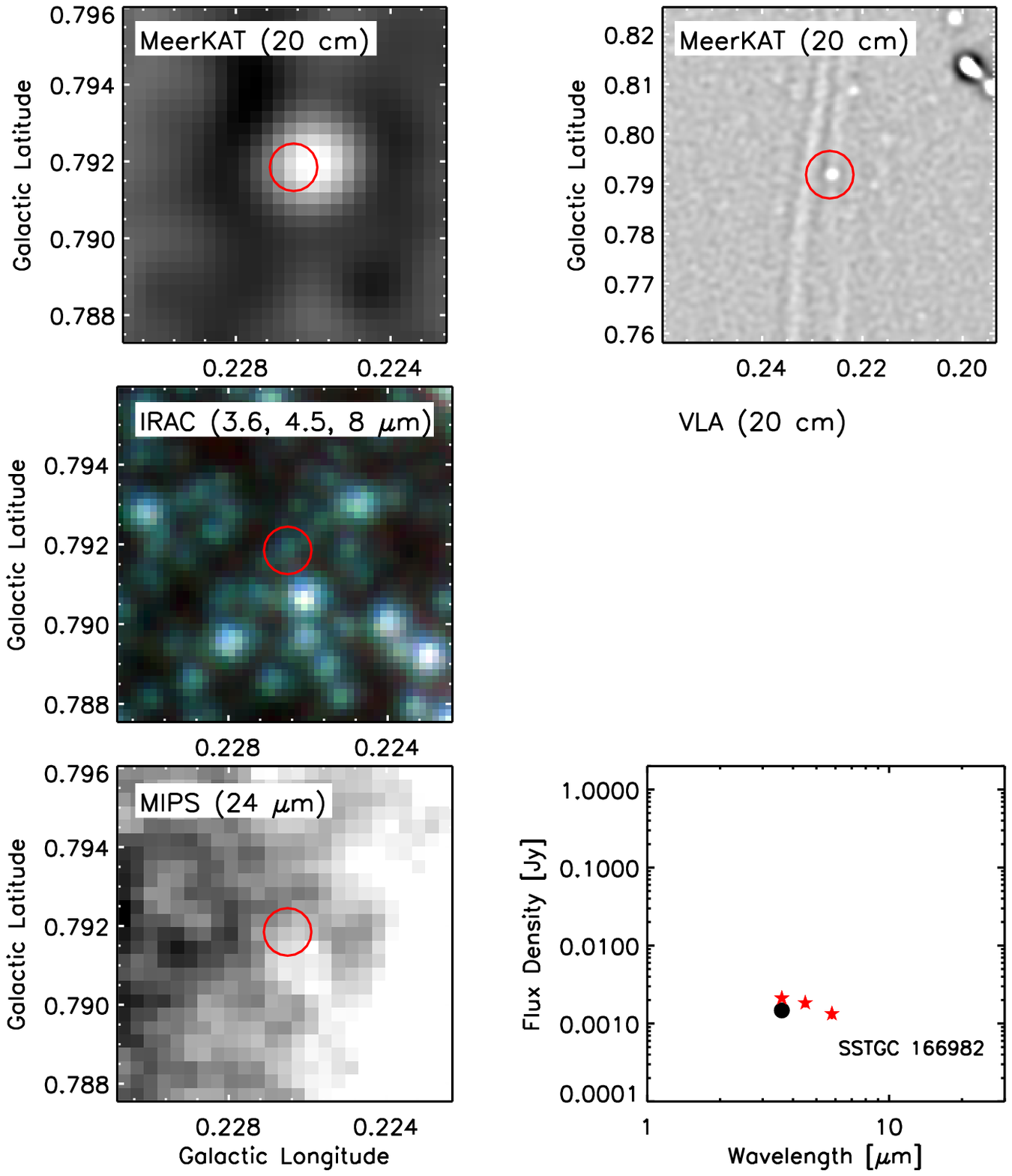}
\caption{
 {\it (zz)}  Same as Fig. 3a except source 52  in Table 1.
}
\end{figure}

\addtocounter{figure}{-1}
\begin{figure}
\center
\includegraphics[]{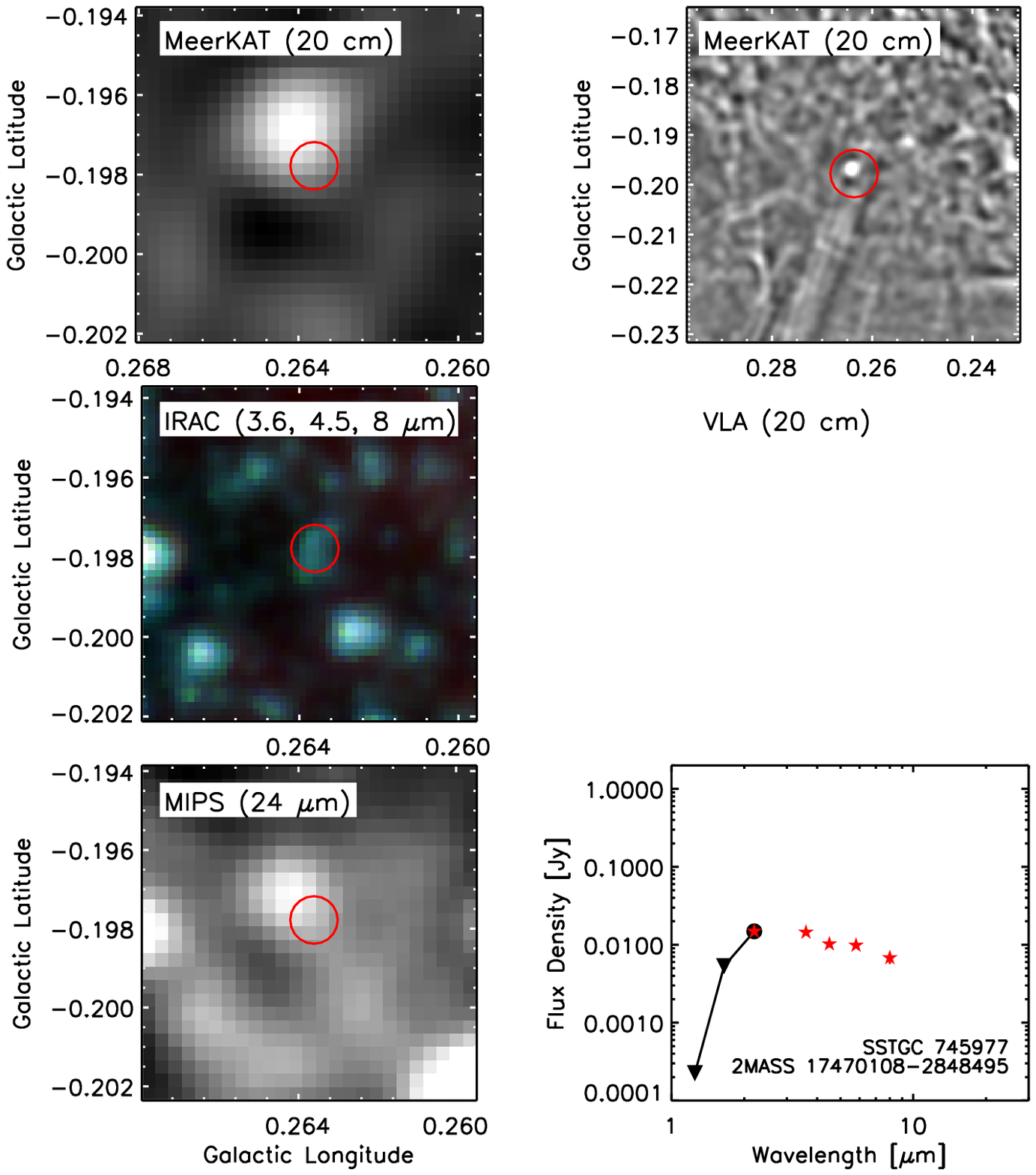}
\caption{
{\it (aaa)}  Same as Fig. 3a except source 53 in Table 1.
}
\end{figure}

\addtocounter{figure}{-1}
\begin{figure}
\center
\includegraphics[]{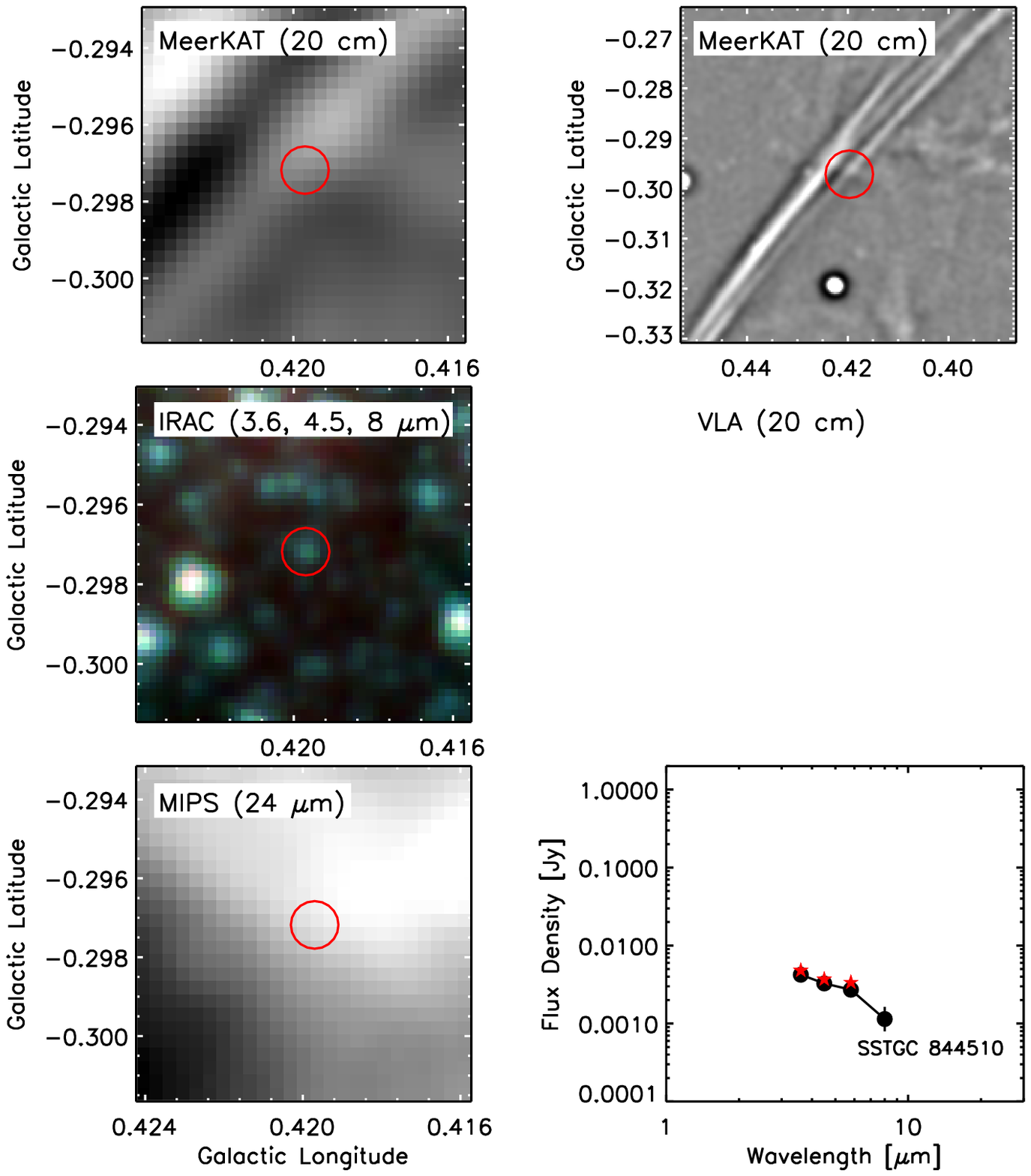}
\caption{
 {\it (bbb)}  Same as Fig. 3a except source 54  in Table 1.
}
\end{figure}

\addtocounter{figure}{-1}
\begin{figure}
\center
\includegraphics[]{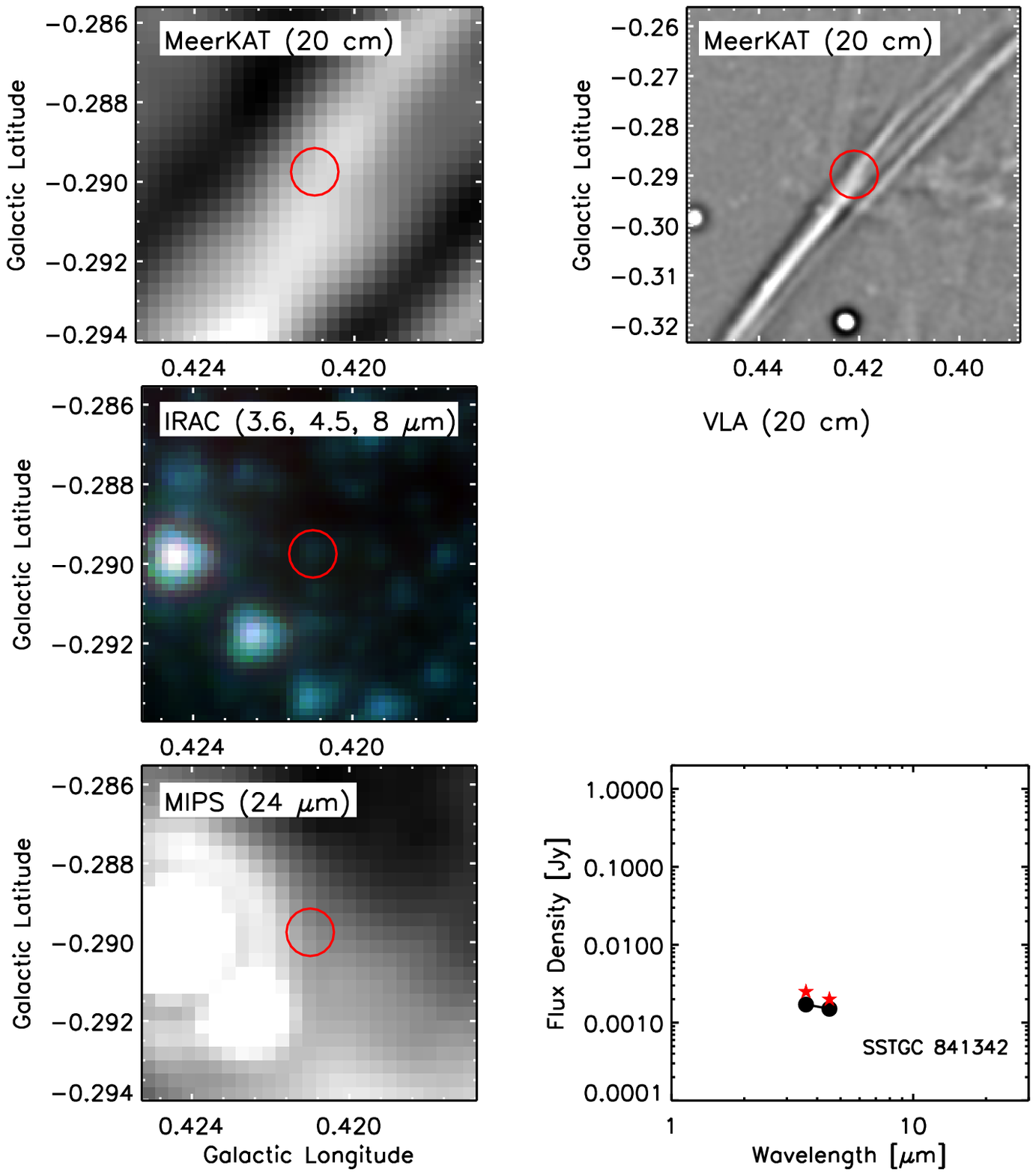}
\caption{
 {\it (ccc)}  Same as Fig. 3a except source 55  in Table 1.
}
\end{figure}

\addtocounter{figure}{-1}
\begin{figure}
\center
\includegraphics[]{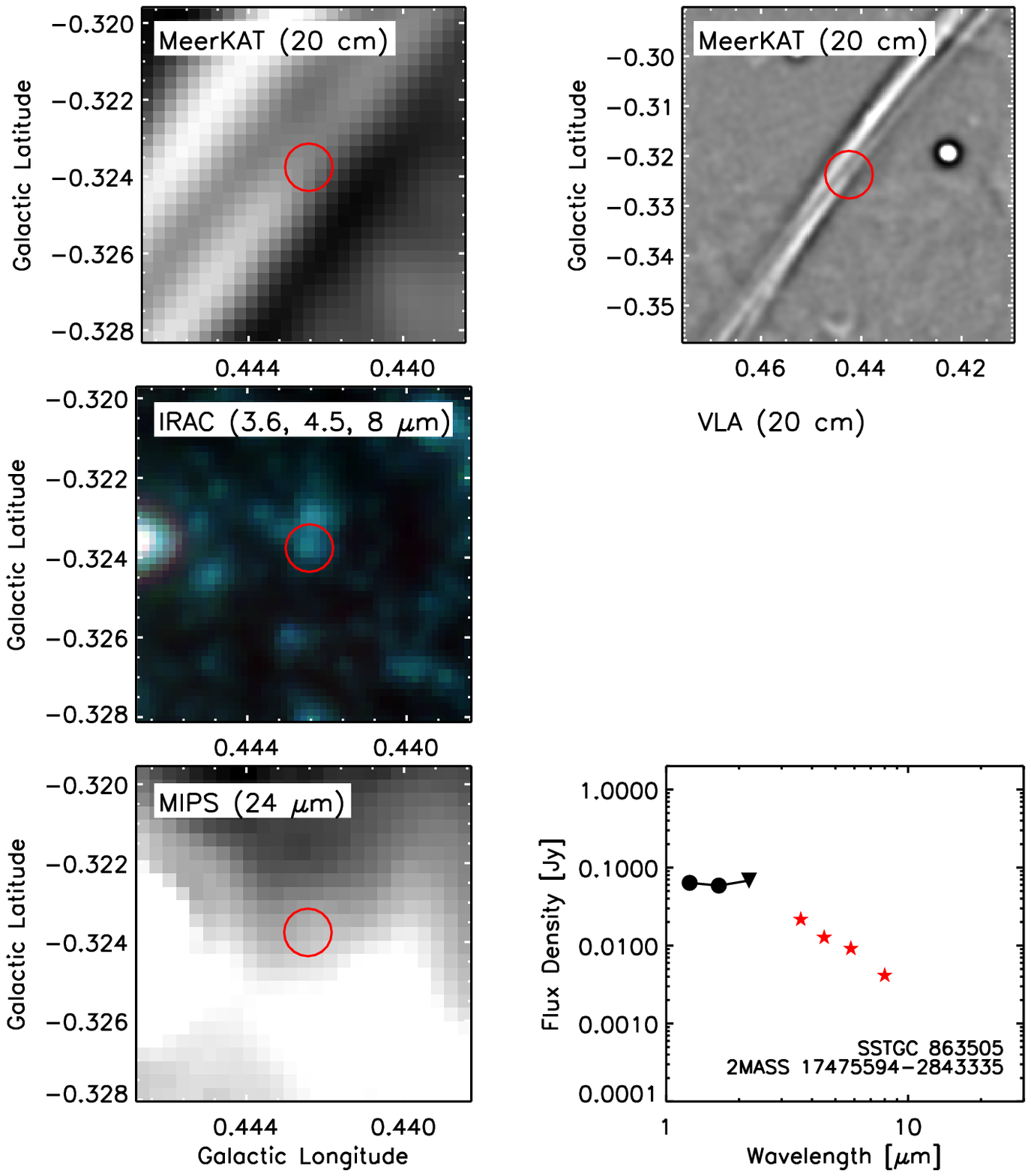}
\caption{
 {\it (ddd)}  Same as Fig. 3a except source 56  in Table 1.
}
\end{figure}

\addtocounter{figure}{-1}
\begin{figure}
\center
\includegraphics[]{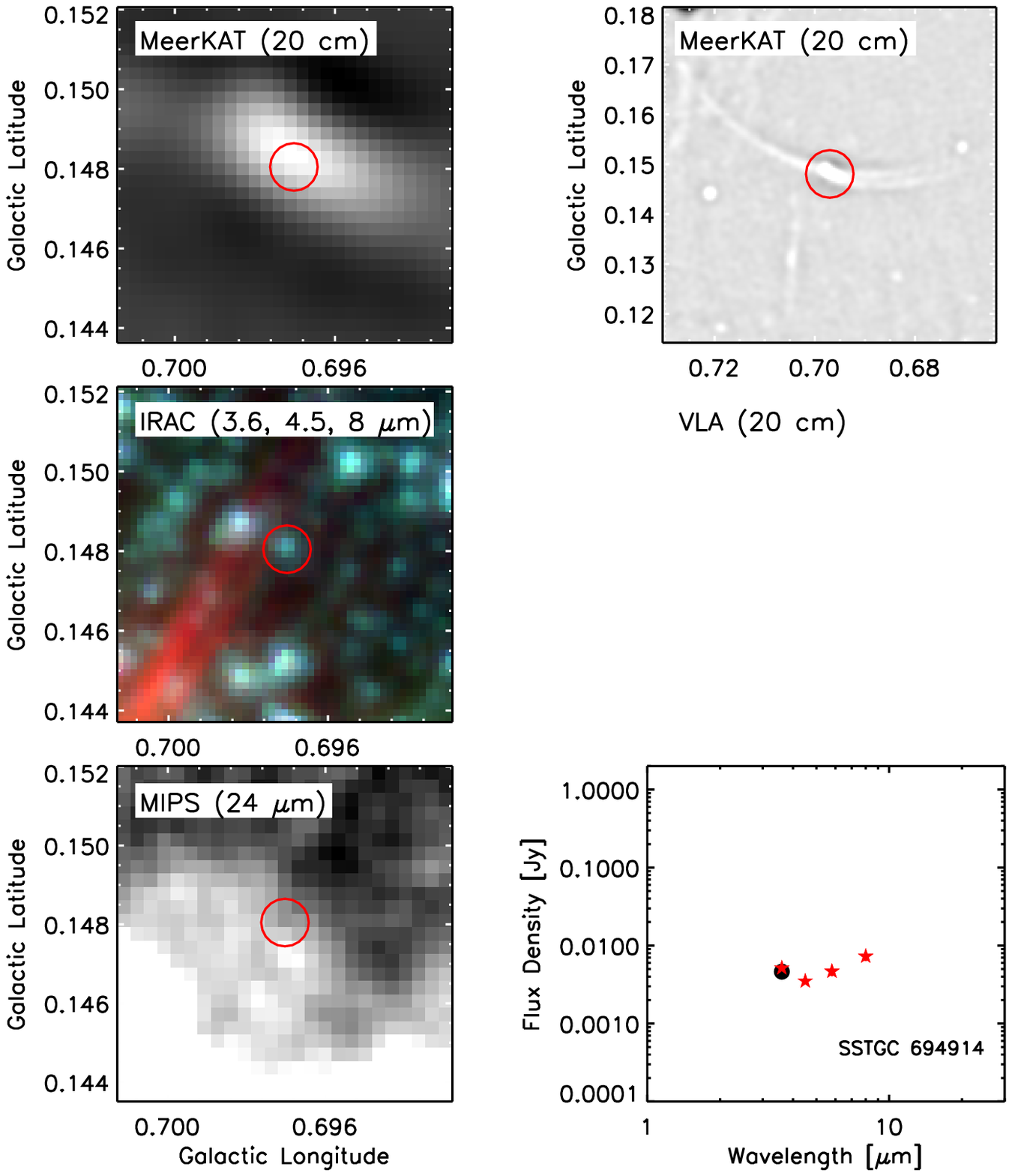}
\caption{
 {\it (eee)}  Same as Fig. 3a except source 57  in Table 1.
}
\end{figure}

\end{document}